 \pdfoutput=1
\documentclass[12pt, letterpaper]{report}
\usepackage{hyperref}   
\usepackage{setspace}
\usepackage{amsmath}
\usepackage{mathrsfs}
\usepackage{verbatim}
\usepackage{graphicx}
\usepackage{epstopdf}
\usepackage{epigraph}
\usepackage{cite}
\usepackage{amsfonts}
\usepackage{amssymb}
\usepackage[]{graphicx}
\usepackage{bm}
\usepackage{stmaryrd}
\usepackage{ragged2e}
\usepackage{booktabs}
\usepackage{caption}
\usepackage{appendix}
\usepackage{subfigure}
\usepackage[subfigure]{tocloft}
\usepackage[utf8]{inputenc}
\usepackage{titlesec}
\usepackage[english]{babel}

\DeclareMathOperator{\sinc}{sinc} 
\DeclareMathOperator{\ceil}{ceil}

\DeclareMathOperator{\PAPR}{PAPR}

\titleformat{\chapter}[display]
  {\normalfont\rmfamily\large\bfseries}	
  {\chaptertitlename\ \thechapter}{14pt}{\large}
\titlespacing*{\chapter}{0in}{-0.25in}{0.25in}
\titlespacing*{\section}{0in}{0.25in}{0.25in}
\titlespacing*{\subsection}{0in}{0.25in}{0.25in} 
        
\titleformat{\section}{\rmfamily\large\bfseries}{\thesection}{1em}{}

\author{David A. Hague}
\title{PhD. Dissertation}
\raggedright
\begin{document}


\setcounter{page}{1}
\pagenumbering{roman}

\thispagestyle{empty}

\begin{center}
University of Massachusetts Dartmouth \\
Department of Electrical and Computer Engineering \\

\vspace*{1.0in}
{\sc The Generalized Sinusoidal Frequency Modulated Waveform for Active Sonar Systems} \\

\vspace*{1.0in}
A Dissertation in 

Electrical Engineering 

by

David A. Hague 

\vspace*{1.0in}
Submitted in Partial Fulfillment of the 

Requirements for the Degree of \\ Doctor of Philosophy

\vspace*{1.0in}
August 2015

\end{center}
\newpage

\thispagestyle{empty} 
I grant the University of Massachusetts Dartmouth the non-exclusive right to use the work for the purpose of making single copies of the work available to the public on a not-for-profit basis if the University's circulating copy is lost or destroyed.

\vskip 2em
\begin{flushright}

\rule{0.395\textwidth}{0.5pt} \\
\vspace*{-1em}
David A. Hague\hspace*{1.625in} 

Date \rule{0.395\textwidth}{0.5pt} 
\end{flushright}
\newpage

\setcounter{page}{2}

 \thispagestyle{empty}
  \singlespacing
  \begin{flushleft}
    \normalsize
    We approve the dissertation of David A. Hague \par

    \vskip 0.1in
    \hspace{4.59in}Date of Signature
    
    \vskip 0.2in
    \rule{0.55\textwidth}{0.5pt} \hspace{1.2in} \rule{1.4in}{0.5pt} 
    \\ John R. Buck 
    \\ Professor, Department of Electrical and Computer Engineering 
    \\ Dissertation Advisor \\
    
    \vskip 0.25in
    \rule{0.55\textwidth}{0.5pt} \hspace{1.2in} \rule{1.4in}{0.5pt} 
    \\ David Brown 
    \\ Professor, Department of Electrical and Computer Engineering 
    \\ Dissertation Committee \\
    
    \vskip 0.25in
    \rule{0.55\textwidth}{0.5pt} \hspace{1.2in} \rule{1.4in}{0.5pt} 
    \\ Paul Gendron 
    \\ Assistant Professor, Department of Electrical and Computer         		   Engineering 
    \\ Dissertation Committee \\ 
    
    \vskip 0.25in
    \rule{0.55\textwidth}{0.5pt} \hspace{1.2in} \rule{1.4in}{0.5pt} 			\\Mary H. Johnson
    \\ Branch Head, Code 1511, Naval Undersea Warfare Center, Newport, RI
    \\ Dissertation Committee \\
    
    \vskip 0.25in
    \rule{0.55\textwidth}{0.5pt} \hspace{1.2in} \rule{1.4in}{0.5pt} 			\\ Christ Richmond
    \\ Senior Technical Staff, MIT Lincoln Laboratory, Lexington, MA
    \\ Dissertation Committee \\
    
    \vskip 0.25in
    \rule{0.55\textwidth}{0.5pt} \hspace{1.2in} \rule{1.4in}{0.5pt}
    \\ Antonio H. Costa 
    \\ Chairperson, Department of Electrical and Computer 		   				Engineering
      
    \vskip 0.25in
    \rule{0.55\textwidth}{0.5pt} \hspace{1.2in} \rule{1.4in}{0.5pt} 
    \\ Robert E. Peck 
    \\ Dean, College of Engineering \\ 
    
    \vskip 0.25in
    \rule{0.55\textwidth}{0.5pt} \hspace{1.2in} \rule{1.4in}{0.5pt}
    \\ Tesfay Meressi 
    \\ Associate Provost for Graduate Studies
  \end{flushleft}
\newpage 


\chapter*{Abstract}

The Generalized Sinusoidal Frequency Modulated Waveform for Active Sonar Systems \newline

\onehalfspacing

by David A. Hague \newline

Pulse Compression (PC) active sonar waveforms provide a significant improvement in range resolution over single frequency sinusoidal waveforms also known as Continuous Wave (CW) waveforms.  Since their inception in the 1940's, a wide variety of PC waveforms have been designed using either Frequency Modulation (FM), phase coding, or frequency hopping to suite particular sonar applications.  The Sinusoidal FM (SFM) waveform modulates its Instantaneous Frequency (IF) by a single frequency sinusoid to achieve high Doppler sensitivity which also aids in suppressing reverberation.  This allows the SFM waveform to resolve target velocities.  While the SFM's resolution in range is inversely proportional to its bandwidth, the SFM's Auto-Correlation Function (ACF) contains many large sidelobes. The periodicity of the SFM's IF creates these sidelobes and impairs the SFM's ability to clearly distinguish multiple targets in range.  This dissertation describes a generalization of the SFM waveform, referred to as the Generalized SFM (GSFM) waveform, that modifies the IF to resemble the time/voltage characteristic of a FM chirp waveform.  As a result of this modification, the Doppler sensitivity of the SFM is preserved while substantially reducing the high range sidelobes producing a waveform whose Ambiguity Function (AF) approaches a thumbtack shape.  This dissertation describes the properties of the GSFM's thumbtack AF shape, compares it to other well known waveforms with a similar AF shape, and additionally considers some of the practical considerations of active sonar systems including transmitting the GSFM on piezoelectric transducers and the GSFM's ability to suppress reverberation.  Lastly, this dissertation also describes designing a family of in-band nearly orthogonal waveforms with potential applications to Continuous Active Sonar (CAS).
\newpage 
\thispagestyle{empty}
\chapter*{Acknowledgments}
First and foremost, I want to thank my family, specifically my parents, David and Catherine Hague, and my brother Alan Hague, for all their love and support.  Their constant encouragement and support over the last six years has been a constant source of motivation especially during the times when I was challenged the most.  I am so lucky to have them all in my life.  
\vspace{1em}

John Buck has been an excellent mentor to me since my senior year at UMass Dartmouth and has had an especially profound impact on me during the last six years as my advisor.  He has provided copious opportunities for me to develop my professional skills inside and outside the classroom and has pushed me to limits far beyond my own expectations.  Taking the plunge and signing up for his Discrete-Time Signal Processing course as a technical elective back in the fall of 2004 has been one of the best decisions of my professional life.
\vspace{1em}

Mary Johnson and Tod Luginbuhl at the Naval Undersea Warfare Center (NUWC) have been terrific mentors to me over the last 5 years.  Mary and Tod's knowledge and professional experience with the Navy has been invaluable to me.  Each summer internship gave me crucial opportunities to learn of the practical applications of the research I've performed at UMass.  I've sincerely enjoyed my time at NUWC each of the last 5 summers and I am looking forward to pursuing new research avenues as a member of NUWC. 
\vspace{1em}

I also want to thank my committee members, David Brown, Paul Gendron, and Christ Richmond for all the helpful suggestions and ideas they contributed to my dissertation.  I have greatly benefited from their expertise and they have all helped to make the technical content of this dissertation stronger.
\vspace{1em}

The Science, Mathematics And Research for Transformation (SMART) Scholarship for Service Program provided the vast majority of my funding during my graduate studies which allowed me to pursue my research interests freely.  I would especially like to thank John Tague and Keith Davidson at the ONR Undersea Signal Processing Program.  John and Keith took an interest in my work very early on in my graduate studies.  In addition to funding my work, they invited me to attend their annual program reviews from 2011 to 2014.  These program reviews gave me the chance to converse with some of the best researchers in my field and to learn about the practical applications of my research, a rare golden opportunity for a graduate student.  Some of the conversations I had at my first meeting in 2011 motivated the development of the GSFM waveform.  I'm excited about continuing our professional relationship and contributing new ideas to their research programs. 
\vspace{1em}
 
I have spent a collective 10 years here at the ECE department at UMass.  During that time, I have interacted with a number of excellent people.  Dayalan Kasilingam and Antonio Costa have both been department chairs during my time here at UMass.  I've had the opportunity to interact with them in and out of the classroom and I deeply appreciate all that they've done for me over the years.  Stephen Nardone, who highly encourages every graduate student in the ECE department to pursue a PhD, aggressively encouraged me as well and heavily influenced my decision to stay and pursue a PhD full time.  Fernanda Botelho has come through for me time and time again for all the logistical tasks that come along with being a graduate student for which I'm deeply appreciative.
\vspace{1em}

I also want to thank my fellow Signal Processing Lab friends, specifically Saurav Tuladhar, Kaushallya Adhikari, Petro Khomchuck, Xiaoli Zhu, Yang Liu, and Ian Rooney for all the great research discussions, laughs, and our always highly anticipated end of semester group lunches.  We all come from diverse backgrounds and I learned so much from all of them.  I will cherish all the times we had together for the rest of my life.
\vspace{1em}

This dissertation is dedicated to the memory of my grandfather, Andrew John Pytel, Sr.  Apart from the special bond of being the oldest of his five grandsons, he had a huge influence on the man I've become.  He saw potential in me, held me to a high standard in most everything that I did, and had a deep respect for me and all of my accomplishments, an honor not easily earned.  When I was deciding to leave a full time job and pursue graduate study in the middle of a difficult recession, he was one of my strongest supporters.  While he never deeply understood my research or the topic of my dissertation, he would be extremely proud that I took on this challenge and saw it through to the very end.  

\vspace{7em}
\emph{We make our world significant by the courage of our questions and the depth of our answers.} \\
\begin{flushright}
\emph{Carl Sagan}
\end{flushright}
\newpage 

\setcounter{tocdepth}{2}
\renewcommand{\contentsname}{Table of Contents}
\renewcommand{\cftbeforechapskip}{\baselineskip} 
\renewcommand{\cfttoctitlefont}{\large\bfseries}
\setlength{\cftbeforetoctitleskip}{-0.2in}
\setlength{\cftaftertoctitleskip}{0.25in}
\renewcommand{\cftchapindent}{0pt}

\renewcommand\cftchappresnum{\chaptername~}
\renewcommand\cftchapaftersnum{~:}
\renewcommand{\cftchapaftersnumb}{} 
\renewcommand{\cftsecleader}{\cftdotfill{\cftsecdotsep}}                  
\renewcommand{\cftchapnumwidth}{1.15in}

\tableofcontents
\newpage 

\addcontentsline{toc}{chapter}{List of Figures}
\renewcommand{\cftloftitlefont}{\large\bfseries}
\setlength{\cftbeforeloftitleskip}{-0.2in}
\setlength{\cftafterloftitleskip}{0.5in}
\renewcommand{\cftfigindent}{0pt}

\listoffigures
\newpage

\addcontentsline{toc}{chapter}{List of Tables}
\renewcommand{\cftlottitlefont}{\large\bfseries}
\setlength{\cftbeforelottitleskip}{-0.2in}
\setlength{\cftafterlottitleskip}{0.5in}
\renewcommand{\cfttabindent}{0pt}

\listoftables
\newpage


\thispagestyle{plain}
\setcounter{page}{1}
\pagenumbering{arabic}

\chapter{Introduction} 
\label{ch:Intro}
This dissertation introduces and evaluates the Generalized Sinusoidal Frequency Modulated waveform GSFM, a novel FM transmit waveform for active sonar.  The GSFM waveform is a modification of the Sinusoidal FM (SFM) waveform which modulates its Instantaneous Frequency (IF) with a sinusoidal function.  The SFM, while Doppler sensitive, contains many high sidelobes in its Auto Correlation Function (ACF), a direct result of the periodicity in the SFM's IF.  The GSFM utilizes an IF that is aperiodic and therefore possesses lower sidelobes in its ACF.  The GSFM waveform possesses a thumbtack Ambiguity Function (AF) allowing for jointly resolving target range and velocity.  The GSFM's AF performance is competitive with other well established thumbtack waveforms.  The GSFM waveform also possesses a constant envelope resulting in a low Peak-to-Average Power Ratio (PAPR) and concentrates the majority of its energy in a tighter band of frequencies than other thumbtack waveforms.  These two properties are very important design considerations when transmitting waveform on piezoelectric transducers.  Lastly, the GSFM has a family of waveforms that are generated using reflections in time and frequency as well as symmetry properties of the GSFM's IF.   This family of waveforms achieve low cross-correlation properties even when occupying the same band of frequencies which can be utilized in Continuous Active Sonar (CAS) systems.
\vspace{1em}  

Sonar systems detect and resolve closely spaced targets in the midst of reverberation and noise by transmitting an acoustic signal and extracting information from the resulting echoes from objects in the medium.  In order to resolve closely spaced objects in range, the acoustic signal, also known as the transmit waveform, must have large bandwidth.  To maximize detection of objects in white Gaussian noise, the waveform should possess high energy.  The amplifiers driving the transducers in a sonar system are peak power limited and operating beyond this peak power limit can either damage the device or drive the amplifier to operate non-linearly  therefore distorting the transmitted waveform.  Additionally, sonar systems cannot transmit at arbitrarily high source levels.  Too high a source level induces cavitation on the head of the sonar's transducers.  These physical constraints are typically countered by transmitting a long duration waveform at a lower source level to provide the necessary detection energy.  Single frequency sinusoidal waveforms, also known as Continuous Wave (CW) waveforms, cannot achieve both high bandwidth and high energy simultaneously.  The CW waveform's bandwidth is inversely proportional to its pulse length.  A longer pulse length will possess more energy but results in less bandwidth and vice versa.  Pulse Compression (PC) waveforms utilize amplitude, phase, or frequency modulation in order to attain large bandwidth in addition to long pulse lengths.  Perhaps the most popular PC waveform is the Linear Frequency Modulated (LFM) waveform, developed in the advent of World War II \cite{Cook}.  By linearly sweeping through a band of frequencies, the LFM achieves the long duration necessary for sufficient energy to detect targets while also providing the large bandwidth required for resolving closely spaced objects. 
\vspace{1em}

Almost every radar and sonar system implements a Matched Filter (MF) or correlation receiver for processing echoes \cite{Cook, Ricker}.  The MF is the ideal detector for the case of a known signal embedded in Additive White Gaussian Noise (AWGN) \cite{Kay}.  The MF's impulse response is the time-reversed complex conjugate of the transmit waveform.  If there is no relative motion between the target and the sonar system platform, then the MF is exactly matched to the resulting echo signal.  However, if the target moves relative to the platform, then the return echo undergoes a Doppler effect.  The Broadband Doppler effect commonly encountered in sonar systems compresses or expands the echo in time.  The time compression of the echoes from moving targets introduces mismatch between the echoes and a MF designed for a stationary target.  This mismatch results in a loss in output SNR and therefore reduced detection performance.  The Ambiguity Function (AF) first proposed by Woodward \cite{Woodward} and then generalized for broadband signals by Kelly and Wishner\cite{KellyWish} and Swick \cite{Swick}, quantifies the mismatch of the MF with constant velocity Doppler scaled echoes and is known as the Broadband Auto Ambiguity Function (BAAF).  Waveforms such as the Hyperbolic FM (HFM) \cite{Jan} experience little SNR loss at their MF’s output due to Doppler scaling and are known as Doppler tolerant.  These waveforms provide high range resolution regardless of the target's velocity.   Waveforms such as the CW that experience substantial SNR loss at their MF's output are known as Doppler sensitive.  There is also a subclass of Doppler sensitive waveforms that can also resolve target range unlike the CW.  These waveforms have an AF shape that has a mainlobe located at the origin whose width in range and velocity are inversely proportional to the waveform's bandwidth and pulse length respectively.  These waveforms are known as Thumbtack waveforms due to their AF shape resembling a thumbtack.
\vspace{1em}

In addition to determining the transmit waveform's MF response to target echoes in target range and velocity, the BAAF also provides an approximate measure of a waveform's ability to suppress reverberation.  Reverberation refers to the unwanted echoes resulting from bubbles, fish, the sea surface/bottom, and any other acoustic scatterers present in the medium.  Assuming that the acoustic scatterers in the environment are stationary relative to the sonar system platform, uniformly distributed in range, and of equal target strength, the response of the waveform's MF to reverberation simplifies to the Q-function.    The Q-function is the integral over time of the squared magnitude of the BAAF and is therefore a function of Doppler.  While realistic sonar environments will have scatterers that are spread in Doppler and non-uniformly distributed in range, the Q-function provides a first order approximation to the level of reverberation suppression a waveform is capable of achieving and allows for a comparison between waveforms that is relatively easy to compute.  
\vspace{1em}

Designing a waveform with a particular AF shape has been studied for over 60 years and is still an open problem.  Refs \cite{Siebert,Wilcox,Sussman} developed a Least Squares approach for designing a waveform with a specified AF.  Later work \cite{Gladkova} expanded upon the Least Squares approach using numerical optimization techniques.  Cook and Bernfield \cite{Cook} suggest an approach intended for the practicing engineer to evaluate the AF amongst other criterion for a collection of potential waveforms and choose the one that best meets their application.  Rihaczek \cite{Rihaczek} also considered the waveform design problem and commented that ``waveform synthesis is commonly done by trial and judicious use of available information, often guided by intuition.  Over the years, a store of information on waveforms and their ambiguity functions has been accumulated.  The designer attempts to select the waveform whose ambiguity surface appears to be best suited for the target environment, using skill and ingenuity in developing modifications leading to ambiguity functions still better suited''  This dissertation embraces Rihaczek's approach to waveform design.  
\vspace{1em}

Sonar waveform design does not focus solely on BAAF and Q-function shape.  There are many practical issues when considering transmitting waveforms on piezoelectric transducers, the most common transmit and receive devices employed by active sonar systems.  From the waveform designer's perspective, the transducer's frequency response is the most important performance measure of the transducer to consider.  Each transducer, whether operating as a projector (transmitter) or receiver, is a resonant device whose frequency response drops off steadily beyond resonance.  The phase of the transducer's frequency response phase is a non-linear function of frequency.  Therefore, the resulting group-delay of the transducer's frequency response is not a constant function of frequency.  When an FM waveform is transmitted or received by a transducer, each frequency component of the waveform further off resonance is attenuated in amplitude and shifted in phase.   The resulting FM acoustic signal transmitted or received by a transducer therefore contains Amplitude Modulation (AM) and Phase Modulation (PM).
\vspace{1em}

Typically, a sonar system utilizing PC waveforms will operate in a band of frequencies centered at resonance to maximize the source level of the transmitted acoustic signal and minimize the AM and FM distortions resulting from the device's frequency response.  Additionally, most sonar receivers will apply a bandpass filter to the return echo signal to remove out of band noise before passing the signal data on to the MF receiver.  It is therefore optimal to design a waveform that contains all or the vast majority of it's energy  in the operational band of frequencies.  Waveforms are typically tapered in time to reduce their spectral leakage, the energy outside the operational band of frequencies of the transducer and driving electronics.  Tapering is commonly applied to phase or frequency coded waveforms that are composed of a train of sub-pulses or chips.  However, tapering the waveform comes at a price.  The electronics driving the transducer are peak power limited and operating beyond this peak power limit either damages the electronics or introduces nonlinear distortions in the transmitted acoustic signal.  In the interest of maximizing the source level of the transmitted acoustic signal, the waveform's average power should be as close as possible to the it's peak power.  In other words, a waveform should possess a low Peak to Average Power Ratio (PAPR).  Tapering waveforms in time increases the PAPR and typically presents the waveform designer with a design tradeoff between spectral containment and PAPR. 

\section{Dissertation Contribution}
\label{sec:DissCont}
A FM waveform of particular recent interest in active sonar is the Sinusoidal FM (SFM).  The SFM is modulated by a sinusoidal function.  The SFM has found extensive use in radar \cite{Levanon} and was first proposed as a sonar waveform in the published literature by Collin and Atkins \cite{Collins}.  The SFM has been shown to resolve target velocities and possess desirable reverberation suppression performance in both theoretical and experimental settings \cite{Collins, Ward}.  However, the SFM has poor range resolution as the Auto-Correlation Function (ACF) contains many high sidelobes due to the periodicity of it's IF.  The poor range resolution for the SFM waveform is reminiscent of the poor range resolution of a CW pulse.  This undesirable property of the CW waveform is due to the periodicity of its time/voltage characteristic and motivated the design of chirp FM waveforms like the Linear FM (LFM) that maintain the same energy while also attaining high range resolution.  This suggests that applying an analogous approach in the IF domain, converting the sinusoidal IF of the SFM to some chirp IF waveform will provide similar mitigation of periodic sidelobes in time while preserving the desirable range resolution and Doppler sensitivity of the SFM waveform.  This work investigates an active sonar waveform whose IF versus time function resembles the voltage versus time function of a chirp waveform.  The proposed new waveform displays many desirable properties including target resolution in range and velocity that is competitive with the performance of other well known waveforms that attain a thumbtack AF.  

\section{Dissertation Outline}
\label{sec:DissOut}	
The rest of this dissertation is organized as follows: Chapter \ref{ch:signalModel} describes the waveform signal model, the AF, and reviews some commonly used transmit waveforms including the SFM.  Chapter \ref{ch:GSFM} describes the GSFM waveform and its main properties.  Chapter \ref{ch:GSFM_Eval_AF} evaluates the performance of the GSFM's AF and compares it's performance to that of other well known waveforms that attain a thumbtack AF.  Chapter \ref{ch:GSFM_Eval_Prac} explores the GSFM reverberation suppression performance and the practical considerations for transmitting the GSFM on piezoelectric transducers.  Chapter \ref{ch:GSFM_CAS} describes generating a family of GSFM's that occupy the same band of frequencies while maintaining low-cross correlation properties and using this family of GSFM waveforms for Continuous Active Sonar (CAS) applications. Finally, Chapter \ref{ch:Conclusion} presents the conclusions.

\chapter{Waveform Signal Model}
\label{ch:signalModel}
In order to evaluate and compare the performance of transmit waveforms, it is necessary to understand not only their signal model but also the main metric of performance comparison, the BAAF.  Unless otherwise specified, this dissertation assumes the sonar system is monostatic (i.e, the transmitter and receiver are co-located). The target of interest is assumed to be a point target undergoing constant velocity motion.  These assumptions greatly simplify analysis of the waveforms and can be extended to more complicated models as design criteria dictate.

\section{Transmit Waveform Model}
\label{sec:waveformModel}
The transmit waveform signal $s\left(t\right)$ is modeled as a complex analytic signal with pulse length $T$ defined either over the interval $0 \leq t \leq T$ or $-T/2 \leq t \leq T/2$ expressed as
\begin{equation}
s\left(t\right) = a\left(t\right)e^{j\phi\left(t\right)} = a\left(t\right)e^{j\varphi\left(t\right)}e^{j2\pi f_c t}
\label{eq:ComplexExpo}
\end{equation}  
where $f_c$ is the carrier frequency, $\phi\left(t\right)$ is the instantaneous phase of the waveform, $\varphi\left(t\right)$ is the phase modulation function of the waveform, and $a\left(t\right)$ is an amplitude tapering functions.  Unless otherwise specified, the amplitude tapering function $a\left(t\right)$ is assumed to be a rectangular function with amplitude $1/\sqrt{T}$ which normalizes the waveform to unit energy.  Utilizing a rectangular taper function results in a waveform whose spectrum does not possess any AM contributions and is solely determined by the modulation function and carrier term.  The IF function of the rectangular tapered waveform is expressed as 
\begin{equation}
f\left(t\right) = \dfrac{1}{2 \pi}\dfrac{\partial \phi \left( t\right)}{\partial t} = \dfrac{1}{2 \pi}\dfrac{\partial \varphi \left( t\right)}{\partial t} + f_c
\end{equation}  
The signal that is transmitted on a transducer is the real component of the complex analytic signal
\begin{eqnarray}
x\left(t\right) = \Re\{s\left(t\right)\} = a\left(t\right)\cos\left(\varphi\left(t\right) + 2\pi f_c t\right)
\label{realSignal}
\end{eqnarray}
The Fourier transform of $s\left(t\right)$, denoted as $S\left(f\right)$, is the right sided version of $X\left(f\right)$, the Fourier transform of $x\left(t\right)$.
While the true signal that is transmitted into the medium is the real valued sinusoid $x\left(t\right)$, the complex analytic model is used throughout this work for two reasons.  First, it is mathematically more convenient to analyze waveform performance as complex functions from which the real signals are derived \cite{Ricker}.  Secondly, many practical sonar systems use IQ modulation when processing echo signals and so the resulting format of the echo signal data is complex valued. 
\vspace{1em}

Two important measures of the waveform when transmitting the waveform on a transducer are Spectral Containment (SC) and PAPR.  For FM waveforms, Carson's bandwidth rule \cite{Couch} states that 98$\%$ of a FM waveform's energy resides in a bandwidth $B$ expressed as $B = 2\left(\Delta f/2 + B_m\right)$ where $\Delta f$ is the peak frequency deviation of the waveform (i.e, swept bandwidth) and $B_m$ is the highest frequency component of the waveform's IF function.  Similar rules exist for Frequency Shift Keying (FSK) and Phase Coded (PHC) waveforms \cite{Couch}.  In order to provide a quantitative measure of SC as means of comparison against different waveforms, this paper defines the SC $\psi$ of a transmit waveform as the ratio of waveform energy in a specific band of frequencies $\Delta F$ to the total energy (here, assumed to be unity) of the waveform across all frequencies expressed as 
\begin{equation}
\psi\left(\Delta F\right) = \dfrac{\int_{-\Delta F/2}^{\Delta F/2}|S\left(f\right)|^2df}{\int_{-\infty}^{\infty}|S\left(f\right)|^2df} = \int_{-\Delta F/2}^{\Delta F/2}|S\left(f\right)|^2df
\label{eq:psi}
\end{equation}
The waveform's PAPR measures the ratio of the peak power of the transmitted acoustic signal $x\left(t\right)$ to it's average power expressed in dB as 
\begin{equation}
\PAPR = 10\log_{10}\Biggl\{\dfrac{\max_t\{|x\left(t\right)|^2\}}{\frac{1}{T}\int_0^T|x\left(t\right)|^2dt}\Biggr\}
\label{PAPR}
\end{equation}
For a given peak power limit, the PAPR is a measure of the waveform's average power.  For waveforms with the same duration $T$, the PAPR provides a measure of the total energy in the waveform.  A low PAPR translates to a high average power and therefore high total energy.  Increasing the PAPR therefore reduces the total energy of the waveform.  An optimal PAPR would be 0 dB from a DC pulse, however active sonar systems transmit sinusoidal waveforms.  Rectangular windowed CW and FM waveforms possess a PAPR of 3.0 dB.  Any tapering of the waveform that might be employed to improve the SC will also increase the PAPR introducing a tradeoff between SC and PAPR.

\section{The Ambiguity Function}
\label{sec:AF}
The most common receiver employed in sonar systems is the Matched Filter (MF), or correlation receiver, as it is the optimal receiver for signal detection in the presence of AWGN \cite{Kay}.  The impulse response of this filter is the time-reversed complex conjugate of the transmit waveform.  Convolving the return signal with the impulse response of the MF is equivalent to correlating the return signal and transmit waveform.  When the target is stationary relative to the sonar platform, the MF is matched exactly to the echo signal which in turn maximizes the output SNR and therefore detection performance.  However, targets undergoing motion relative to the sonar transmitter and receiver introduce a Doppler effect to the echo signal.  The Doppler effect compresses or expands the signal in the time domain when the target is closing or receding respectively.   The constant velocity Doppler scaling factor is expressed as \cite{Ricker, Lin}
\begin{equation}
\eta \cong \dfrac{1+v/c}{1-v/c}
\label{eq:eta}
\end{equation} 
where $v$ is the relative velocity or range rate of the target and $c$ is the speed of sound in the medium.  The Broadband Auto-AF (BAAF) measures the response of the MF to a single echo and is defined as \cite{Ricker}
\begin{equation}
\chi\left(\tau_0, \eta_0, \tau, \eta\right) = \sqrt{\eta_0\eta} \int_{-\infty}^{\infty}s^*\left(\eta_0\left(t-\tau_0\right)\right)s\left(\eta \left(t+\tau \right) \right) dt
\label{eq:BAAF_0}
\end{equation} 
where $\tau_0$ and $\eta_0$ are the hypothesized time-delay and Doppler scaling factor of the echo and $\tau$ and $\eta$ are the true time-delay and Doppler scaling factor of the echo.  
\vspace{1em}

The magnitude-squared of the BAAF is a function of time-delay $\tau$ and Doppler scaling factor $\eta$.  The peak of the BAAF is unity for waveforms normalized to unit energy and occurs when $\tau = \tau_0$ and $\eta = \eta_0$.  This means that the MF is maximally correlated with the echo when the MF's time-delay and Doppler scaling factor equal that of the echo.  Therefore, in addition to the MF being the optimal detector for known signal in AWGN, the MF also provides an estimate of the echo time-delay and Doppler scaling factor (target velocity). Without loss of generality, the BAAF can be simplified by setting $\tau_0 = 0$ and $\eta_0 = 1.0$ \cite{Ricker} which simplifies \eqref{eq:BAAF_0} to 
\begin{equation}
\chi\left(\tau, \eta\right) = \sqrt{\eta} \int_{-\infty}^{\infty}s\left(t\right)s^*\left(\eta \left(t+\tau \right) \right) dt.
\label{BAAF}
\end{equation}
The expression in (\ref{BAAF}) simply shifts the peak response to $\left(\tau = 0, \eta = 1.0\right)$  and is the standard BAAF expression encountered in the literature \cite{Lin}.  The BAAF can be further simplified to a narrowband model assuming the target velocity is much lower than the speed of the medium and that the waveform's fractional bandwidth, $B/2f_c$ is very low (i.e. $\leq 1/10$) which means that the signal can be well approximated as narrowband.  The Doppler Effect for a narrowband waveform is a shift in frequency known as a Doppler shift given by \cite{Ricker, Rihaczek}
\begin{equation}
\phi = \left(2v/c\right)f_c.
\end{equation}
The Narrowband Auto-AF (NAAF) is then expressed as \cite{Ricker, Rihaczek, Cook, Levanon}
\begin{equation}
\chi \left(\tau, \phi\right) = \int_{-\infty}^{\infty} s\left(t\right)s^*\left(t+\tau\right) e^{j2\pi \phi t}dt.
\label{NAAF}
\end{equation}
The NAAF is useful in waveform design problems mainly because some sonar systems and many radar systems transmit narrowband waveforms.  Additionally, the NAAF is closely related to the Fourier Transform and Wigner Ville Distribution \cite{Cohen} and shares many of their properties. This greatly simplifies deriving exact closed form expressions for a waveform's NAAF.  Deriving exact closed form expressions for the BAAF is typically more difficult than for the NAAF \cite{Lin}.   It is important to note that there are minor differences in terminology of the AF from a wide variety of sources \cite{Cook, Ricker, Rihaczek, Levanon, Richards}.  Many references define the AF as $|\chi\left(\tau, \eta\right)|^2$ and refer to either $\chi\left(\tau, \eta\right)$ or $|\chi\left(\tau, \eta\right)|$ as the uncertainty function \cite{Ricker}.  Other references \cite{Rihaczek} however will call all three relations the AF.  In the interest of simplicity, this dissertation adopts the terminology used by \cite{Rihaczek} which applies the AF term to all three relations while specifying whether the AF is of the broadband or narrowband variety. 
\vspace{5em}

The BAAF can be generalized to the cross-correlation between one waveform $s_1\left(t\right)$ and the Doppler scaled or shifted echoes of another waveform $s_2\left(t\right)$ known as the Broadband Cross AF (BCAF) and Narrowband Cross AF (NCAF) expressed as 
\begin{equation}
\chi_{1,2}\left(\tau, \eta\right) = \sqrt{\eta}\int_{-\infty}^{\infty}s_1\left(t\right)s_2^*\left(\eta\left(t + \tau\right)\right)dt
\label{eq:BCAF_I}
\end{equation}
\begin{equation}
\chi_{1,2}\left(\tau, \phi\right) = \int_{-\infty}^{\infty}s_1\left(t\right)s_2^*\left(t + \tau\right)e^{j2\pi \phi t}dt
\label{eq:NCAF_I}
\end{equation}
which becomes the BAAF/NAAF when $s_1\left(t\right) = s_2\left(t\right)$.  The BCAF/NCAF is useful for analyzing the cross-talk between transmit waveforms of sonar systems that may be operating in the same environment.  Another interpretation relevant to this work is that the CAF measures the cross correlation between a transmit waveform and its Mis-Matched Filter (MMF).  An MMF is a detection filter that is not matched to the transmit waveform.  MMF's are employed to reduce the peak sidelobe levels of a waveform's CAF in exchange for reduced output SNR and a widened mainlobe.  For FM waveforms, MMF's are typically implemented by tapering the waveform in frequency and time to reduce the range and Doppler sidelobes respectively \cite{Cook}. 
\vspace{1em} 

An echo whose Doppler scale does not match with the MF's Doppler scale results in a SNR loss at the output of the MF.  A loss in output SNR results in a reduction in detection performance.  The amount of SNR loss depends upon the transmit waveform and how it responds to the Doppler Effect.  Sonar waveforms fall under two broad categories concerning the Doppler effect.  Waveforms which possess a small SNR loss at the output of their MF from Doppler scaling are known as Doppler tolerant.  Waveforms that experience substantial MF output SNR loss are Doppler sensitive.  Doppler tolerant waveforms simplify system implementation as only one MF is required to process all Doppler scaled echoes with minimal reduction in output SNR and therefore minimal reduction in detection performance.  Doppler sensitive waveforms are well suited to target velocity estimation.  Target velocity estimation is implemented with a bank of MF's with each MF being tuned to a particular Doppler scale factor.  The MF that is the best match to the Doppler scaled echo will generate the strongest correlation to the echo.  As a result this best matched MF response will have the largest output.  The Doppler scaling factor for that MF is then taken as the estimate of the target's Doppler scaling factor and therefore velocity.  
\vspace{1em}

If the waveform designer wishes to resolve multiple echoes in range and velocity they should choose a waveform whose AF is ideally a delta function centered at the origin with zero energy in the remainder of the range-Doppler plane.  This results in infinite resolution in both range and velocity.  Such an AF shape is a theoretical idealization and not a realizable AF shape for finite duration and bandwidth waveforms.  However, waveforms can closely approximate the ideal AF.  These waveforms attain an approximation of the ideal AF possessing a mainlobe whose width in range and velocity is inversely proportional to the bandwidth and pulse length respectively. The rest of the AF's volume is spread as uniformly as possible in the range-velocity plane \cite{Cook, Klauder}.  A waveform with a thumbtack AF can estimate and resolve target velocity like a CW or SFM waveform but also has the added benefit of providing high range resolution which a CW waveform cannot achieve.   
   
\section{Reverberation Suppression and the Q-function}
\label{sec:qFunc}
The MF is optimal for detecting targets in the presence of AWGN and is the standard detector for noise limited conditions.  Increasing the energy of the transmitted pulse will improve the output SNR of the MF and therefore detection performance. However, the majority of active sonar systems operate in reverberation limited conditions.  Reverberation refers to the unwanted echoes resulting from bubbles, fish, the sea surface and bottom, and any other acoustic scatterers present in the medium \cite{Urick}.  Assuming that the acoustic scatterers in the environment are stationary relative to the sonar system platform, uniformly distributed in range, and of equal target strength, the response of the waveform's MF to reverberation quantified by the Q-function \cite{Cook, Zabal} expressed as 
\begin{eqnarray}
Q\left(\eta\right) = \int_{-\infty}^{\infty}|\chi\left(\tau, \eta\right)|^2 d\tau.
\label{eq:QFunc}
\end{eqnarray}
Note that the Q-function described here should not be confused with the cumulative distribution function of a Gaussian random variable which is also referred to as the Q-function.  Rather, the Q-function in (\ref{eq:QFunc}) evaluates the total energy from reverberation for a particular Doppler scaling factor and is used to compare reverberation suppression performance between various waveforms. As with BAAF shape, different waveforms possess different Q-function shapes that a system designer can use to assess waveform performance.  The CW waveform's Q-function possesses a high peak at zero Doppler but drops off steadily with increasing Doppler.  Doppler tolerant and thumbtack waveforms possess a nearly uniform Q-function across Doppler whose height is inversely proportional to the waveform's time-bandwidth product $TBP$ \cite{Collins}.  Comb waveforms with their "bed of nails" BAAF possess a Q-function that has high peaks in Doppler at the locations of the BAAF's grating lobes and deep valleys between these peaks.  This Q-function shape makes such waveforms ideal for suppressing reverberation over a broad range of Doppler values \cite{Cook, Rihaczek, PecknoldI}.

\section{Commonly Employed Transmit Waveforms}
\label{sec:waveforms}
This section describes several well known transmit waveforms and their ambiguity functions.  This section also introduces the SFM waveform and describes its performance in detail.

\subsection{The Continuous Wave (CW) Waveform}
\label{subsec:CW}
\begin{figure}[h]
\centering
\includegraphics[width=1.0\textwidth]{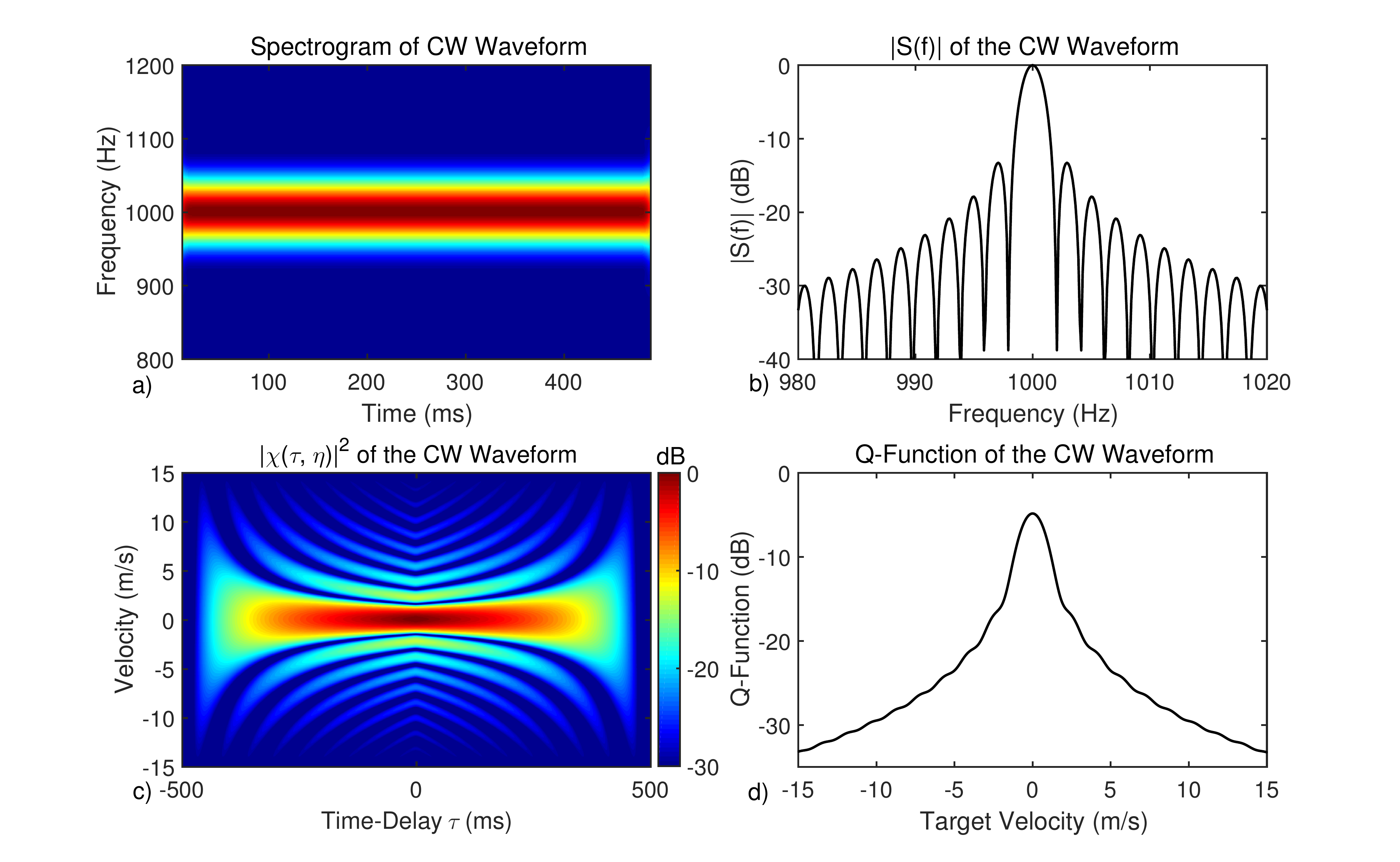}
\caption[Spectrogram (a), Spectrum (b), BAAF (c), and Q-function (d) of a CW waveform with duration $T = 500$ ms and $f_c$ = 1000 Hz.]{Spectrogram (a), Spectrum (b), BAAF (c), and Q-function (d) of a CW waveform with duration $T = 500$ ms  and $f_c$ = 1000 Hz.  The CW waveform is Doppler sensitive but possesses poor range resolution while achieving increasing reverberation suppression with increasing target velocity (Doppler).}
\label{fig:CW}
\end{figure}

The CW is simply a constant frequency sinusoid with amplitude tapering function $a\left(t\right)$ expressed as 
\begin{equation}
s_{CW}\left(t\right) = a\left(t\right)e^{j 2 \pi f_c t}.
\end{equation}
Figure \ref{fig:CW} shows the spectrogram, spectrum, BAAF, and Q-function of the CW waveform.  Of particular interest is the CW's BAAF and Q-function.  The CW's AF has the shape of a triangular function in time-delay (target range) and a sinc function in Doppler (target velocity).  As a result of this AF shape, the CW possesses poor range resolution but high Doppler resolution.  The Q-function shape, a direct result of the AF shape, drops off steadily in Doppler meaning that the CW waveform is better at suppressing reverberation at higher Doppler values.  The CW is typically employed for resolving multiple targets in velocity and suppressing reverberation \cite{Baggenstoss, PecknoldI}, but possesses poor range resolution.  

\vspace{10em}
\subsection{The Linear FM (LFM) Waveform}
\label{subsec:LFM}
\begin{figure}[h]
\centering
\includegraphics[width=1.0\textwidth]{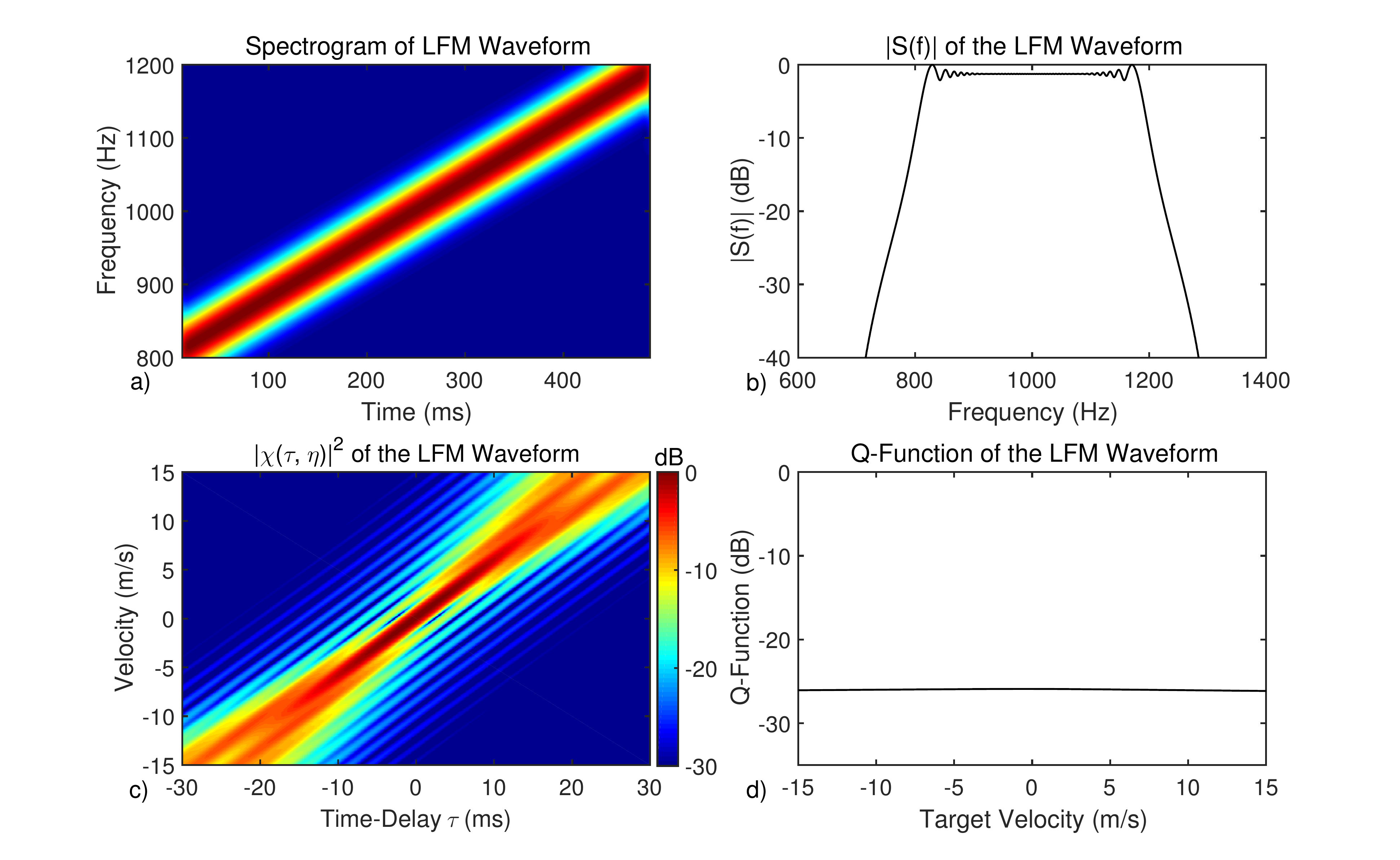}
\caption[Spectrogram (a), Spectrum (b), BAAF (c), and Q-function (d) of a LFM waveform with duration $T = 500$ ms , $f_c$ = 1000 Hz, and $\Delta f$ of 400 Hz.]{Spectrogram (a), Spectrum (b), BAAF (zoomed to $\pm 30$ ms) (c), and Q-function (d) of a LFM waveform with duration $T = 500$ ms, $f_c$ = 1000 Hz, and swept bandwidth $\Delta f$ of 400 Hz.  By linearly sweeping through a band of frequencies $\Delta f$ throughout its duration, the LFM achieves the long duration and large bandwidth required for detecting and resolving closely spaced objects.}
\label{fig:LFM}
\end{figure}

The LFM waveform is the first and possibly the most widely used PC waveform \cite{Levanon}.  The LFM was designed to mitigate the range resolution limitations of the CW waveform by linearly sweeping across a band of frequencies $\Delta f$.  The LFM's phase and IF functions are expressed as 
\begin{equation}
\varphi_{LFM}\left(t\right) = \pi\left(\dfrac{\Delta f}{T}\right)t^2,
\end{equation}
\begin{equation}
f_{LFM}\left(t\right) = \left(\dfrac{\Delta f}{T}\right)t + f_c
\end{equation}
for time $t$ defined as $-T/2 \leq t \leq T/2$.  Figure \ref{fig:LFM} shows the spectrogram, spectrum, BAAF, and Q-function of the CW waveform.  As seen from the spectrogram and spectrum, the LFM sweeps linearly across the band of frequencies $\Delta f$ and therefore places nearly equal energy across that band.  The LFM's AF has narrow mainlobe in time-delay whose width is inversely proportional the waveform's bandwidth $\Delta f$.  For non-zero target velocities, the AF's peak occurs at non-zero time-delays introducing a bias in the joint estimation of a target's range and velocity.  This bias, also known as range-Doppler coupling, limits the LFM to being used in applications where range resolution is the system's main design goal.  When the LFM's fractional bandwidth is sufficiently small (i.e, $\leq 1/10$), the LFM is Doppler tolerant.  However, as the fractional bandwidth increases, the LFM becomes increasingly Doppler sensitive \cite{Kramer}.  The Q-function is nearly constant across Doppler with a magnitude that is inversely proportional to the waveform's bandwidth \cite{Collins} meaning the LFM suppresses reverberation from all Doppler values nearly equally.  Additionally, increasing the waveform's bandwidth increases its ability to suppress reverberation.  The LFM has found extensive use in both radar and sonar systems due to its range resolution and reverberation suppression properties and its relative ease of implementation \cite{Ricker, Cook, Levanon, Rihaczek}.

\vspace{25em}
\subsection{The Hyperbolic FM (HFM) Waveform}
\label{subsec:HFM}
\begin{figure}[h]
\centering
\includegraphics[width=1.0\textwidth]{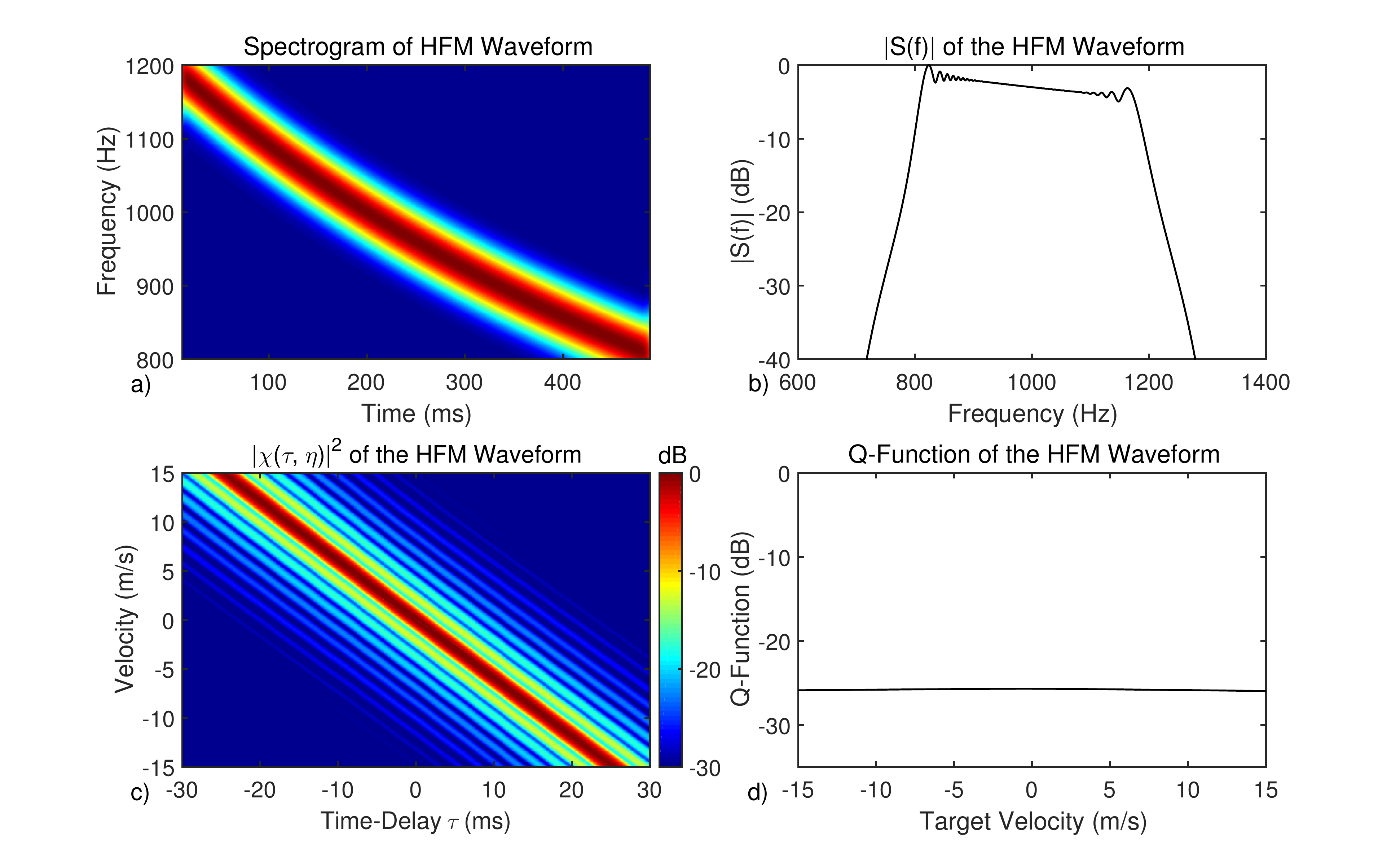}
\caption[Spectrogram (a), Spectrum (b), BAAF (c), and Q-function (d) of a HFM waveform with duration $T = 500$ ms, $f_c$ = 1000 Hz, and $\Delta f$ of 400 Hz.]{Spectrogram (a), Spectrum (b), BAAF (zoomed to $\pm 30$ ms) (c), and Q-function (d) of a HFM waveform with duration $T = 500$ ms, $f_c$ = 1000 Hz, and swept bandwidth $\Delta f$ of 400 Hz.  The HFM is optimally Doppler tolerant.}
\label{fig:HFM}
\end{figure}

The HFM waveform, first proposed in the literature by \cite{Jan}, uses hyperbolic FM and closely resembles the types of signals emitted by various species of echo-locating bats \cite{Lin, Jan, Altes}.  The HFM's phase and IF functions are expressed as 
\begin{equation}
\varphi\left(t\right) = 2 \pi a \ln\left(t+b\right),
\end{equation}
\begin{equation}
f\left(t\right) = \dfrac{a}{t+b}
\end{equation}
where $b = T\left(\frac{f_c - \Delta f/2}{\Delta f}\right)$ and $a = \left(f_c + \frac{\Delta f}{2}\right)b$.  Unlike the LFM which becomes increasingly Doppler sensitive with increasing fractional bandwidth, the HFM is optimally Doppler tolerant for both the narrowband and broadband Doppler models \cite{AltesI}.  Figure \ref{fig:HFM} (c) and (d) shows the AF and Q-function of the HFM.  Like the LFM, the HFM's AF possesses range-Doppler coupling.  Additionally, the HFM's AF has a very strong peak value for all target velocities.  The HFM's Q-function very closely resembles that of the LFM.  This means that the HFM also suppresses reverberation nearly equally for all Doppler values and that again increasing the waveform's bandwidth improves it's ability to suppress reverberation.  The HFM has been widely employed on broadband active sonar systems due to its optimal Doppler tolerance and reverberation suppression performance \cite{Ricker}.

\vspace{15em}
\subsection{The Costas Waveform}
\label{subsec:Costas}
\begin{figure}[ht]
\centering
\includegraphics[width=1.0\textwidth]{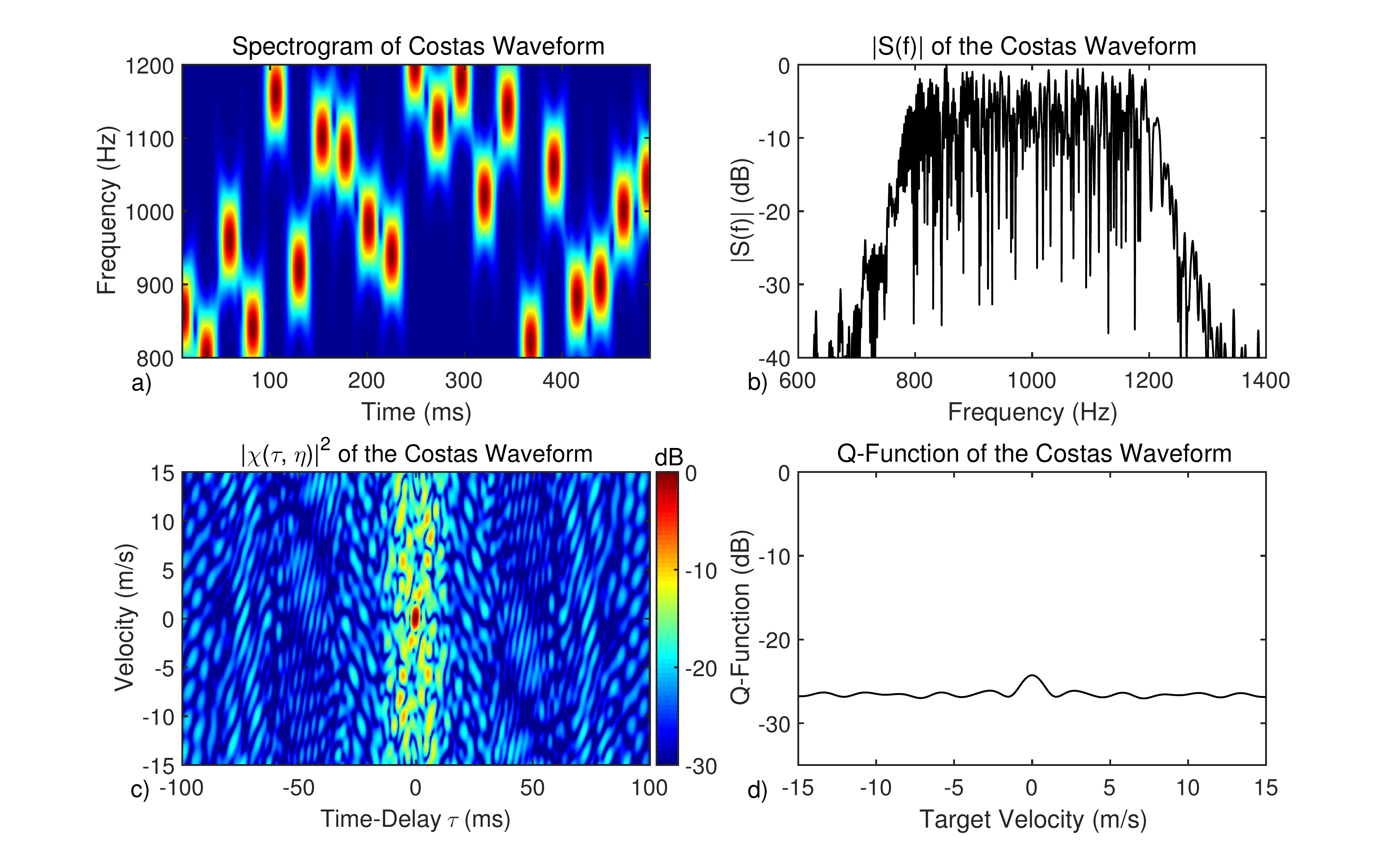}
\caption[Spectrogram (a), Spectrum (b), BAAF (c), and Q-function (d) of a Costas waveform with duration $T = 500$ ms, $f_c$ = 1000 Hz, and $\Delta f$ of 400 Hz.]{Spectrogram (a), Spectrum (b), BAAF (zoomed to $\pm 100$ ms) (c), and Q-function (d) of a Costas waveform with duration $T = 500$ ms, $f_c$ = 1000 Hz, and modulated bandwidth $\Delta f$ of 400 Hz.}
\label{fig:Costas}
\end{figure}

The Costas waveform is a Frequency Shift Keying (FSK) waveform comprised of N contiguous amplitude tapered CW pulses, called chips.  Each chip has a duration $T/N$ where $T$ is the waveform's duration and a different center frequency.  The Costas waveform is expressed as
\begin{equation}
s_{Costas}\left(t\right) = \sum_{i=1}^{N} a\left(t - iT/N\right)e^{j \left(2 \pi f_i\left(t - iT/N\right) + \theta_i\right)}
\end{equation}
where $N$ is the number of chips in the waveform, $a\left(t\right)$ is the chip's amplitude tapering function, $f_i$ is the frequency of the $i^{th}$ chip, and the phase term $\theta_i$ is included to ensure phase continuity between the chips in the waveform.  The frequency shift sequence for each chip is given by a Costas code \cite{Costas}.  The Costas code minimizes the waveform's AF sidelobes and achieves a thumbtack AF.  For a given TBP and a rectangular tapering function applied to each chip, the Costas waveform requires at least $\ceil \left(\sqrt{TB}\right)$ chips \cite{Costas}.  As seen in Figure \ref{fig:Costas}, the Costas waveform achieves a thumbtack AF and a Q-fuction that is nearly constant across Doppler with a magnitude inversely proportional to the waveform's bandwidth.


\subsection{The Binary-Phase Shift Keying (BPSK) Waveform}
\label{subsec:BPSK}
\begin{figure}[h]
\centering
\includegraphics[width=1.0\textwidth]{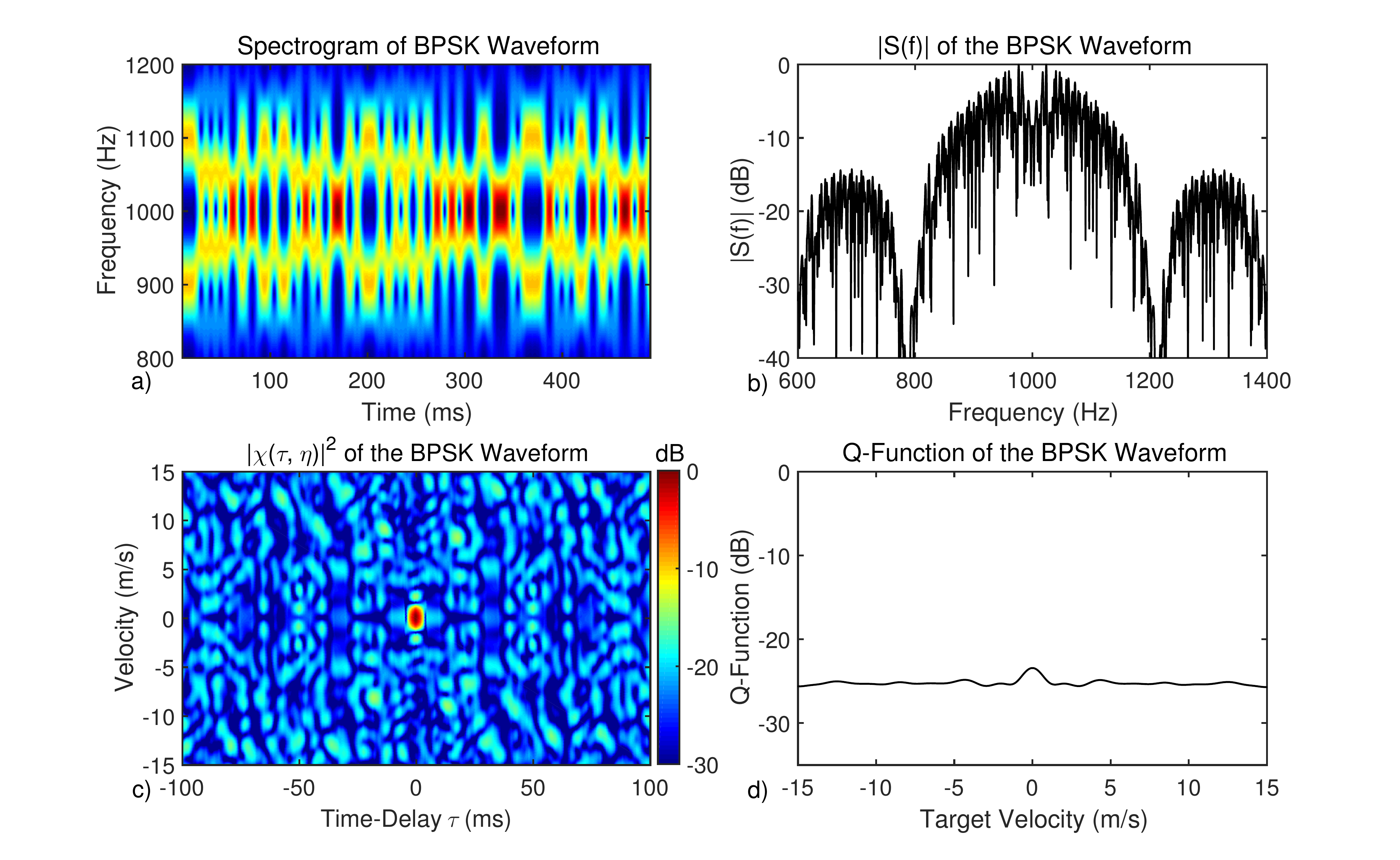}
\caption[Spectrogram (a), Spectrum (b), BAAF (c), and Q-function (d) of a BPSK waveform with duration $T = 500$ ms, $f_c$ = 1000 Hz, and $\Delta f$ of 400 Hz.]{Spectrogram (a), Spectrum (b), BAAF (zoomed to $\pm 100$ ms) (c), and Q-function (d) of a BPSK waveform with duration $T = 500$ ms, $f_c$ = 1000 Hz, and modulated bandwidth (null-to-null) $\Delta f$ of 400 Hz.  The BPSK possesses a thumbtack AF, but suffers from poor spectral efficiency as it's spectral sidelobes roll off at 6 dB per octave.}
\label{fig:BPSK}
\end{figure}

The BPSK waveform is similar to the Costas waveform in that it is a collection of individual CW chips except the BPSK's chips are all the same frequency and the instantaneous phase $\theta_i$ of each chip is changed according to a binary sequence.  The BPSK waveform is expressed as 
\begin{equation}
s_{BPSK}\left(t\right) = \sum_{i=1}^{N} a\left(t - iT/N\right)e^{j \left(2 \pi f_ct + \theta_i\right)}
\end{equation}
The phase sequence $\theta_i$ controls the AF shape of the BPSK waveform and a number of phase sequences have been designed to achieve desirable auto-correlation properties \cite{Levanon, Li}.  Some of the most commonly used phase sequences are pseudo random sequences known as Maximum Length Shift Register (MLSR) sequences \cite{Ricker}.  The resulting BPSK waveform is Doppler sensitive due its CW nature and the MLSR sequence helps spread the waveform's AF volume as evenly as possible resulting in a thumbtack AF.  One limitation of the BPSK waveform is that it contains substantial energy across frequency \cite{Levanon}.  The spectral sidelobes, visible in Figure \ref{fig:BPSK} (b), fall off at a rate of 6 dB per octave.  As a result of this, the BPSK attains poor SC.  Applying an amplitude tapering function to the chips reduces the spectral sidelobes thus improving the BPSK's SC, but the tapering in turn increases the BPSK's PAPR.  When using a BPSK waveform, the waveform designer must strike a compromise between spectral efficiency and low PAPR.

\vspace{-1.5em}
\subsection{The Quadri-Phase Shift Keying (QPSK) Waveform}
\label{subsec:QPSK}
\begin{figure}[h]
\centering
\includegraphics[width=1.0\textwidth]{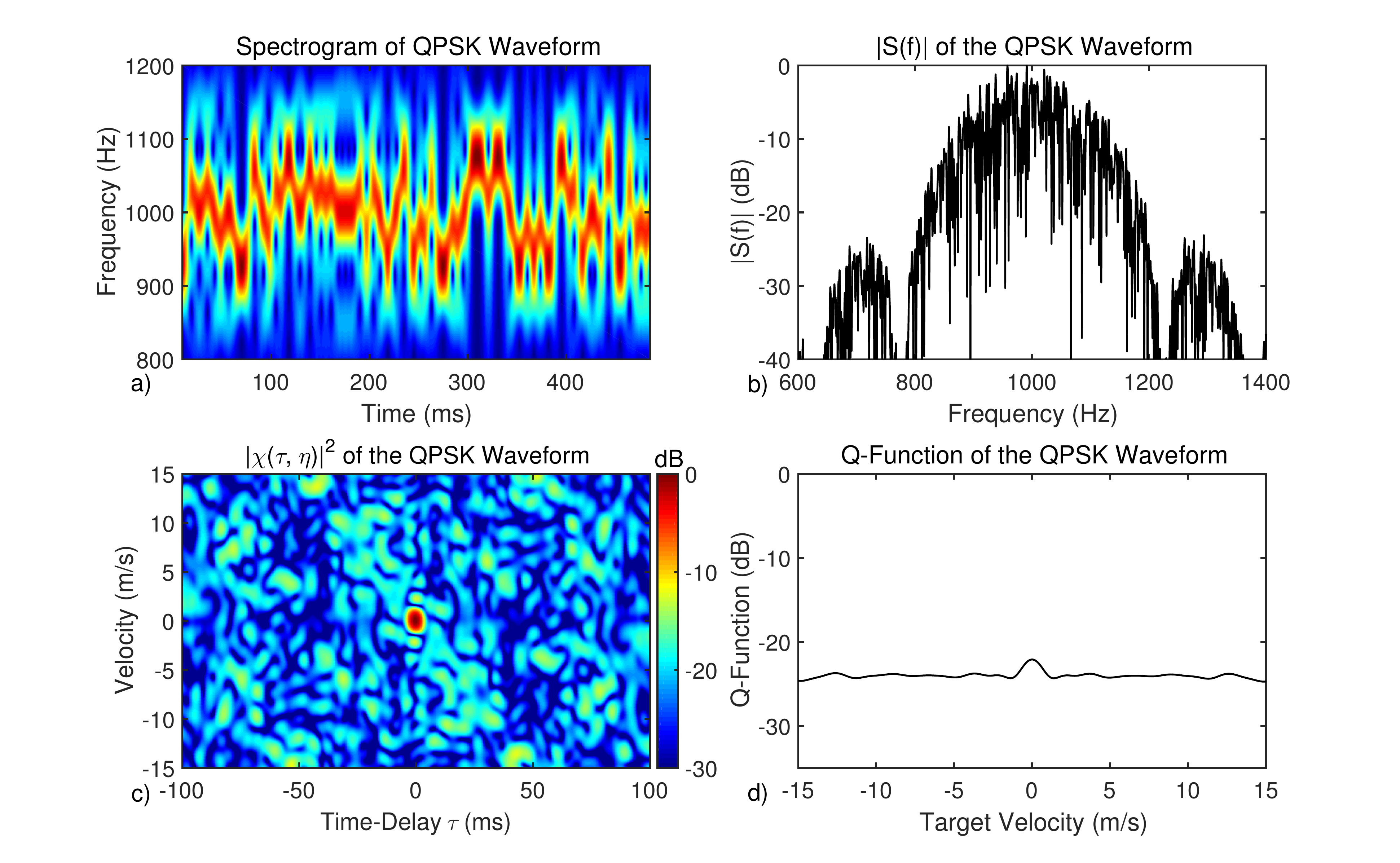}
\caption[Spectrogram (a), Spectrum (b), BAAF (c), and Q-function (d) of a QPSK waveform with duration $T = 500$ ms, $f_c$ = 1000 Hz, and $\Delta f$ of 400 Hz.]{Spectrogram (a), Spectrum (b), BAAF (zoomed to $\pm 100$ ms) (c), and Q-function (d) of a QPSK waveform with duration $T = 500$ ms, $f_c$ = 1000 Hz, and modulated bandwidth (null-to-null) $\Delta f$ of 400 Hz.  The QPSK waveform largely preserved the AF properties of the BPSK waveform while also possessing improved spectral efficiency.}
\label{fig:QPSK}
\end{figure}

The Quadriphase Shift Keying (QPSK) waveform, developed by Taylor and Blinchikoff \cite{QuadPhase} utilizes a binary-to-quadriphase transformation that produces a waveform that maintains nearly the same AF shape as its binary counterpart, reduced spectral sidelobes, and a nearly constant amplitude response.  The binary-to-quadriphase transformation is expressed as
\begin{equation}
q_i = j^{\pm\left(i-1\right)}e^{j\theta_i}
\label{eq:biquad}
\end{equation}
where $\theta_i$ is the phase sequence.  Therefore, applying the transformation in \eqref{eq:biquad} to a MLSR sequence produces a thumbtack waveform with improved spectral efficiency over a BPSK and a constant envelope resulting in a low PAPR.  Figure \ref{fig:QPSK} shows the spectrogram, spectrum, BAAF, and Q-function for a QPSK waveform generated by transforming the MLSR sequence for the BPSK waveform in Figure \ref{fig:BPSK}.  Note the resulting waveform's spectral sidelobes in Figure \ref{fig:QPSK} (b) are substantially lower than those of the BPSK in Figure \ref{fig:BPSK} (b).  

\vspace{20em}

\subsection{The Sinusoidal FM (SFM) Waveform}
\label{subsec:SFM}
The SFM is a FM waveform whose IF function is itself a CW sinusoid.  Its phase and IF functions are expressed as [14, 15]
\begin{equation}
\varphi_{SFM} \left( t \right) = \beta \sin \left(2 \pi f_m t\right) = \left(\dfrac{\Delta f}{2 f_m}\right) \sin \left(2 \pi f_m t\right)
\label{eq:SFM_Phase}
\end{equation}
\begin{equation}
f_{SFM} \left( t \right) = \beta f_m \cos \left(2 \pi f_m t \right) =\left(\dfrac{\Delta f}{2}\right)\cos \left(2 \pi f_m t\right)
\label{eq:SFM_IF}
\end{equation}
where $\beta$ is the modulation index given as $\beta = \Delta f/2 f_m$, $f_m$ is the modulation frequency and $\Delta f$ is the swept bandwidth.  There also exists the cosine phase counterpart of the SFM, the Cosine FM (CFM) whose instantaneous phase and frequency functions are shifted by $\pi/2$ radians and maintains the same waveform characteristics of the SFM. The spectrum of the SFM, derived in Appendix \ref{ch:appendix_A}, is expressed as 
\begin{equation}
S_{SFM}\left(f\right) = \sqrt{T}\sum_{n=-\infty}^{\infty}J_n\{\beta\} \sinc\left[\pi T \left(f - f_c - f_m n \right) \right]
\label{eq:SFM_Spectrum}
\end{equation}
where $J_n\{\beta\}$ is the $n^{th}$ order Cylindrical Bessel Function of the First Kind.  The expression in \eqref{eq:SFM_Spectrum} can be used to derive Carson's Bandwidth Rule \cite{Couch} for the SFM and is expressed as 
\begin{equation}
2\left( \beta + 1\right)f_m = \Delta f + f_m
\label{eq:SFM_Carson}
\end{equation}
When the SFM's swept bandwidth $\Delta f$ is much larger than the modulation frequency $f_m$ (i.e, $\Delta f > 10 f_m)$, the vast majority of the waveform's energy is concentrated in the swept bandwidth.  Additionally, the SFM has a constant envelope and requires minimal tapering for transmission on piezoelectric tranducers and therefore attains a low PAPR. 
\vspace{1em}

\begin{figure}[h]
\centering
\includegraphics[width=1.0\textwidth]{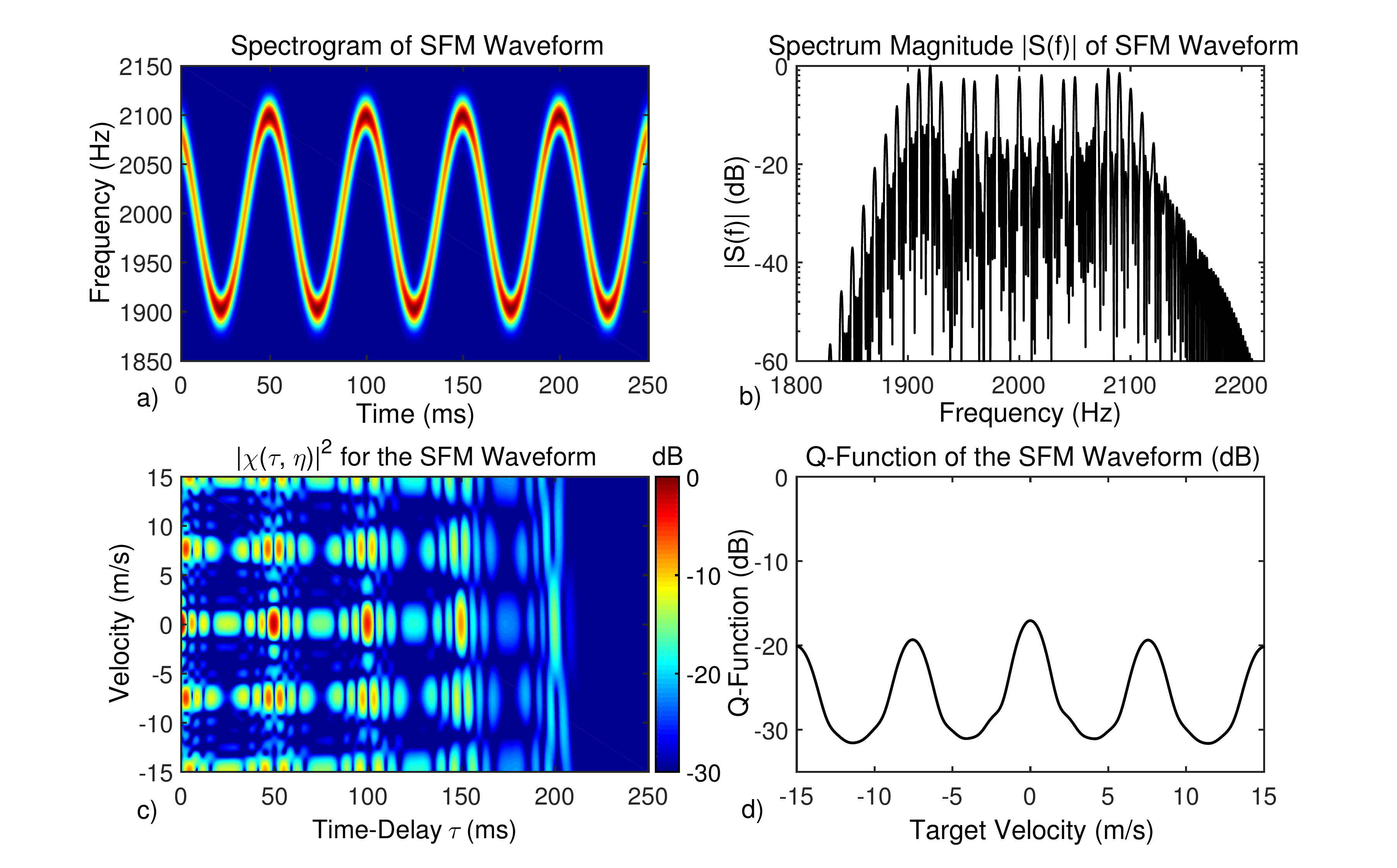}
\caption[Spectrogram (a), Spectrum (b), BAAF (c), and Q-function (d) of a SFM waveform with duration $T = 250$ ms,  $\Delta f = 200$ Hz, $f_m = 20$ Hz, and $f_c = 2000$ Hzz.]{Spectrogram (a), Spectrum (b), BAAF (c), and Q-function (d) of a SFM waveform with duration $T = 250$ ms,  $\Delta f = 200$ Hz, $f_m = 20$ Hz, and $f_c = 2000$ Hz.  The SFM possesses a comb spectrum and as a result is Doppler sensitive, but attains poor range resolution performance due to the high sidelobes in time-delay.  The SFM's Q-function contains peaks in the locations of its BAAF's Doppler grating lobes and deep valleys between these grating lobes which are used for suppressing reverberation at those Doppler values.}
\label{fig:SFM_Fig}
\end{figure}

Figure \ref{fig:SFM_Fig} shows the spectrogram, spectrum, BAAF, and Q-function for an SFM of duration $T = 250$ ms, a modulation frequency $f_m = 20$ Hz, a bandwidth $\Delta f = 200$ Hz, and a center frequency $f_c = 2000$  Hz.  The SFM's IF function is clearly visible in the spectrogram and the SFM's spectrum is of the comb variety.  Each spectral component is equally spaced $f_m$ Hz apart in frequency.  The BAAF is not a thumbtack shape but is of the ``bed of nails'' \cite{Rihaczek} variety and possesses a distinct mainlobe at the origin whose width in time-delay and velocity is inversely proportional to the waveform's bandwidth and pulse length respectively with multiple grating lobes in range and Doppler.  The Q-function has peaks in Doppler that are a result of the grating lobes of the SFM's BAAF.  The region between the BAAF grating lobes are low in magnitude and therefore translate to a valley in the Q-function.  

The SFM's NAAF and BAAF, derived in Appendix \ref{ch:appendix_B}, are expressed as
\begin{equation}
|\chi\left(\tau, \phi\right)| = \left| \dfrac{\left(T-|\tau|\right)}{T}\sum_{n=- \infty}^{\infty}  J_n\{2 \beta \sin \left(\pi f_m \tau \right) \} \sinc \left[\pi\left(f_m n + \phi\right)\left(T - |\tau|\right)\right] \right|
\label{eq:SFM_NAAF}
\end{equation}
\begin{multline}
|\chi\left(\tau, \eta\right)| \cong \dfrac{\sqrt{\eta}\left(T-|\tau|\right)}{T} \left|\sum_{n=- \infty}^{\infty}  J_n\{2 \beta \sin \left(\pi f_m \eta \tau \right) \} \right. \times
\\ \left. \sinc \left[ \pi\left(\left(\eta-1\right)f_c -\dfrac{ f_m n \left(1+\eta \right)}{2} \right)\left(T-|\tau|\right) \right] \right|
\label{eq:SFM_BAAF}
\end{multline}
where $J_n\{\}$ is the $n^{th}$ order cylindrical Bessel function of the first kind and the $\left(T-|\tau|\right)/T$ term is a triangular function.  The result in \eqref{eq:SFM_NAAF} generalizes the result obtained by Cook and Bernfield in [1].  Their result assumed a modulation frequency $f_m$ of $1/T$ (one period of IF per pulse) whereas the result in \eqref{eq:SFM_NAAF} holds for any modulation frequency $f_m$.  The result in \eqref{eq:SFM_BAAF} is an approximation of the BAAF that assumes the ratio of the target velocity to sound speed is small (i.e $\leq$ 1/100) and appears to be novel.  
\vspace{1em}

\begin{figure}[h]
\centering
\includegraphics[width=1.0\textwidth]{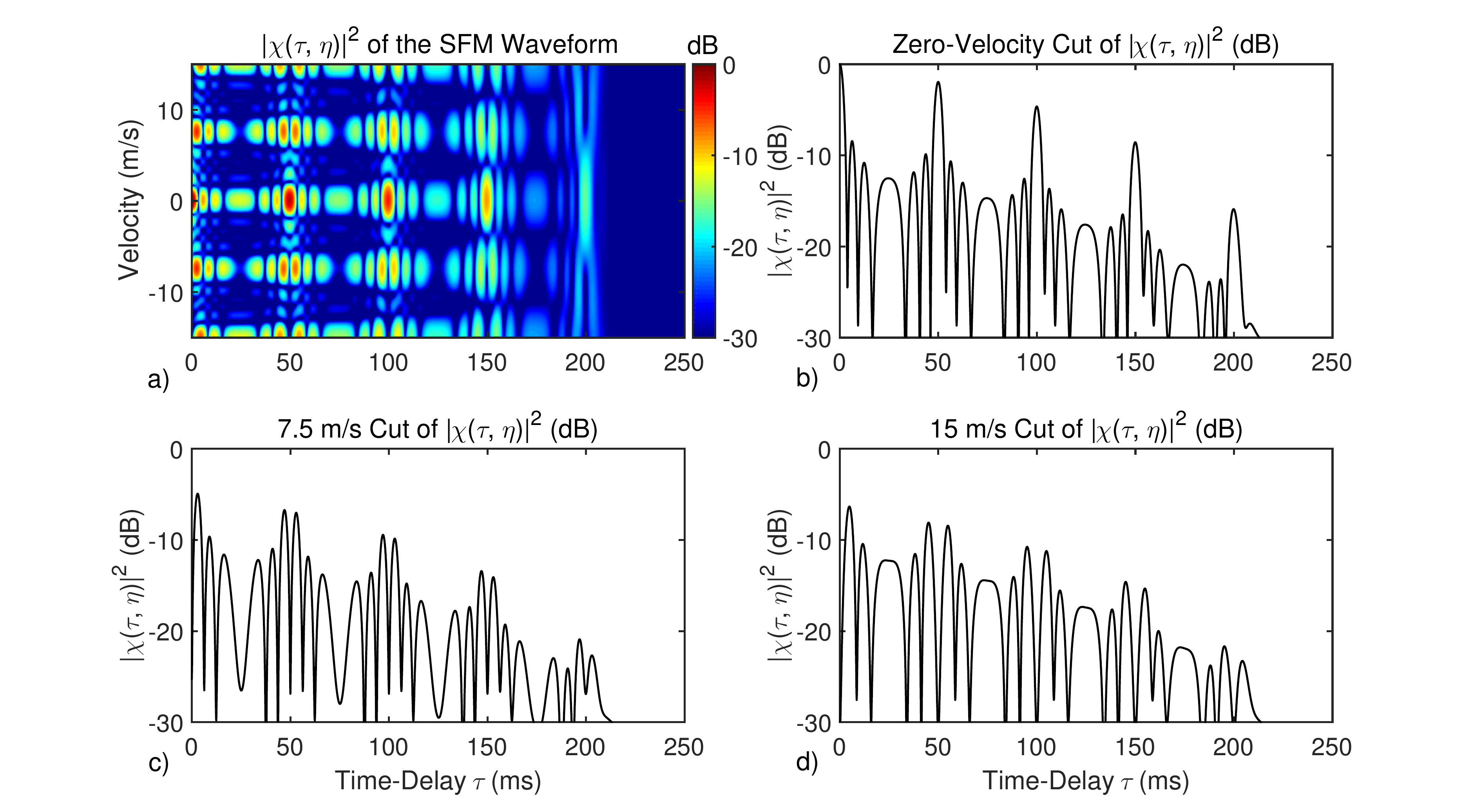}
\caption[BAAF (a), zero-velocity (b), 7.5 m/s (c), and 15 m/s (d) cuts of the BAAF of a SFM waveform.]{BAAF (a), zero-velocity (b), 7.5 m/s (c), and 15 m/s (d) cuts of the BAAF of a SFM with duration $ T = 250$ ms, $f_m$ = 20 Hz, $f_c = 2000$ Hz, and a bandwidth $\Delta f$ of 200 Hz.  The SFM is Doppler sensitive, but attains poor range resolution due to the high sidelobes in time-delay.}
\label{fig:SFM_Cuts}
\end{figure}

The SFM's AF behavior in time-delay (range) is largely determined by the Bessel and triangular functions.  Its Doppler behavior (velocity) is determined by the $\sinc$ term.  Figure \ref{fig:SFM_Cuts} shows the AF of the SFM from Figure \ref{fig:SFM_Fig} along with cuts across time-delay at 0, 7.5, and 15 m/s. The SFM's BAAF possesses a distinct mainlobe at the origin whose width in time-delay and velocity is inversely proportional to the waveform's bandwidth and pulse length respectively.  The zero-velocity cut of the BAAF corresponds to the triangular function multiplied by a $0^{th}$ order Bessel function.  Note that the argument passed to the Bessel function in (\ref{eq:SFM_NAAF}) and (\ref{eq:SFM_BAAF}) is a periodic function of $\tau$ with period $2/f_m$ whose amplitude varies from $ \pm  2\beta$.  The zero-velocity cut shows the Bessel function repeats every 25 ms and is attenuated by the triangular function.  The same can be said of the 7.5 m/s and 15 m/s velocity cuts of the BAAF except now the Bessel function orders are 1 and 2 respectively.  The locations of these cuts in Doppler can be calculated by setting the $\left(nf_m + \phi\right)$ argument in (\ref{eq:SFM_NAAF}) or $\left[\left(1+\eta\right)nf_m + \left(\eta-1\right)f_c\right]$ argument in (\ref{eq:SFM_BAAF}) to zero and solving for target velocity.  This means that the SFM's modulation frequency $f_m$ determines the locations of the AF Doppler sidelobes.  Additionally, $f_m$ can be set such that the AF Doppler sidelobes appear at high velocities beyond what is realistically expected for a sonar target.  This coupled with a narrow mainlobe proportional to the carrier frequency $f_c$ and inversely proportional to the pulse length $T$ allows the SFM to provide an accurate estimate of target velocity.  However, the zero velocity cut of the AF, the waveform's ACF, contains many high sidelobes.  While the SFM is able to discriminate between targets that have different velocities, it attains poor range resolution due to the high sidelobes in time delay. 
\vspace{1em}

\begin{figure}[p]
\centering
\includegraphics[width=1.0\textwidth]{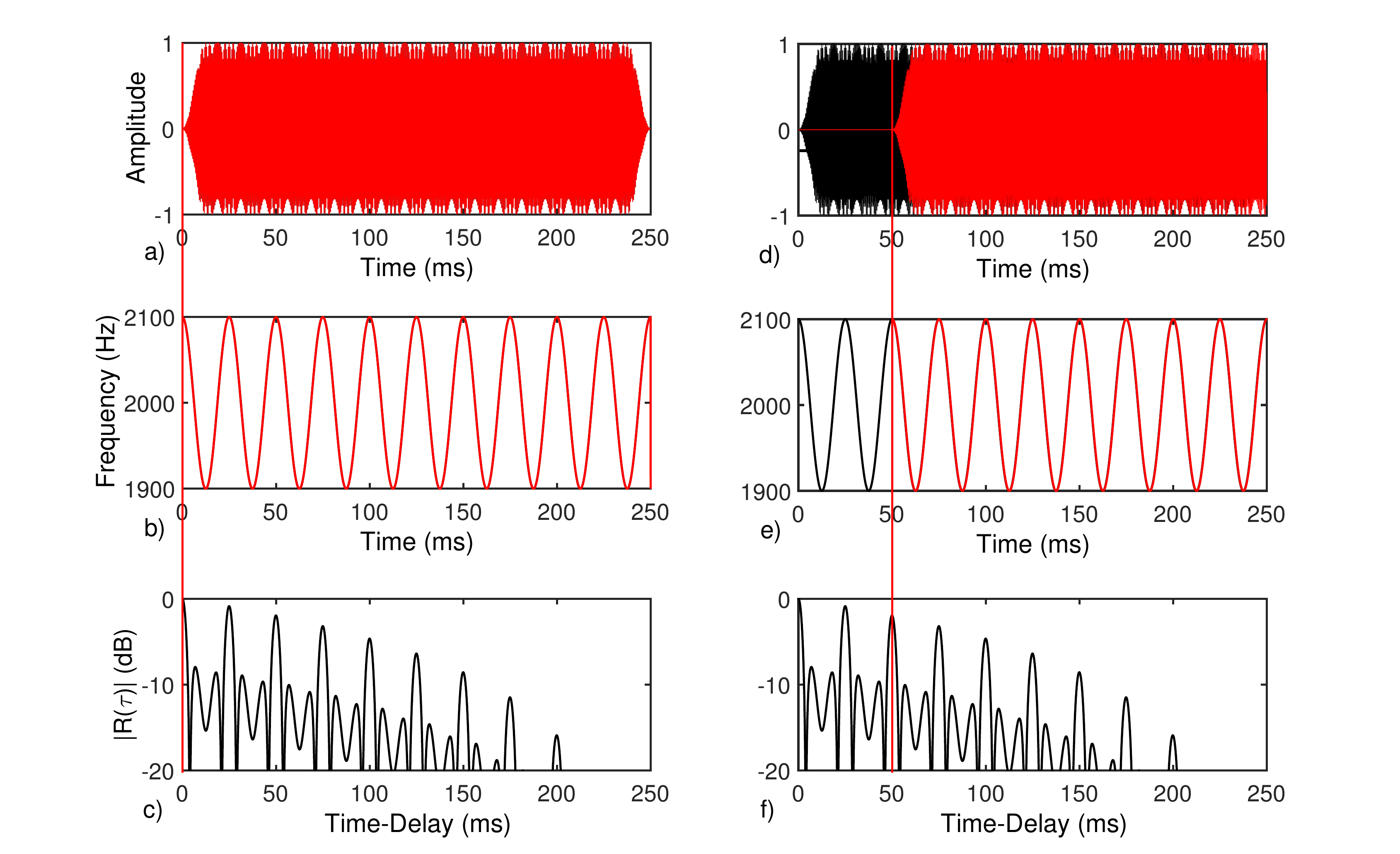}
\caption[Illustration of the origin of the SFM Auto Correlation sidelobes for two time delays.]{Illustration of the origin of the SFM Auto Correlation sidelobes for two time delays.  The first column shows the correlation operation in time (a), the waveforms in IF (b), and the location of the ACF peak (mainlobe) in (c).  In the second column, correlating the waveform at time-delay of 50 ms (d) results in 8 of the 10 SFM cycles to overlap in IF (e) and therefore a sidelobe height of 8/10 or -1 dB (f).}
\label{fig:SFM_IF_Fig}
\end{figure}

The high sidelobes of the SFM's ACF are a direct result of the periodicity in the SFM’s IF.  This is illustrated in Figure \ref{fig:SFM_IF_Fig}.  When convolving a zero Doppler SFM echo with a zero Doppler MF, the echo's IF completely overlaps with the MF's IF function at zero time-delay yielding the peak of the mainlobe.  When the time-delay equals an integer multiple $q$ of the modulation period $(q/f_m)$, the spectral content will overlap with all but $C-q$ cycles where $C$ is the number of cycles in the IF expressed as $f_mT$.  This results in a sidelobe with height $\left(C-q\right)/C$ .  The periodic range sidelobes can be removed by designing an SFM with a single cycle in its IF, which is equivalent to reducing the modulation frequency $f_m$ to $1/T$.  However, there is a cost in reducing the modulation frequency.  As described earlier, reducing the modulation frequency will result in shifting the high sidelobes in Doppler given by (\ref{eq:SFM_NAAF}) and (\ref{eq:SFM_BAAF}) closer to the origin.  As a result of this, the sidelobes may be located in the range of velocities where a realistic sonar target is expected and therefore reduces the ability to resolve multiple targets in velocity.  The SFM can be designed to estimate and resolve target range or target velocity, but not simultaneously. 

\chapter{The Generalized Sinusoidal FM (GSFM) Waveform}
\label{ch:GSFM}
The GSFM waveform is a novel FM transmit waveform for active sonar that possesses a BAAF shape that closely resembles a thumbtack.  The GSFM waveform is a modification of the SFM waveform that uses an IF function that resembles the time-voltage characteristic of a LFM chirp waveform.  Utilizing this "chirped" IF function removes the periodicity of the SFM's IF in order to mitigate periodic sidelobes in time-delay while preserving the desirable bandwidth and Doppler sensitivity properties of the SFM.  There are a multitude of ways to generate the phase and IF functions of the GSFM and each approach has their relative merits.  This chapter defines the three principle phase and IF functions of the GSFM, describes the GSFM's properties, and explains why the GSFM waveform possesses a thumbtack AF.

\section{The GSFM's Phase and IF Functions}
\label{sec:GSFM_IF}

The first two of three GSFM waveforms are defined using the phase and IF functions expressed as \cite{HagueII}
\begin{equation}
\varphi_{I} \left( t \right) = \left[\dfrac{\pi \Delta f}{\rho \left(2 \pi \alpha\right)^{\frac{1}{\rho}}} \right] S \{2 \pi \alpha t^\rho, 1/\rho \},
\label{GSFM_Phi_I}
\end{equation}

\begin{equation}
\varphi_{II} \left( t \right) = \left[\dfrac{\pi \Delta f}{\rho \left(2 \pi \alpha\right)^{\frac{1}{\rho}}} \right] C \{2 \pi \alpha t^\rho, 1/\rho \},
\label{GSFM_Phi_II}
\end{equation}

\begin{equation}
f_{I} \left( t \right) = \left(\dfrac{\Delta f}{2}\right) \sin \left(2 \pi \alpha t^\rho\right),
\label{GSFM_IF_I}
\end{equation}

\begin{equation}
f_{II} \left( t \right) = \left(\dfrac{\Delta f}{2}\right) \cos \left(2 \pi \alpha t^\rho\right)
\label{GSFM_IF_II}
\end{equation}
where $\Delta f$ is the waveform's swept bandwidth, $S\{\}$ and $C\{\}$ are the Generalized Sine/Cosine Fresnel Integrals (GSFI/GCFI) \cite{NHMF, Fresnel}, $\rho$ is a unitless exponent parameter that must be greater than or equal to 1, and $\alpha$ is a modulation term with units $s^{-\rho}$ that is loosely analogous to the SFM's modulation frequency $f_m$.  Like the SFM, there are sine and cosine IF function versions.  While these two GSFM waveforms both possess a thumbtack AF and largely share the same properties and performance characteristics, the later chapters of this dissertation will demonstrate that there are particular situations where their respective performance characteristics notably differ.  The third GSFM definition utilizes an approximation to the GCFI and is expressed as \cite{HagueI}
\begin{equation}
\varphi_{III} \left( t \right) = \dfrac{\tilde{\beta}}{t^{\left(\rho-1\right)}} \sin\left(2\pi\alpha t^{\rho}\right),
\label{GSFM_Phi_III}
\end{equation}
\begin{multline}
f_{III} \left( t \right) = \tilde{\beta} \alpha \rho \Biggl[\cos\left(2\pi\alpha t^{\rho}\right) - \left(\dfrac{\rho-1}{\rho}\right) \sinc\left(2\pi\alpha t^{\rho}\right)  \Biggr] \\ = \left(\dfrac{\Delta f}{2}\right) \Biggl[\cos\left(2\pi\alpha t^{\rho}\right) - \left(\dfrac{\rho-1}{\rho}\right) \sinc\left(2\pi\alpha t^{\rho}\right)  \Biggr] 
\label{GSFM_IF_III}
\end{multline}
where $\alpha$ and $\rho$ are defined as above, $\tilde{\beta}$ is the waveform's frequency deviation ratio, the ratio of the swept bandwidth $\Delta f$ to the IF function's bandwidth \cite{Couch}.  The deviation ratio $\tilde{\beta}$ is loosely analogous to the SFM's modulation index $\beta$.
The GSFM waveform defined by (\ref{GSFM_Phi_III}), while an approximation to the GCFI phase (\ref{GSFM_Phi_II}) and attains similar performance characteristics,
\footnote{The author wishes to point out that the phase and IF functions defined in (\ref{GSFM_Phi_III}) and (\ref{GSFM_IF_III}) were the original GSFM phase and IF functions \cite{HagueI} resulting from this dissertation.  While these equations were largely convenient to work with mathematically and easy to implement, they were particularly unwieldy for the analysis presented in Chapter \ref{ch:GSFM_Eval_AF}.  This fact in turn led to the development of the GSFM waveforms described in (\ref{GSFM_Phi_I})-(\ref{GSFM_IF_II}) \cite{HagueII}.  Later analysis showed that while the GSFMs defined using the (\ref{GSFM_Phi_II}) and (\ref{GSFM_Phi_III}) were nearly identical in terms of performance, the GSFM defined (\ref{GSFM_Phi_I}) had substantially different performance characteristics under certain conditions.  These differences will be explained in greater detail in Chapter \ref{ch:GSFM_Eval_AF}.}is more convenient to work with mathematically under certain situations.  
\vspace{1em}

\begin{figure}[h]
\centering
\includegraphics[width=1.0\textwidth]{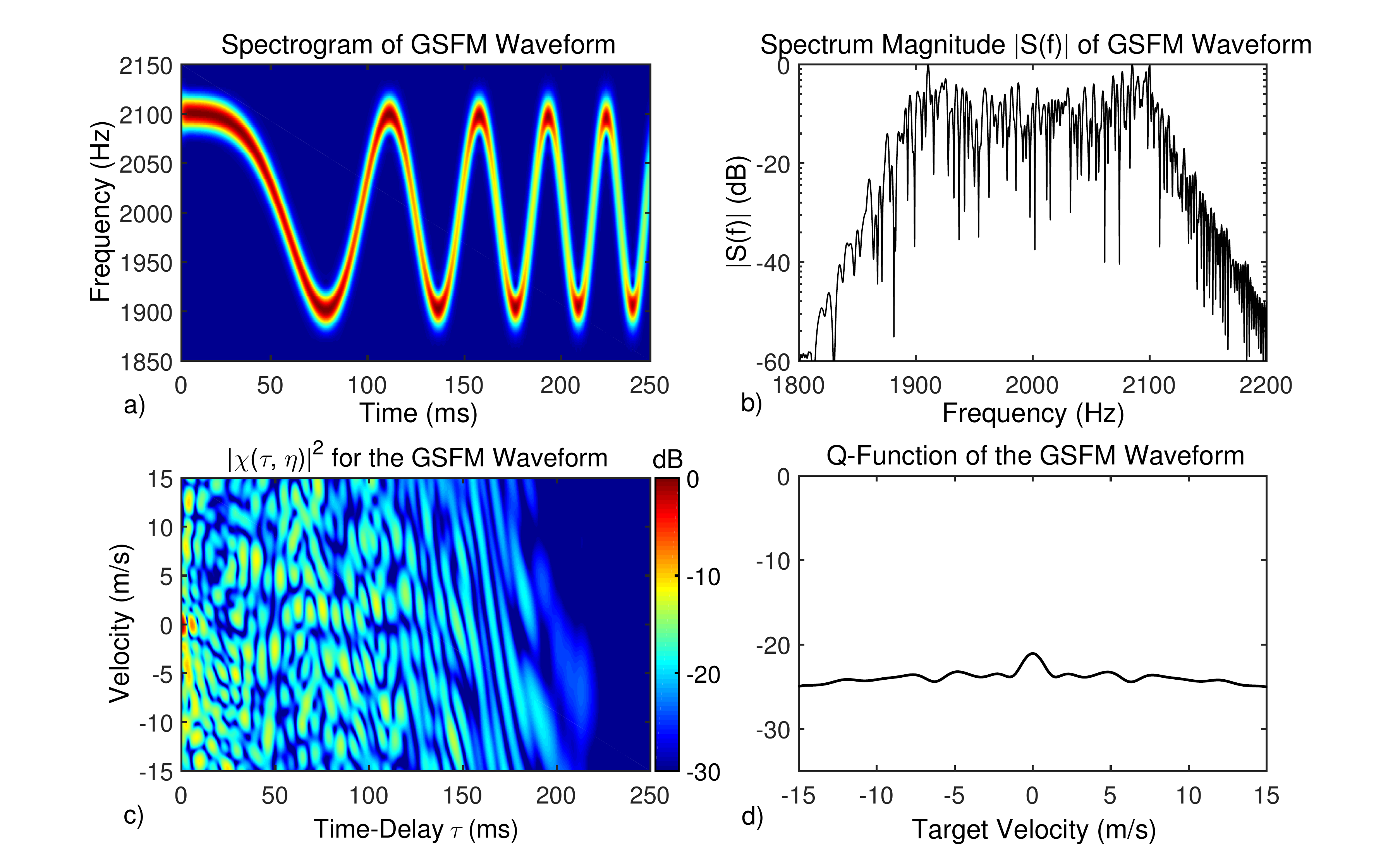}
\caption[Spectrogram (a), Spectrum (b), BAAF (c) Q-function (d) of a GSFM with duration $T = 250$ ms, $f_c = 2000$ Hz, and a bandwidth $\Delta f= 200$ Hz.]{Spectrogram (a), Spectrum (b), BAAF (c), and Q-function (d) of a GSFM with duration $T = 250$ ms, $\Delta f = 200$ Hz, $\rho = 2$, $\alpha = 80$ s$^{-2}$, and $f_c = 2000$ Hz.  Because the IF of this variant of the GSFM has a time varying period, its AF possesses low range sidelobes while maintaining the Doppler sensitivity of the SFM.}
\label{fig:GSFM}
\end{figure}

Defining time to be $0 \leq t \leq T$ generates a waveform whose IF function resembles the time-voltage characteristic of an up-sweeping chirp for $\rho > 1$.  This waveform has a non-symmetric IF.  Defining time to be $-T/2 \leq t \leq T/2$ and replacing the $t^{\rho}$ term with $|t|^{\rho}$ generates a waveform with an even-symmetric IF function that resembles the time-voltage characteristic of a base-banded chirp waveform.  The frequency modulation term $\alpha$ determines the number of cycles $C$ in the IF of the GSFM and is expressed as $C = \alpha T^{\rho}$ for a non-symmetric IF function and $C = 2\alpha\left(T/2\right)^{\rho}$ for an even-symmetric IF function.  The exponent parameter $\rho$ determines the overall shape of the IF function.  When $\rho = 1$, the GSFM's phase (\ref{GSFM_Phi_II} $\&$ \ref{GSFM_Phi_III}) and IF functions (\ref{GSFM_IF_II} $\&$ \ref{GSFM_IF_II}) become equivalent to the SFM waveform’s phase (\ref{eq:SFM_Phase}) and IF (\ref{eq:SFM_IF}) functions respectively.  When $\rho = 2$ the resulting waveform's phase and IF functions resemble the time/voltage characteristic of the LFM chirp waveform.  The LFM sinusoid IF variant of the GSFM does not exhibit the strict periodicity of the SFM's IF.  For any non-zero time-delay the spectral energy of the echo will not have substantial alignment with the IF of the MF replica resulting in much lower delay sidelobes in the ACF.  Figure \ref{fig:GSFM} shows the IF function and BAAF of the GSFM with duration $T = 250$ ms, $\Delta f = 200$ Hz, $\rho = 2$, $\alpha = 80s^{-2}$ (or $C = 5$), and $f_c = 2000$ Hz.  Unlike the SFM, the BAAF of this variant of the GSFM exhibits a single distinct mainlobe centered at the origin with low sidelobes in range while preserving the Doppler sensitivity of the SFM.  The GSFM's AF closely approximates a thumbtack AF, the design goal of this dissertation.

\section{The GSFM's Spectrum and Ambiguity Function}
\label{sec:GSFM_AF}
The spectrum of the GSFM with a non-symmetric IF function, derived in Appendix \ref{ch:appendix_A}, is expressed as
\begin{multline}
\left|S_{GSFM}\left(f\right)\right| = \left|\sqrt{T}\sum_{n=-\infty}^{\infty} \mathcal{J}_n^{1:\infty}\Bigl\{ \frac{\widetilde{a}_m\Delta f T}{2}; \frac{\widetilde{b}_m\Delta f T}{2} \Bigr\} \times \right. \\ \left. \sinc\left[\pi T \left(f - f_c - \frac{a_0 \Delta f}{4} - \frac{ n}{T} \right)\right]\right|
\label{eq:GSFM_SpectrumI}
\end{multline} 
where $\mathcal{J}_n^{1:\infty}\Bigl\{ \frac{\widetilde{a}_m\Delta f T}{2}; \frac{\widetilde{b}_m\Delta f T}{2} \Bigr\}$ is the $n^{th}$ order infinite dimensional Generalized Bessel Function (GBF) of the mixed type \cite{Lorenzutta, Dattoli, DattoliII}, $T$, the pulse length, is also the fundamental period of the Fourier harmonics, and $\widetilde{a}_m$ and $\widetilde{b}_m$ are the Fourier coefficients of $\varphi\left(t\right)$.  The NAAF of the GSFM with a non-symmetric IF function, derived in Appendix \ref{ch:appendix_C}, is
\begin{multline}
\left|\chi \left(\tau, \phi\right)\right| = \left(\dfrac{T - |\tau|}{T} \right)\times \\  \Biggl| \sum_{n=-\infty}^{\infty}\mathcal{J}_n^{1:\infty}\Biggl\{\Delta f T \widetilde{b}_m \sin \left(\frac{\pi m  \tau}{T}\right); \Delta f T \widetilde{a}_m \sin \left(\frac{\pi m  \tau}{T}\right)\Biggr\} \times \biggr. \\ \biggl.  \sinc \left[\left(\frac{\pi n}{T} + \phi \right)\left(T - |\tau|\right)\right] \Biggr|
\label{eq:GSFM_NAAF}
\end{multline}
where $\mathcal{J}_n^{1:\infty}\{\}$ is again the $n^{th}$ order, infinite dimensional GBF of the mixed type \cite{Lorenzutta, Dattoli, DattoliII}, $\widetilde{a}_m$ and $\widetilde{b}_m$ are again the Fourier coefficients of $\varphi\left(t\right)$, and $T$ is the period of the Fourier harmonics.  An approximation of the BAAF of the GSFM with a non-symmetric IF function, also derived in Appendix \ref{ch:appendix_C}, is expressed as
\begin{multline}
\left|\chi \left(\tau, \eta\right)\right| \cong \dfrac{\sqrt{\eta}\left(T - |\tau|\right)}{T} \times \\ \Biggl| \sum_{n=-\infty}^{\infty}\mathcal{J}_n^{1:\infty}\Biggl\{\Delta f T \widetilde{a}_m \sin\left(\frac{\pi m \eta \tau}{T}\right); \Delta f T \widetilde{b}_m \sin\left(\frac{\pi m \eta \tau}{T}\right) \Biggr\}\times  \Biggr. \\ \Biggl. \sinc \left[\pi \left( \left(\eta - 1\right) \left(f_c + \Delta f a_0/4\right) - \frac{\left(1+\eta\right) n}{2T} \right)\left(T - |\tau|\right)\right] \Biggr|.
\label{eq:GSFM_BAAF}
\end{multline}

\begin{figure}[ht]
\centering
\includegraphics[width=1.0\textwidth]{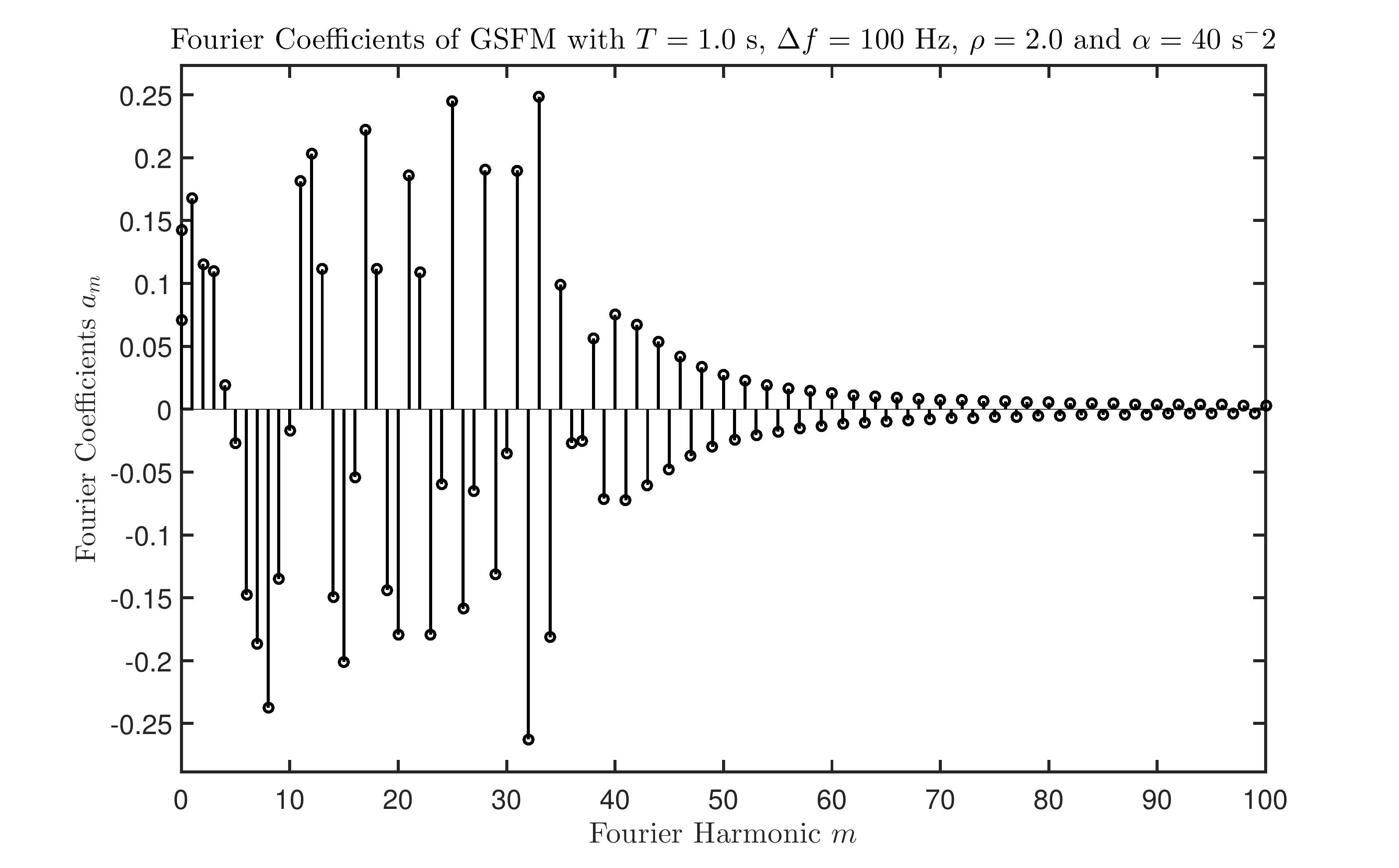}
\caption[Fourier Series coefficients $a_m$ of a GSFM waveform with an even-symmetric IF.]{Fourier Series coefficients $a_m$ of a GSFM waveform with an even symmetric IF of duration $T = 1.0$ s, $\Delta f = 100$ Hz, $\rho = 2.0$, and $\alpha = 40$ s$^{-2}$.  Unlike the SFM's IF, the GSFM's IF contains contributions from many Fourier harmonics resulting in a non-periodic IF function.  This non-periodic IF function removes the high sidelobes in time-delay and Doppler seen in the SFM's AF while preserving the SFM's desirable bandwidth and Doppler sensitivity properties.}
\label{fig:GSFM_Fourier}
\end{figure}

The Fourier Series coefficients of the GSFM's IF function play a crucial role in determining the GSFM's AF shape.  Setting $\rho = 1.0$ produces an SFM waveform.  The resulting Fourier series for that SFM's IF function is $a_m = 1$ for $m = 1$ and $0$ elsewhere with the fundamental harmonic being $f_m$.  Plugging these values into \eqref{eq:GSFM_NAAF} and \eqref{eq:GSFM_BAAF} result in the special cases of the NAAF (\ref{eq:SFM_NAAF}) and BAAF (\ref{eq:SFM_BAAF}) of the SFM and exhibit periodicity in both time-delay and Doppler as explained earlier.  The Fourier series coefficients of a GSFM with a non-symmetric IF function with $\rho = 2.0$, derived in Appendix \ref{subsec:GSFM_Fourier_Coeff_I}, are expressed as 
\begin{equation}
a_0 = \dfrac{S\{2\sqrt{\alpha}T\}}{2\sqrt{\alpha}T},
\end{equation}
\vspace{-2em}
\begin{multline}
a_m = \dfrac{1}{2\sqrt{\alpha}T}\biggl\{\cos\left(2\pi\alpha \left(\dfrac{m}{2T\alpha}\right)^2 \right)\biggl[S\bigl\{2\sqrt{\alpha} z_1 \bigr\} -  S\bigl\{2\sqrt{\alpha}z_2 \bigr\}\biggr]  \\  -\sin\left(2\pi\alpha \left(\dfrac{m}{2T\alpha}\right)^2 \right)\biggl[C\bigl\{2\sqrt{\alpha} z_1 \bigr\} -  C\bigl\{2\sqrt{\alpha}z_2 \bigr\} \biggr] \biggr\},
\end{multline}
\vspace{-2em}
\begin{multline}
b_m = \dfrac{1}{2\sqrt{\alpha}T}\biggl\{\sin\left(2\pi\alpha \left(\dfrac{m}{2T\alpha}\right)^2 \right)\biggl[S\bigl\{2\sqrt{\alpha} z_1 \bigr\} - S\bigl\{2\sqrt{\alpha}z_2 \bigr\} \biggr]  \\  - \cos\left(2\pi\alpha \left(\dfrac{m}{2T\alpha}\right)^2 \right)\biggl[C\bigl\{2\sqrt{\alpha} z_1 \bigr\} -  C\bigl\{2\sqrt{\alpha}z_2 \bigr\}\biggr]\biggr\}
\end{multline}
	where $C\{\}$ and $S\{\}$ are the Fresnel Integrals, $z_1 = T + \left(\frac{m}{2T \alpha}\right)$, and $z_2 = T - \left(\frac{m}{2T\alpha}\right)$.  As an illustrative example, Figure \ref{fig:GSFM_Fourier} shows the Fourier series coefficients $a_m$ for a GSFM with an even-symmetric IF function (derived in Appendix \ref{subsec:GSFM_Fourier_Coeff_II}) with duration $T = 1.0$ s, $\Delta f = 100$ Hz, $\rho = 2.0$, $\alpha = 40$ s$^{-2}$.  Unlike the SFM IF's Fourier series which contains only a single harmonic, the GSFM IF's Fourier series $a_m$ contains contributions from many harmonics that decay in magnitude with increasing $m$.  Each differently weighted harmonic contribution in the GBF arguments in (\ref{eq:GSFM_NAAF}) and (\ref{eq:GSFM_BAAF}) destructively interfere with one another for time-delay and Doppler values outside the mainlobe resulting in reduced sidelobe levels.  
\vspace{1em}
	
\begin{figure}[ht]
\centering
\includegraphics[width=1.0\textwidth]{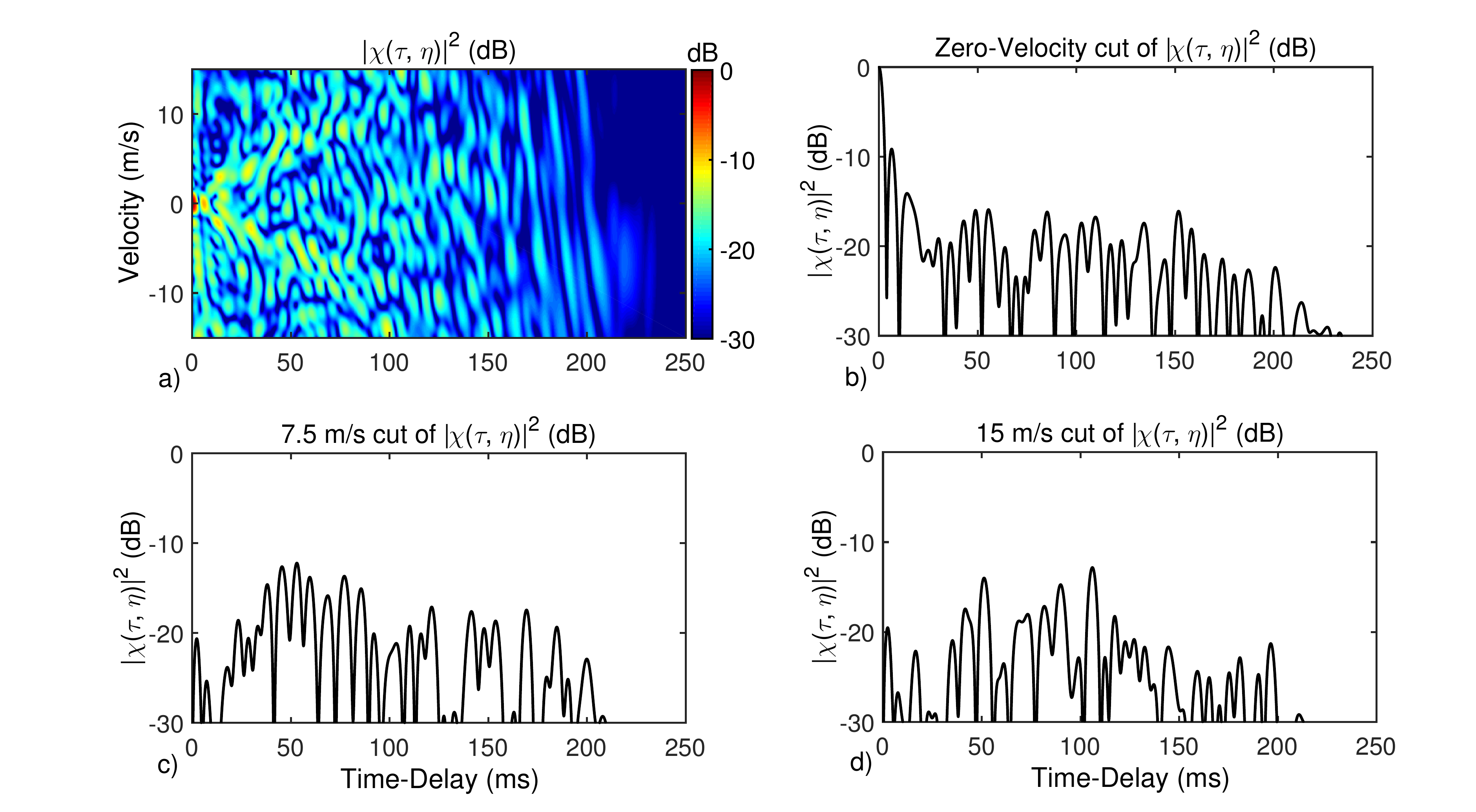}
\caption[BAAF (a), zero-velocity (b), 7.5 m/s (c), and 15 m/s (d) cuts of the BAAF of a GSFM waveform.]{BAAF (a), zero-velocity (b), 7.5 m/s (c), and 15 m/s (d) cuts of the BAAF of a GSFM with duration $T = 250$ ms, $\Delta f = 200$ Hz, $\rho = 2.0$, $\alpha$ = 40 s$^{-2}$, $f_c$ = 2000 Hz, and a swept bandwidth $\Delta f$ of 200 Hz.  Because the IF of the GSFM has a time varying period, this waveform attains high range resolution while maintaining the Doppler sensitivity of the SFM.}
\label{fig:GSFM_Cuts}
\end{figure}

\begin{figure}[ht]
\centering
\includegraphics[width=1.0\textwidth]{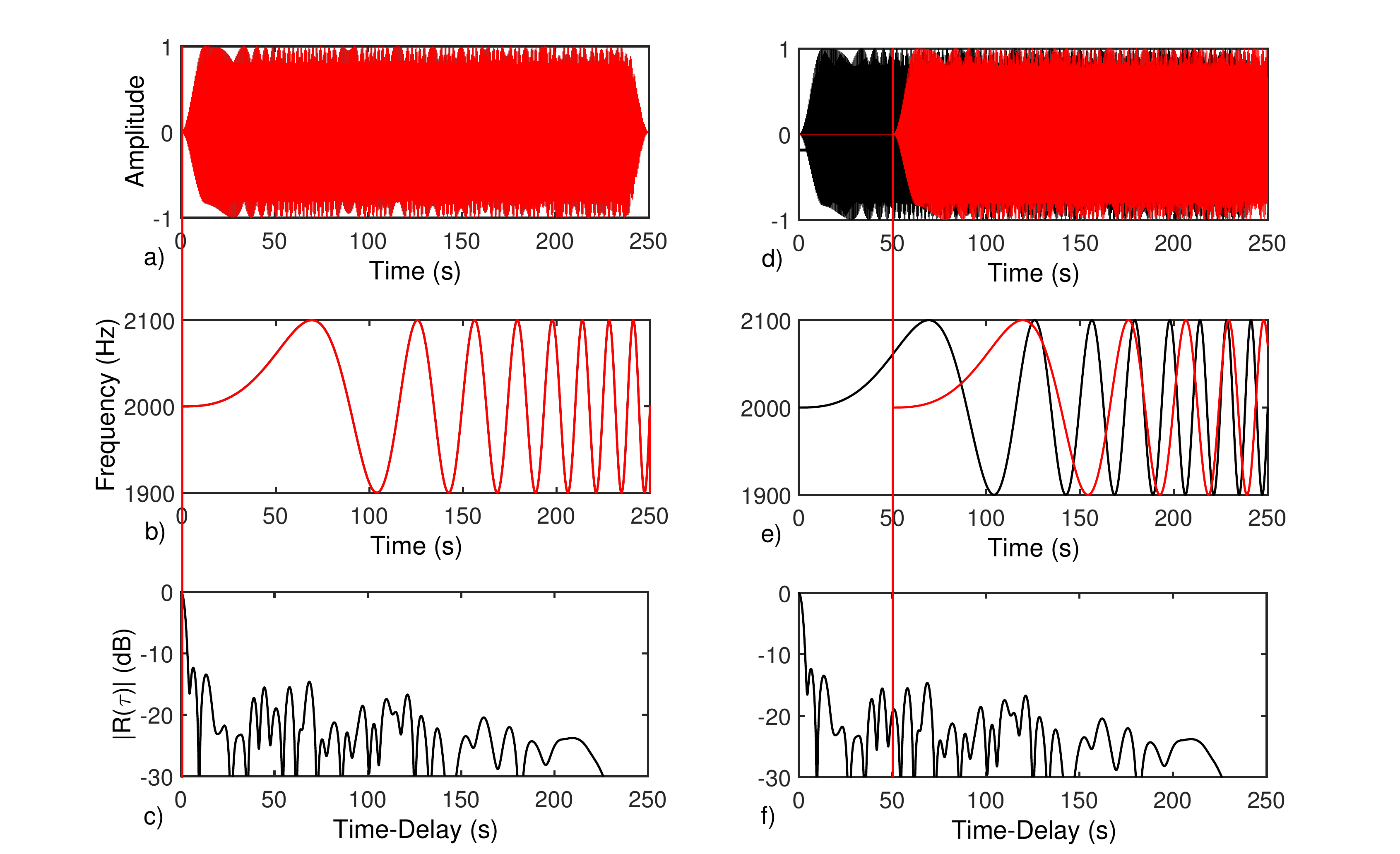}
\caption[Illustration of the origin of the GSFM Auto Correlation sidelobes for two time delays.]{Illustration of the origin of the GSFM Auto Correlation sidelobes for two time delays.  The first column shows the correlation operation in time (a), the waveforms in IF (b), and the location of the ACF peak (mainlobe) in (c).  In the second column, when correlating the waveform at time-delay of 50 ms (d), the IF functions do not overlap in time and frequency (e) resulting in a low sidelobe level (f).  The IF functions do not substantially overlap in time and frequency for any time-delay outside the mainlobe.}
\label{fig:GSFM_IF_Fig}
\end{figure}

These reduced sidelobe characteristics are illustrated in Figure \ref{fig:GSFM_Cuts} which displays the AF and 0, 7.5, and 15 m/s cuts of the AF for a the GSFM pictured in Figure \ref{fig:GSFM}.  The harmonic extent of the Fourier series for the GSFM's IF compared to that of the SFM's is loosely analogous to the spectral content of the LFM waveform compared to the CW waveform.  The CW has its energy concentrated about its center frequency $f_c$ and therefore possesses a periodic time-voltage characteristic.  The LFM's spectrum on the other hand contains energy across a wide band of frequencies thus removing the periodicity of the LFM's time-voltage characteristic.  Removing the periodicity in the LFM's time-voltage characteristic results in range resolution and sidelobe levels that are vastly superior to that of the periodic CW waveform.  As a result of the GSFM's non-periodic IF function, the GSFM's AF does not contain any large periodic sidelobes like the SFM's AF. This is illustrated in Figure \ref{fig:GSFM_IF_Fig}.  Now, any time-delay greater than the extent of the GSFM's mainlobe in time-delay results in sidelobes much lower than that of the SFM.  

\chapter{Performance Evaluation of the GSFM Waveform: Ambiguity Function Shape}
\label{ch:GSFM_Eval_AF}

The main design goal of the GSFM waveform is to achieve a thumbtack AF shape in order to resolve closely spaced targets in time-delay (range) and Doppler (velocity).  The results from Chapter \ref{ch:GSFM} showed that GSFM does indeed possess a thumbtack AF.  However, there are a number of known waveforms that achieve a thumbtack AF \cite{Cook, Levanon, Gladkova, Li} raising the question of whether the GSFM waveform's design and resulting AF is an improvement over other well known thumbtack waveforms.  This chapter looks at three of the most well-known and better performing waveforms which attain a thumbtack AF; the Costas \cite{Costas}, BPSK \cite{Ricker}, and QPSK \cite{QuadPhase} waveforms and compares their AF performance to that of the GSFM's.  Performance is characterized by the waveform's AF mainlobe shape (both width and range-Doppler coupling) and Peak Sidelobe Level (PSL) for Matched Filtering (MF) and Mis-Matched Filtering (MMF).

\section{Measures of Performance}
\label{sec:GSFM_Eval_Methods}
The return echoes from a collection of point targets distributed in range and velocity creates a return echo signal with copies of the transmit waveform at their respective delays and Doppler values.  When this echo signal is processed with a bank of MF's tuned to different Doppler scaling values, the resulting output is a 2-D function of the target distribution.  This 2-D function can be loosely interpreted as a superposition of the target's AF's scaled in magnitude by the target's echo strength and delayed in time and Doppler scaling factor by the target's range and velocity respectively.  Resolving multiple targets in range and Doppler is the two-dimensional analogue of resolving multiple sinusoids in frequency encountered in spectral analysis.  The mainlobe width determines the waveform's ability to resolve closely spaced targets in range and Doppler in the same way that the mainlobe width of the frequency response of a spectral analysis window determines that window's ability to resolve sinusoids closely spaced in frequency.  The thumbtack AF analysis must also account for the mainlobe possessing range-Doppler coupling, the bias in the range estimate of the target resulting from the Doppler effect of the target's motion.  A thumbtack AF must therefore possess minimal range-Doppler coupling in order to minimize the bias in estimating the range and Doppler of the target and maximize the waveform's ability to resolve two closely spaced targets in range and Doppler \cite{Rihaczek}.  The thumbtack AF's volume, which is bounded, must be spread out as evenly as possible.  Much like in spectral analysis where lower sidelobes allow for detecting weak sinusoids in the presence of a strong sinusoid, lower sidelobe levels in a thumbtack AF allow for detecting a weak target in the presence of a stronger target.  Increasing the waveform's time bandwidth product TB spreads the thumbtack AF's bounded volume evenly over a larger region of range and Doppler values thereby reducing the average sidelobe level.  The mainlobe widths in delay and Doppler, the range-Doppler coupling, and the sidelobe levels are the main performance characteristics of the thumbtack AF.  

\section{Mainlobe Performance}
\label{mainlobe}
A thumbtack AF's mainlobe determines the waveform's ability to estimate the range and velocity of a target and to resolve multiple targets in range and velocity.  This section focuses on the mainlobe widths in time-delay (range) and velocity and the mainlobe's range-Doppler coupling.  The BAAF and NAAF mainlobe can be approximated by a second order Taylor series expansion \cite{Wilcox, Altes}.  The contour of the mainlobe at some height $1-\epsilon$ is always an ellipse known as the Ellipse Of Ambiguity (EOA).  The EOA for the BAAF and NAAF are expressed as \cite{Altes}
\begin{equation}
1 - |\chi\left(\tau, \eta\right)|^2 = \epsilon = \beta_{rms}^2\tau^2 + 2\gamma_{B}\tau\left(\eta-1\right) + \lambda_{B}^2\left(\eta-1\right)^2
\label{eq:BAAF_EOA}
\end{equation}
\begin{equation}
1 - |\chi\left(\tau, \phi\right)|^2 = \epsilon = \beta_{rms}^2\tau^2 + 2\gamma_{N}\tau\phi + \lambda_{N}^2\phi^2
\label{eq:NAAF_EOA}
\end{equation}
where $\beta_{rms}^2$ is the Root Mean Square (RMS) bandwidth of the waveform and determines the time-delay (range) sensitivity of the waveform, $\lambda_B^2$ and $\lambda_N^2$ are the RMS broadband and narrowband Doppler sensitivity respectively, and $\gamma_B$ and $\gamma_N$ are the broadband and narrowband range-Doppler coupling factors for the AF mainlobe. The expressions in \eqref{eq:BAAF_EOA} and \eqref{eq:NAAF_EOA} were first derived in Refs. \cite{Wilcox, Altes}.  The RMS bandwidth is the same for the NAAF and the BAAF and is expressed as \cite{Wilcox, Altes}
\begin{equation}
\beta_{rms}^2 = \dfrac{1}{2\pi}\int_{\infty}^{\infty}\left(\omega - \omega_0\right)^2 |S\left(\omega\right)|^2 d\omega = \int_{\Omega_t} | \dot{s}\left(t\right)|^2 dt - \left| \int_{\Omega_t} s\left(t \right) \dot{s}^*\left(t\right) dt \right|^2
\label{eq:RMSBand}
\end{equation}
where $\omega_0$ is the wavform's spectral centroid, $S\left(\omega\right)$ is the waveform's Fourier transform, $\dot{s}\left(t\right)$ is the first time derivative of the waveform $s\left(t\right)$ and $\Omega_t$ represents the region of support in time of the waveform.
The broadband and narrowband Doppler sensitivity terms are expressed below as \cite{Wilcox, Altes}
\begin{equation}
\lambda_{B}^2 = \int_{\Omega_t} t^2 |\dot{s}\left(t\right)|^2 dt - \left| \int_{\Omega_t} t s\left(t\right) \dot{s}^*\left(t\right)dt \right|^2
\label{eq:RMSDopB}
\end{equation}
\begin{equation}
\lambda_{N}^2 = 4\pi^2\int_{\Omega_t} \left(t-t_0\right)^2 |s\left(t\right)|^2 dt
\label{eq:RMSDopN}
\end{equation}
The broadband and narrowband coupling terms are expressed as \cite{Wilcox, Altes}
\begin{equation}
\gamma_B = \int_{\Omega_t} t |\dot{s}\left(t\right)|^2 dt - \Re \Biggl\{ \int_{\Omega_t} \dot{s}\left(t\right)s^*\left(t\right)dt \times \int_{\Omega_t} t s\left(t\right)\dot{s}^*\left(t\right) dt \Biggr\}
\label{eq:gamma_B}
\end{equation}
\begin{equation}
\gamma_N = -2\pi \Im \Biggl\{\int_{\Omega_t} ts\left(t\right)\dot{s}^*\left(t\right) dt \Biggr\}
\label{eq:gamma_N}
\end{equation}
where $\Re \{\}$ in (\ref{eq:gamma_B}) denotes the real component of the two integrals and $\Im \{\}$ denotes the imaginary component of the integral in (\ref{eq:gamma_N}).  The estimation variances for time-delay and Doppler are \cite{Ricker, Nehorai}
\begin{equation}
\sigma_{\tau}^2\geq \left(\dfrac{1+SNR}{2 SNR^2}\right)\left(\dfrac{\lambda^2}{\beta_{rms}^2\lambda^2 - \gamma^2}\right)
\end{equation}
\begin{equation}
\sigma_{\eta}^2\geq \left(\dfrac{1+SNR}{2 SNR^2}\right)\left(\dfrac{\beta_{rms}^2}{\beta_{rms}^2\lambda^2 - \gamma^2}\right)
\end{equation}
where $SNR$ is the signal to noise ratio at the output of the MF.  For fixed $SNR$, $\beta_{rms}^2$, and $\lambda^2$ the only way to minimize the estimation variances is to minimize $\gamma$.  The minimum value $\gamma$ can take is zero and the resulting minimum variances are given by \\[0.5em]
\begin{equation}
\sigma_{\tau}^2 = \left(\dfrac{1+SNR}{2 SNR^2}\right)\left(\dfrac{1}{\beta_{rms}^2}\right)
\end{equation}
\begin{equation}
\sigma_{\eta}^2 = \left(\dfrac{1+SNR}{2 SNR^2}\right)\left(\dfrac{1}{\lambda^2}\right) 
\end{equation} \\[0.5em]
The design objective for the AF mainlobe is now clear; design a waveform whose IF function yields zero range Doppler coupling.  Having zero range Doppler coupling in the mainlobe of the waveform's AF minimizes the estimation variance for a target's time-delay (range) and Doppler (velocity) and also maximizes the waveform's ability to resolve closely spaced targets. 
\vspace{1em} 

The EOA parameters of the GSFI and GCFI GSFM waveforms for both the broadband and narrowband models are derived in Appendix \ref{ch:appendix_D}.  Both waveforms have comparable mainlobe performance.  To avoid redundancy this section focuses solely on the GSFI GSFM waveform and will simply refer to it as the GSFM.  The RMS bandwidth of the GSFM for both the broadband and narrowband models are expressed as 
\begin{equation}
\beta_{rms}^2 = \dfrac{\pi^2 {\Delta f}^2}{2}\left[1 - \dfrac{2 C\{4 \pi \alpha {\left(T/2\right)}^{\rho}, 1/\rho \}}{\rho T \left(4 \pi \alpha \right)^{\frac{1}{p}}} -\dfrac{8 S^2\{2 \pi \alpha {\left(T/2\right)}^{\rho}, 1/\rho \}}{{\left(\rho T\right)}^2 \left(2 \pi \alpha \right)^{\frac{2}{p}}} \right]
\end{equation}
where $C\{4 \pi \alpha {\left(T/2\right)}^{\rho}, 1/\rho \}$ and $S\{2 \pi \alpha {\left(T/2\right)}^{\rho}, 1/\rho \}$ are the GCFI and GSFI.  The Doppler sensitivity parameters are given below
\begin{multline}
\lambda_B^2 = \dfrac{\pi^2 f_c^2 T^2}{3} + \dfrac{\pi^2 {\Delta f}^2 T^2}{6} - \dfrac{\pi^2 {\Delta f}^2 C\{4\pi \alpha T^{\rho}, 1/\rho\}}{2 T \left(4\pi \alpha\right)^{\frac{1}{p}}} \\ + \dfrac{2 \pi \Delta f f_c}{T \left(2\pi \alpha\right)^{\frac{3}{\rho}}}S\{2\pi \alpha T^{\rho}, 3/\rho\}
\label{lambda_B}
\end{multline}
\begin{equation}
\lambda_N^2 = \dfrac{\pi^2 T^2}{3}
\label{lambda_N}
\end{equation}
As shown in Appendix \ref{ch:appendix_D}, as the waveform becomes more narrowband (i.e, the waveform's fractional bandwidth decreases) and the narrowband AF assumptions are invoked, the result in (\ref{lambda_B}) converges to the narrowband Doppler sensitivity term in (\ref{lambda_N}).
\vspace{1em}

As described earlier, the range-Doppler coupling factor is the parameter we wish to minimize.  Generally speaking it may be possible to design a waveform whose modulation function is distributed in time and frequency in such a way such that the integrals in (\ref{eq:gamma_B}) and (\ref{eq:gamma_N}) are zero.  However, the most straightforward way to achieve this is to employ a waveform with an even-symmetric IF function.  The range-Doppler coupling factor $\gamma$ for the GSFM, as shown in Appendix \ref{ch:appendix_D}, is exactly zero for the even symmetric IF version of the GSFM.  This means the GSFM's mainlobe is perfectly symmetric in range and Doppler.  Therefore, the GSFM, like any waveform with an even-symmetric IF function, achieves the minimum estimation variance for time-delay and Doppler and optimal resolution of closely spaced targets in range and Doppler for a given $\beta_{rms}^2$ and $\lambda^2$.    
\vspace{1em}

\begin{figure}[h]
\centering\includegraphics[width=1.0\textwidth]{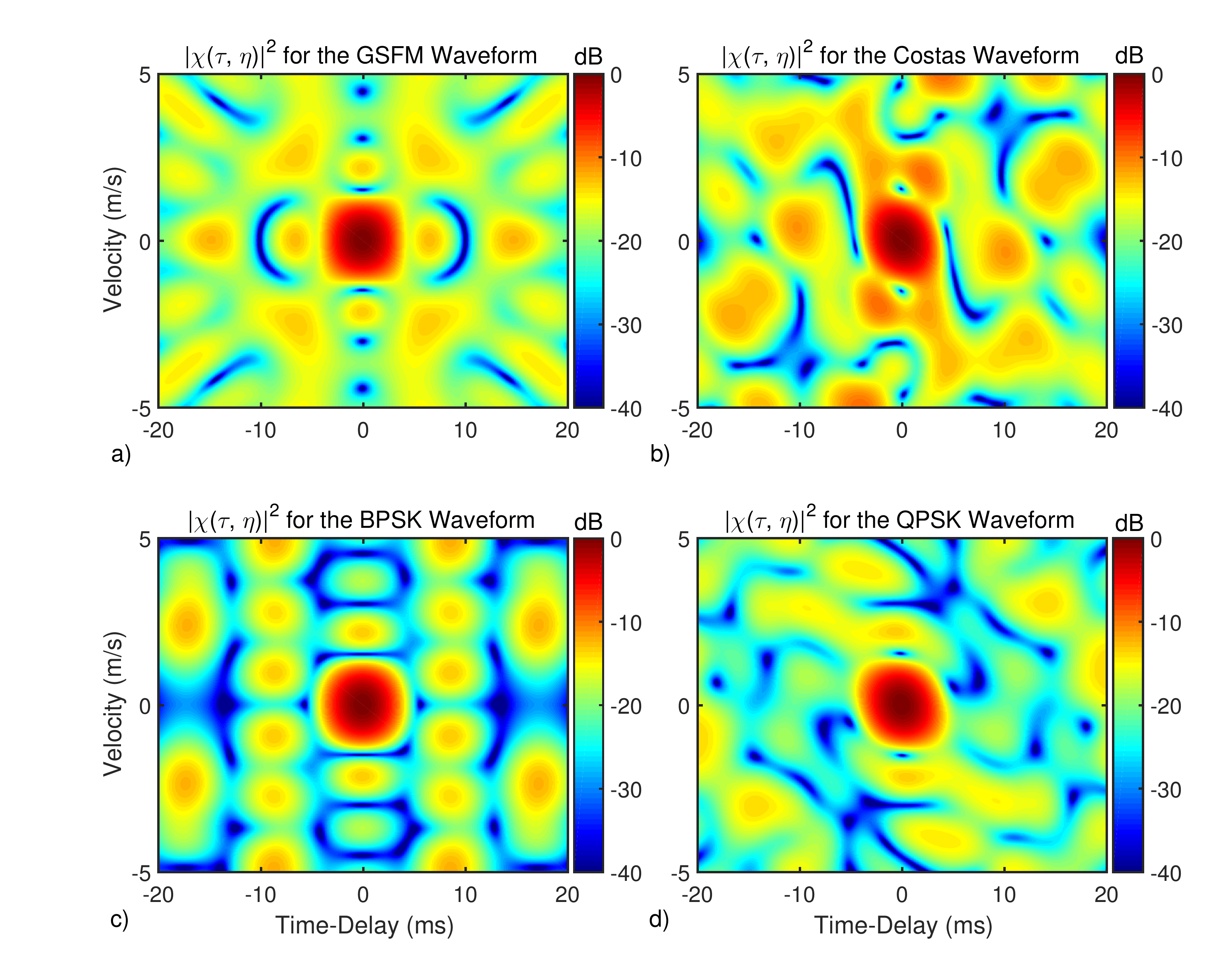}
\caption[Illustration of the AF mainlobes of the GSFM (a), Costas (b), BPSK (c), and QPSK (d) waveforms.]{Illustration of the AF mainlobes of the GSFM (a), Costas (b), BPSK (c), and QPSK (d) waveforms.  Both the BPSK and GSFM waveforms attain a zero range-Doppler coupling factor while the Costas and QPSK waveforms attain a small but non-zero range-Doppler coupling factor.}
\label{fig:MainlobeI}
\end{figure}

Figure \ref{fig:MainlobeI} shows the AF mainlobes for a design example of the GSFM, Costas, and BPSK waveforms.  The waveforms have a pulse length $T = 250$ ms, bandwidth $\Delta f = 200$ Hz, and carrier frequency $f_c = 2000$ Hz.  The Costas waveform had $N = 15$ chips spaced $13.3$ Hz with each chip being tapered by a Tukey window \cite{Harris} using an $85\%$ taper.  The BPSK waveform used $N = 34$ chips with each chip tapered by a Hanning window.  The QPSK waveform was realized by performing a binary-phase to quaternary-phase transformation of a 70-bit binary MLSR code.  The code sizes and tapering of the Costas, BPSK, and QPSK waveforms were empirically determined to produce the same RMS bandwidths and Doppler sensitivities as the GSFM such that all the waveforms' resulting AF's had the same mainlobe widths in time-delay and Doppler as the GSFM for the same pulse length, bandwidth, and carrier frequency values described above.  Upon visual inspection of Figure \ref{fig:MainlobeI}, the GSFM and BPSK waveforms clearly have symmetric mainlobes while the Costas and QPSK waveforms possess small but non-zero range-Doppler coupling in their mainlobes.  
\vspace{1em}

\begin{figure}[h]
\centering
\includegraphics[width=1.0\textwidth]{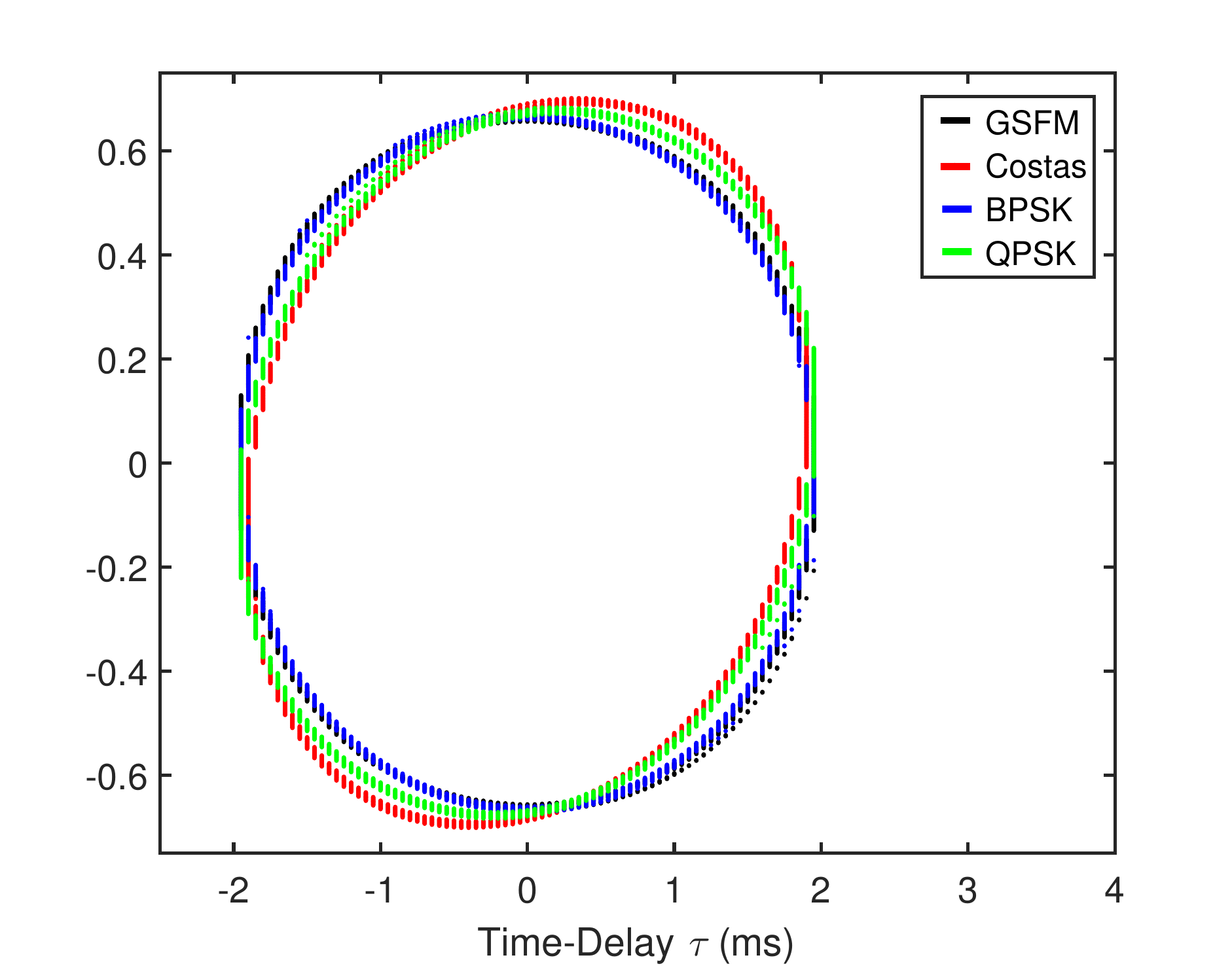}
\caption[Illustration of the AF mainlobes of the GSFM, Costas, BPSK, and QPSK waveforms from Figure \ref{fig:MainlobeI} for $\epsilon=0.5$.]{Illustration of the AF mainlobes of the GSFM, Costas, BPSK, and QPSK waveforms from Figure \ref{fig:MainlobeI} for $\epsilon=0.5$.  The GSFM and BPSK EOA's completely overlap one another as they are perfectly symmetric ellipses.  The Costas and QPSK EOA's are clearly small but non-zero range-Doppler coupling with the Costas EOA possessing the most range-Doppler coupling.}
\label{fig:MainlobeII}
\end{figure}

Figure \ref{fig:MainlobeII} shows the EOA of each mainlobe for $\epsilon = 0.5$ shows that this is indeed the case. The EOA's of the GSFM and BPSK waveforms are perfectly symmetric ellipses and overlap each other in the figure.  This is a result of the waveforms having an even-symmetric IF function.  The Costas and QPSK waveforms have small but non-zero range-Doppler coupling.  The QPSK waveform's AF, while still attaining a thumbtack shape, is notably different from that of the BPSK.  This result is not surprising.  Taylor and Blinchikoff \cite{QuadPhase} noted that adjacent sub-pulses of the QPSK waveform overlap in time and can cause a slight degradation in the waveform's ACF performance.  Additionally, work by Levanon and Freedman \cite{LevanonI} built upon the results in \cite{QuadPhase} and showed that AF of a QPSK can at time substantially differ from its binary-phase counterpart.  For this particular example, the Costas waveform's range-Doppler coupling is greatest.  The non-zero range-Doppler coupling of the Costas waveform is due to the IF function which is determined by the Costas code for the frequency hopping sequence.  Costas codes are Unit Allocation (UA) codes meaning that one frequency slot of the waveform is occupied at one time slot.  Therefore, a Costas code can never be even symmetric.  However, as the number of chips in the Costas waveform is increased, the Costas code generates an IF function that  evaluates the integrals in (\ref{eq:gamma_B}) and (\ref{eq:gamma_N}) to a value that asymptotically approaches zero and has a $1/N^2$ dependence \cite{RickerTW}.  While the Costas and QPSK waveforms range-Doppler coupling closely approaches zero, the BPSK and GSFM waveform attain exactly zero range-Doppler coupling. 
\vspace{1em}
 
\vspace{5em}

\section{Sidelobe Performance}
\label{sidelobes}
A thumbtack AF's sidelobe levels determine the waveform's ability to detect a weak target in the presence of a stronger target.  Therefore, in addition to the waveform's AF possessing an uncoupled mainlobe, the waveform must also achieve the lowest sidelobe levels possible.  However, one cannot reduce a waveform's AF sidelobes to arbitrarily low levels.  The volume of the AF is bounded, and so reducing the sidelobe levels in one region requires that the sidelobe levels increase in another region.  The best one can do to reduce is employ a waveform with a thumbtack AF.  The thumbtack AF will have its bounded volume spread as evenly as possible.  Consider a transmit waveform with bandwidth $\Delta f$ and pulse length $T$ with unit energy.  The volume of the NAAF of that waveform is bounded to the square of the energy and therefore attains unity AF volume.  The AF will extend in time-delay from $-T\leq \tau\leq T$ and in Doppler shift $-\Delta f \leq \phi \leq \Delta f$ \cite{Rihaczek}.  The thumbtack AF will distribute the AF's volume evenly and so the average sidelobe level of the AF will be $\frac{1}{4T\Delta f}$ where $T\Delta f$ is the waveform's TBP.  Increasing the waveform's TBP spreads the AF's bounded volume across a larger region in time-delay and Doppler thus reducing the average sidelobe level of the AF.  While the average sidelobe level is reduced by increasing the TBP, the AF will have peak sidelobes which are a direct result of the waveform being time and band-limited.  The PSL of a waveform's AF does not necessarily reduce proportionally with TBP and is generally a function of the spread of the waveform's energy in frequency and time \cite{Rihaczek}.  Any two waveforms with the same TBP will have the same average sidelobe level but not necessarily the same PSL.
\vspace{1em}

The mainlobe and sidelobe levels of a waveform's AF can be quantified by analyzing the ratio of the area of the AF's mainlobe in time-delay and Doppler to Woodward's resolution constants, which for unit energy waveforms, is defined for time-delay and Doppler as 
\begin{equation}
A_{\tau} = \dfrac{\int_{-\infty}^{\infty}|\chi\left(\tau, 0\right)|^2 d\tau}{|\chi\left(0,0\right)|^2} = \int_{-\infty}^{\infty}|\chi\left(\tau, 0\right)|^2 d\tau
\label{eq:WoodRange}
\end{equation} \\ [0.5em]
\begin{equation}
A_{\eta} = \dfrac{\int_{-\infty}^{\infty}|\chi\left(0, \eta\right)|^2 d\eta}{|\chi\left(0,0\right)|^2} = \int_{-\infty}^{\infty}|\chi\left(0, \eta\right)|^2 d\eta
\label{eq:WoodDoppler}
\end{equation}\\ [0.5em]
Note that (\ref{eq:WoodRange}) and (\ref{eq:WoodDoppler}) focus on the zero Doppler and Time-Delay cuts respectively of the AF.  Rihaczek showed\cite{Rihaczek} that the respective ratios of (\ref{eq:WoodRange}) and (\ref{eq:WoodDoppler}) to their mainlobe widths $A_{\tau_0}$ and $A_{\eta_0}$ are proportional to
\begin{equation}
\dfrac{A_{\tau}}{A_{\tau_0}} \propto \beta_{rms} 
\label{eq:RangeRatio}
\end{equation} 
\begin{equation}
\dfrac{A_{\eta}}{A_{\eta_0}} \propto \lambda_{B} 
\label{eq:DopRatio}
\end{equation} 
where $\beta_{rms}$ is the waveform's RMS bandwidth given in (\ref{eq:RMSBand}) and $\lambda_B$ is the waveform's Doppler sensitivity given in in (\ref{eq:RMSDopB}).  An expression similar to (\ref{eq:DopRatio}) holds for the NAAF.  The time-delay and Doppler mainlobe widths and sidelobe levels behave in the same manner and therefore this discussion will for simplicity focus solely on time-delay resolution with the understanding that the same analysis applies for the Doppler domain.  
\vspace{1em}

\begin{figure}[p]
\centering
\includegraphics[width=1.0\textwidth]{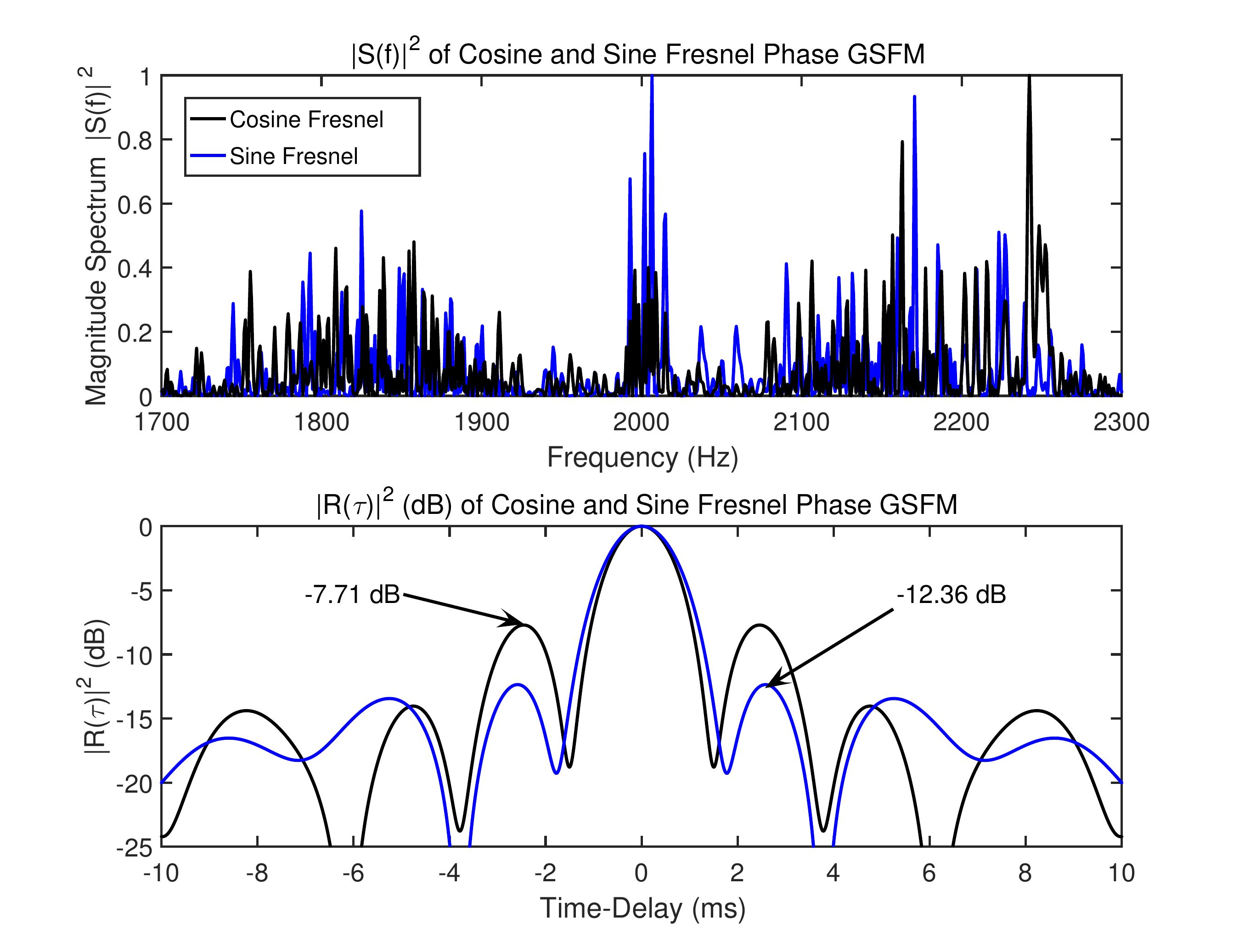}
\caption[Spectra (a) and ACF's (b) of two GSFM waveforms, one using the GCFI phase (\ref{GSFM_Phi_II}) and the other using the GSFI phase (\ref{GSFM_Phi_I}).]{Spectra (a) and ACF's (b) of two GSFM waveforms, one using the GCFI phase (\ref{GSFM_Phi_II}) and the other using the GSFI phase (\ref{GSFM_Phi_I}).  Both waveforms have a duration $T = 0.5$ s, $f_c = 2$ KHz, swept bandwidth $\Delta f = 500$ Hz, $\rho = 2.75$, and cycles $C = 35$.  The GCFI GSFM has a strong concentration of spectral energy at higher frequencies than the GSFI GSFM does resulting in a larger RMS bandwidth.  As a result of this, the GCFI GSFM waveform has a narrower mainlobe and higher sidelobes than the GSFI GSFM waveform.}
\label{fig:CosineSineACF}
\end{figure}

As $\beta_{rms}$ increases, there is increasing area in $A_{\tau}$ (the ACF) and decreasing area in the mainlobe.  This means that the waveform's time-delay resolution has improved at the expense of higher and a greater number of sidelobes.  When $\beta_{rms}$ decreases, there is decreasing area in $A_{\tau}$ (the ACF) and increasing area in the mainlobe implying that the sidelobe levels are lower and less in number at the expense of reduced time-delay resolution.  Recall that $\beta_{rms}$ is largely determined by the shape of the waveform's spectrum.  Concentrations of energy at higher frequencies get more heavily weighted by the $\omega^2$ term in \eqref{eq:RMSBand} and therefore will result in greater $\beta_{rms}$ and vice-versa.  Therefore, a collection of different waveforms with the same swept bandwidth $\Delta f$ but with different spectrum shapes will have have different mainlobe widths and sidelobe levels in their ACF.  This is illustrated in Figure \ref{fig:CosineSineACF}.  The GCFI GSFM has a stronger concentration of spectral energy at higher frequencies than the GSFI GSFM does which results in a larger RMS bandwidth.  This in turn results in a narrower mainlobe but a much higher PSL of -7.71 dB than the GSFI which has a $-3$ dB mainlobe width that is $11.4\%$ wider but with a PSL of -12.36 dB.  
\vspace{1em}

The mainlobe/sidelobe analysis presented above considered only the zero Doppler and Time-Delay axis of the AF.  For the GSFM waveform employing an even-symmetric IF function, the PSL's of the GSFM's AF typically occur either at the zero Doppler axis (ACF) or the zero time-delay axis.  However, a waveform's AF sidelobe behavior off-axis can substantially vary from the zero Doppler and Time-Delay axis of the AF \cite{Rihaczek, Ricker}.  When evaluating the performance of thumbtack AF sidelobe behavior, the entire range-Doppler plane should be analyzed.  This section analyzes the PSL's of the GSFM's AF and compares them to the Costas, BPSK, and QPSK waveforms over a range of TBP's.  In addition to analyzing PSL's using the MF, this section also analyzes PSL's when using a MMF to reduce sidelobes in exchange for a wider mainlobe width and loss in output SNR.  

\clearpage
\subsection{Matched Filter Sidelobe Performance}
\label{subsubsec:MF}
\begin{figure}[p]
\centering
\includegraphics[width=1.0\textwidth]{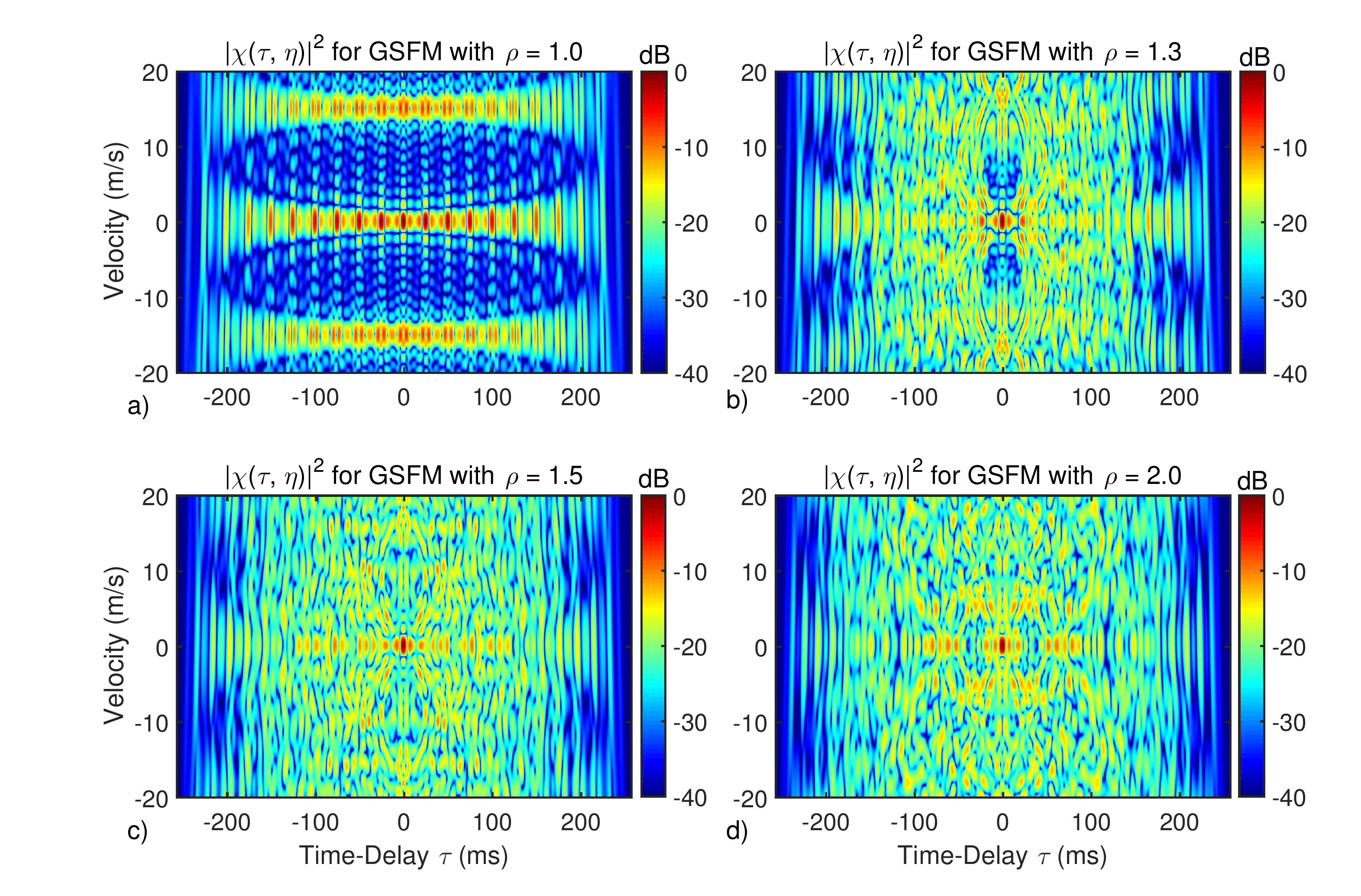}
\caption[BAAF's of four GSFM waveforms with fixed modulation parameter $\alpha$ and varying parameter $\rho$ values of 1.0 (a), 1.3 (b), 1.5 (c), and 2.0 (d).]{BAAF's of four GSFM waveforms with fixed modulation parameter $\alpha$ and varying parameter $\rho$ values of 1.0 (a), 1.3 (b), 1.5 (c), and 2.0 (d).  Each AF has a distinct mainlobe at the origin but the sidelobe levels and locations change drastically with $\rho$.  }
\label{fig:Four_Rho}
\end{figure}

\begin{figure}[h]
\centering
\includegraphics[width=1.0\textwidth]{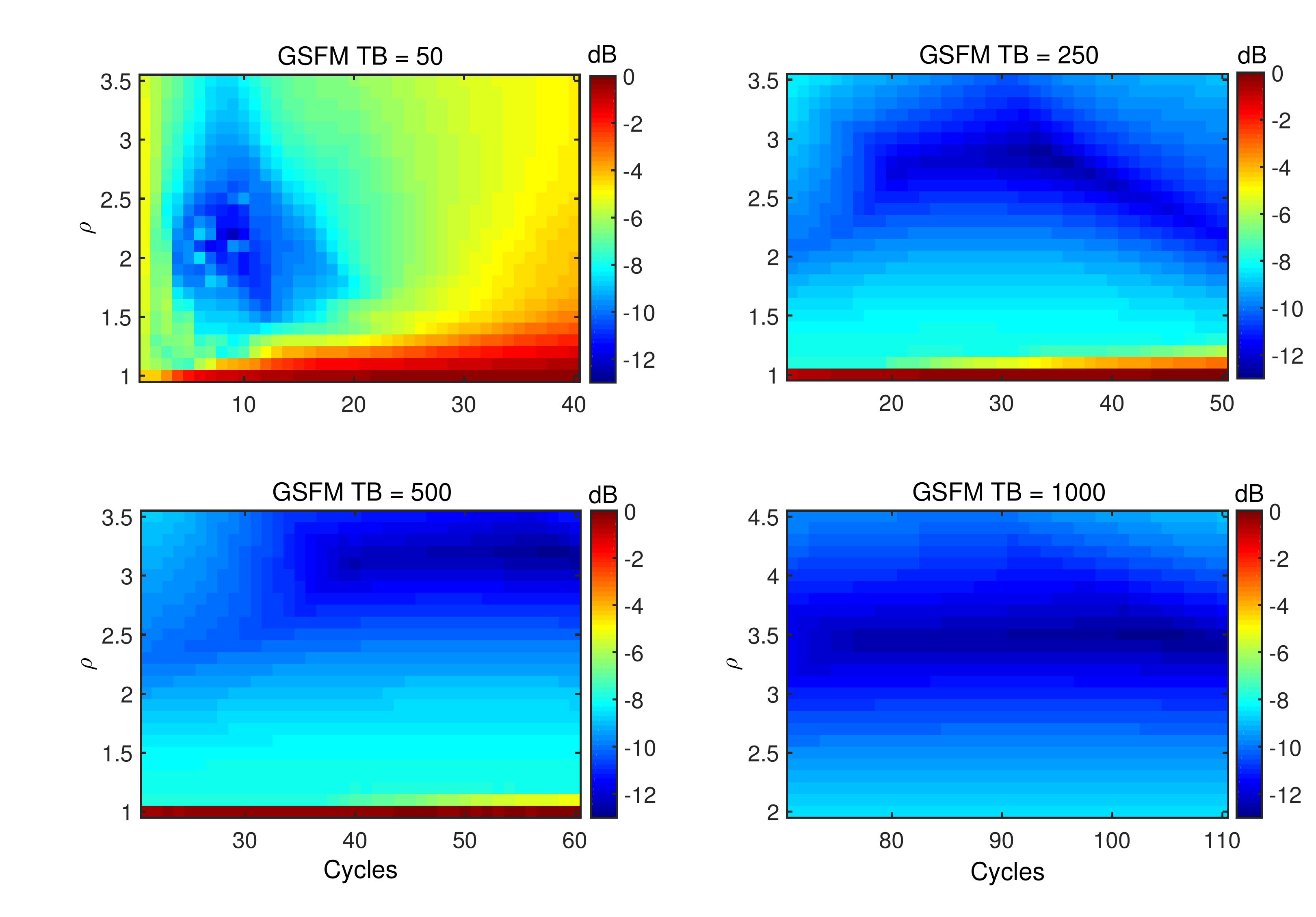}
\caption[PSL values for the GSFM with TBPs of (a) 50, (b) 250, (c) 500, and (d) 1000 as a function of GSFM parameters $\alpha$.]{PSL values for the GSFM with TBPs of (a) 50, (b) 250, (c) 500, and (d) 1000 as a function of GSFM parameters $\alpha$ (expressed in number of cycles in the IF function) and $\rho$.  Note the change of axis in each panel.  For each case there is a relatively flat region near the minimum PSL value, showing there are a collection of $\alpha$ and $\rho$ values that generate GSFM waveforms with low PSL's.}
\label{fig:PSL}
\end{figure}

The GSFM's parameters $\alpha$ and $\rho$ give the waveform designer flexibility in choosing the waveform that is most "thumbtack-like".  The previous section showed that the GSFM's AF possesses a perfectly symmetric mainlobe simply by utilizing an even-symmetric IF function.  However, the PSL of the GSFM's AF changes substantially with different $\alpha$ and $\rho$ values which is illustrated in Figure \ref{fig:Four_Rho}.  For each value of $\rho$, there is a distinct mainlobe at the origin whose width in range and Doppler does not vary substantially.  However, the height and locations of the sidelobes do vary substantially with changing $\rho$ suggesting there exists an optimum value for $\rho$ that produces a minimum PSL.  This depedence on $\rho$ is indeed confirmed in Figure \ref{fig:PSL} which shows the PSL of the GSFM AF for a wide range of $\alpha$ (expressed as number of cycles C) and $\rho$ values for TBP's of 50, 250, 500, and 1000.  In all cases, when $\rho = 1$ (the SFM variant of the GSFM), the PSL of the AF is highest.  In other words, to achieve a thumbtack like AF with the lowest sidelobe levels possible, the least desirable variant of the GSFM to use is in fact the SFM.  For the region where $\rho > 1.0$ , there is a single local minimum.  Additionally, the area near the local minimum is largely flat meaning there are there several combinations of $\alpha$ and $\rho$ that nearly achieve the same minimal PSL's in their AF.  Rather than needing a particular combination of $\alpha$ and $\rho$ to achieve thumbtack AF’s with low sidelobes, a wide variety of values for $\alpha$ and $\rho$ can be used to achieve the desired AF. This multitude of design options leaves the waveform designer with more flexibility in waveform selection.
\vspace{1em}

\clearpage

\clearpage

\begin{figure}[!htb]
\centering
\includegraphics[width=1.0\textwidth]{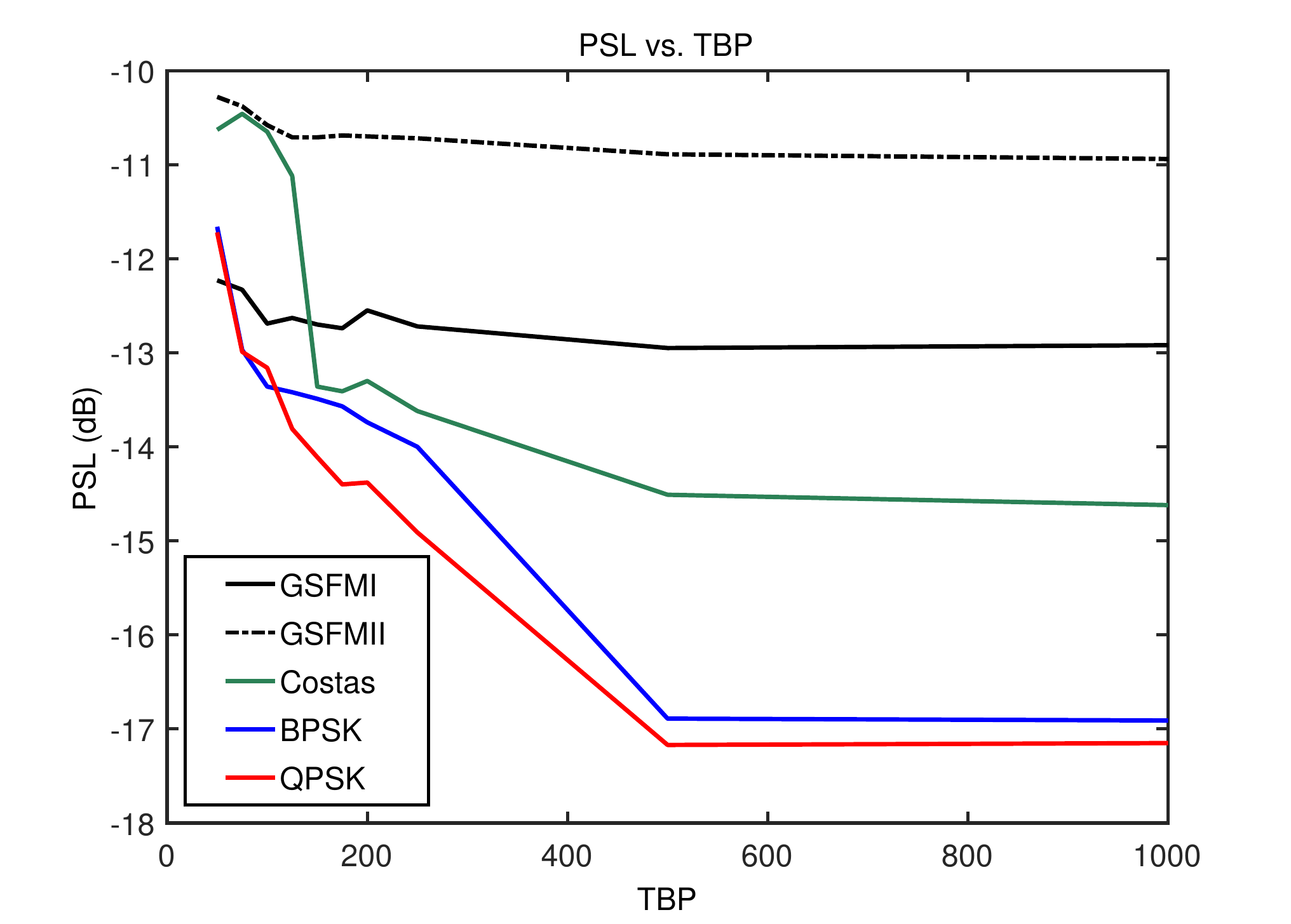}
\caption[Minimum PSL values (in dB) for the GSFM using the GSFI (I) and the GCFI (II), Costas, BPSK, and QPSK waveforms for a broad range of TBP's.]{Minimum PSL values (in dB) for the GSFM using the GSFI (I) and the GCFI (II), Costas, BPSK, and QPSK waveforms for a broad range of TBP's.  The GSFM possesses the lowest PSL of all waveforms when the TBP is 50 and the BPSK/QPSK waveforms perform best overall for TBPs greater than 50.}
\label{fig:PSLPlotDiss}
\end{figure}

The next consideration is to compare the PSL values of these variants of the GSFM to the Costas and BPSK waveforms.  For each of the ten TBP's tested, 1000 Costas and BPSK waveforms were generated and the PSL's from their resulting AF were computed.  Figure \ref{fig:PSLPlotDiss} below lists the lowest PSL values of each waveform for each time bandwidth product.  The first thing to note is the difference in PSL values between the GSFM's using the GSFI phase (labeled GSFM I) and the GCFI phase (labeled as GSFM II).  For each TBP, GSFMI possesses a lower PSL by roughly 2 dB.  Inspection of the waveforms' AFs showed that the PSL's were almost always the ACF sidelobes as well.  GSFMII possesses a larger RMS bandwidth meaning it's ACF mainlobe is narrower in width but will have higher sidelobes than that of GSFMII.  Both GSFM waveforms also have lower sidelobes than the Costas waveform for TBP's less than 150.  When comparing the GSFM to the BPSK and QPSK waveforms, the GSFM only possesses a lower PSL when the TBP is 50.  Overall, the BPSK and QPSK waveforms performed the best for TBPs greater than 50.  We conjecture that the reason why the BPSK and QPSK waveforms performed best for larger TBPs is a direct result of large fractional bandwidth, the ratio of the waveform's bandwidth $\Delta f$ to carrier frequency $f_c$.  As a waveform's fractional bandwidth increases, the Doppler scaling effect introduces substantial time-compression of the waveform.  The adjacent chips in the waveform destructively interfere with one another resulting in reduced PSLs.  However, the amount of trials required to confirm this conjecture is extreme and beyond the scope of this dissertation.

\subsection{Mis-Matched Filter Sidelobe Performance}
\label{subsubsec:MMF}
While the MF, which maximizes the output SNR, is the ideal filter for detection, the PSL of a waveform's MF can be less than optimal for practical applications.  Reducing the AF sidelobes can be accomplished in a manner similar to what's done in spectral analysis.  Applying a taper function in time to a sinusoidal signal results in a spectrum whose PSL is reduced at the cost of a widened mainlobe.  Correspondingly, reducing the sidelobes in time-delay and Doppler is accomplished by tapering the edges of the waveform's energy density spectrum $|S\left(f\right)|^2$ and time energy density $|s\left(t\right)|^2$ respectively \cite{Rihaczek, Levanon}.  The tapering reduces $\beta_{rms}$ and $\lambda_B$ respectively which results in reducing (\ref{eq:RangeRatio}) and (\ref{eq:DopRatio}).  This means that the tapering reduces sidelobe levels while widening the mainlobes in time-delay and Doppler.  Typically, the tapering is applied to the detection filter rather than the waveform itself to minimize the waveform's PAPR \cite{Cook}.  The new detection filter is no longer matched to the transmit waveform and is therefore known as a Mis-Match Filter (MMF).  The resulting AF is now a CAF between the waveform and the MMF.  This mis-match between the transmit waveform and the detection filter results in a reduction in the output SNR defined here as the SNR Loss (SNRL).  MMF design introduces a tradeoff between reducing PSL in exchange for SNRL and a widened mainlobe in time-delay and Doppler \cite{Rihaczek, Cook}.  
\vspace{1em}

There is extensive literature on MMF processing for a variety of waveforms, including \cite{Cook, Rihaczek, Levanon}.  Specifically, for phase coded waveforms like the BPSK/QPSK, there exist a multitude of approaches to MMF design to reduce Cross-Correlation Function (CCF) or CAF sidelobes \cite{Li, Stoica, Baden}.  Work by \cite{Zejak} designed MMF's for Costas waveforms that reduced the PSL of the CCF.  Of particular interest to the author was \cite{SFM_MMF} which investigated using MMF design to reduce the PSL of the SFM waveform.  One MMF design from \cite{SFM_MMF} reduced the SFM's ACF PSL to $-33$ dB in exchange for an SNRL of $20.5$ dB.  The MMF design was unable to substantially reduce Doppler sidelobes however, attaining 1.5 dB PSL.  While the PSL reduction in time-delay is impressive, the SNRL is likely too great a cost to implement on a practical active sonar system.  Many active sonar systems operate in low SNR scenarios where a $18$ dB reduction in output SNR would severely degrade detection performance.  However, if waveform MMF's can be designed to achieve even moderate PSL reduction at the cost of no more than a few dB of SNRL, then the resulting design trade-off might be worth serious consideration.  
\vspace{1em}

\begin{figure}[!h]
\includegraphics[width=1.0\textwidth]{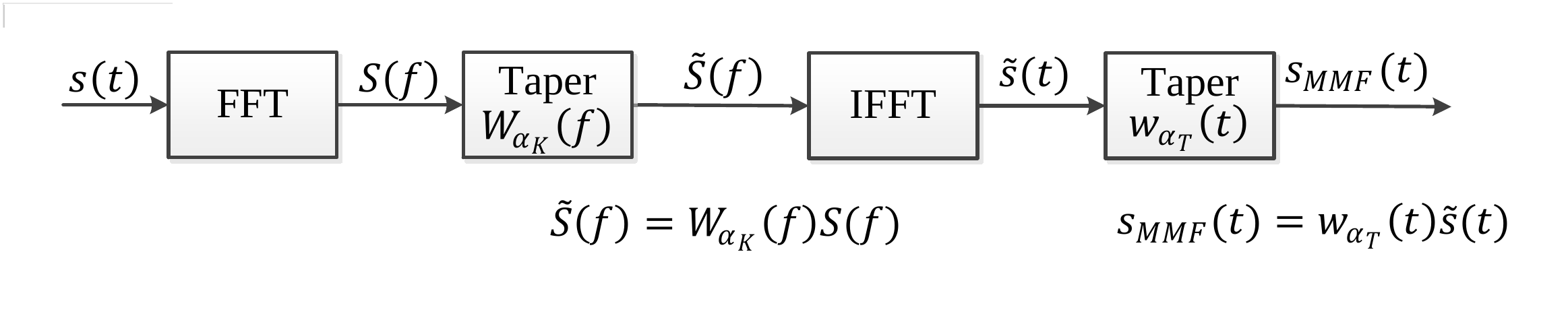}
\caption[MMF Processing chain.]{MMF Processing chain.  The waveform's MF is first tapered in frequency using a Kaiser-Bessel window with parameter $\alpha_K$.  The frequency tapered MMF is then tapered in time using a Tukey window with tapering parameter $\alpha_T$.  The resulting MMF, $\tilde{s}\left(t\right)$, has been tapered in frequency to reduce the time-delay sidelobes and in time to reduce the Doppler sidelobes.}
\label{fig:MMF_Processing}
\end{figure}

Figure \ref{fig:MMF_Processing} shows the processing used to generate a MMF for the GSFM waveform.  The original waveform $s\left(t\right)$ is tapered in the frequency domain by a Kaiser window \cite{Kaiser} with shape parameter $\alpha_K$ to reduce time-delay sidelobes.  The frequency tapered MMF is then transformed back to the time domain where it is then tapered in time by a Tukey window with shape parameter $\alpha_T$.  The Kaiser window is an approximation to the Slepian window which optimizes the ratio given by (\ref{eq:RangeRatio}).  Therefore, this window will reduce the time-delay sidelobes and maximize the area of the ACF's mainlobe.  Concentrating area in the mainlobe helps to minimize the SNRL that results from applying the MMF.  The Tukey window was chosen for mitigating the Doppler sidelobes for two reasons.  First, the Tukey window smoothly transitions from a Rectangular to Hanning window by increasing the shape parameter $\alpha_T$ from 0 to 1 which provides sufficient Doppler sidelobe suppression for the TBP's tested.  Additionally, a Tukey window is a commonly employed amplitude tapering function for tapering the acoustic signal that is transmitted on a piezoelectric transducer.  Therefore, utilizing a Tukey window for suppressing Doppler sidelobes is well matched to the transmit waveform which helps to minimize the SNRL resulting from applying a MMF.  
\vspace{1em}

\begin{figure}[!h]
\includegraphics[width=1.0\textwidth]{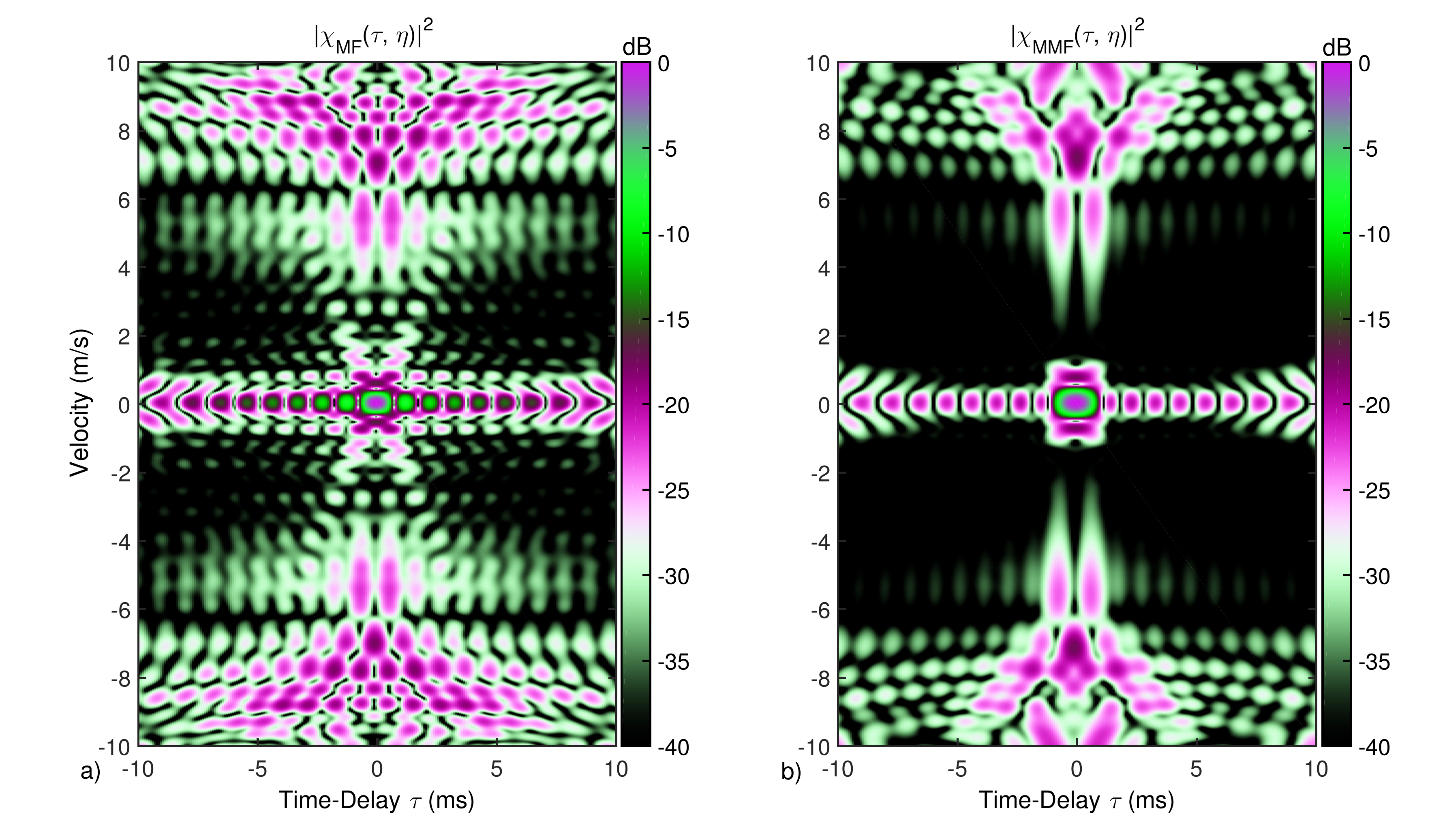}
\caption[BAAF using the MF (a) and MMF (b) of a GSFM with with parameters $T = 1.0$ s, $f_c = 2000$ Hz, $\Delta f = 1000$Hz, $\rho = 1.25$ and $C = 27$ cycles.]{BAAF using the MF (a) and MMF (b) of a GSFM with with parameters $T = 1.0$ s, $f_c = 2000$ Hz, swept bandwidth $\Delta f = 1000$ Hz, $\rho = 1.25$ and $C = 27$ cycles.  The Kaiser and Tukey windows used for this MMF had parameter values of $\alpha_K = 14$ and $\alpha_T = 0.6$.  The resulting MMF reduces the waveform's PSL from -7.06 dB to -17.81 dB in exchange for a SNRL of 2.05 dB and a mainlobe that is $27\%$ wider in time-delay and $32\%$ wider in Doppler.}
\label{fig:MMF_AF_Illustration}
\end{figure}

As mentioned earlier, the ratios given by \eqref{eq:RangeRatio} and \eqref{eq:DopRatio} apply to the zero-Doppler and zero-Time-Delay cuts of the AF respectively.  Therefore, the tapering utilized by the MMF has a direct impact on mainlobe/sidelobe behavior of the zero-Doppler and zero-Time-Delay AF cuts and is not guaranteed to suppress off-axis sidelobes in the AF.  However, the GSFM's strongest sidelobes are located in the zero-Doppler and zero-Time-Delay AF cuts.  This means that the GSFM MMF's should have a direct impact on sidelobe suppression.  Figure \ref{fig:MMF_AF_Illustration} demonstrates the difference between applying an MF and a MMF to a GSFM waveform with a TBP of 1000.  In the MF's AF, there is a strong concentration of high sidelobes in range (zero-Doppler cut), the strongest of which is 7.06 dB below the mainlobe. There is also a peak sidelobe in Doppler that is apprximately 13.2 dB below the mainlobe response.  In the MMF CAF, the tapering has strongly suppressed the high time-delay and Doppler sidelobes.  The resulting PSL is now 17.81 dB below the mainlobe peak.  The sidelobe reduction was accomplished at the cost of a 2.05 dB reduction in the mainlobe height (SNRL) and a mainlobe that is $27\%$ wider in time-delay and $32\%$ wider in Doppler.  For a modest SNRL, the MMF applied to the GSFM waveform was able to reduce the PSL in time-delay and Doppler by 10.75 dB. 
\vspace{1em}

The MMF design for the GSFM can also be applied to the Costas, BPSK, and QPSK waveforms and provides a basis of comparison to the GSFM's MMF performance.  MMF's with a range of $\alpha_K$ and $\alpha_T$ values were applied to the waveforms from the MF PSL simulations for TBP's of 125, 250, 500, and 1000.  For each TBP, the SNRL and mainlobe widening in time-delay and Doppler were computed for the waveform and $\alpha_K$ and $\alpha_T$ values that generated the minimum PSL.  The results of these simulations are shown in tables \ref{table:MMF_PSL_GSFM}-\ref{table:MMF_PSL_BQPSK}.  These tables display the PSL, SNRL, -3 dB mainlobe width in time-delay $\Delta\tau$, -3 dB mainlobe width in Doppler $\Delta\eta$, and the products of the -3 dB mainlobe widths to provide an overall measure of mainlobe extent in both time-delay and Doppler.  It is insightful to first compare between GSFM's using the GSFI and GCFI phase versions of the GSFM, again denoted as GSFMI and GSFMII respectively.  These results are shown in Table \ref{table:MMF_PSL_GSFM}.  GSFMI achieved lower PSL's for all TBP's except for 1000 and lower SNRL's for TBP's of 125 and 1000.  Additionally, GSFMI's mainlobe width is less than that of GSFMII except for when the TBP is 500.  The MMF results for the Costas waveform is shown in Table \ref{table:MMF_PSL_Costas}.  For all TBP's tested, the Costas waveform has a lower SNRL and mainlobe extent than both of the GSFM waveforms, but its PSLs are higher than the GSFM waveforms.  In fact, the MMF only provided an improvement in PSL for a TBP of 1000.  Analysis of the individual waveforms showed that the PSL's were off axis sidelobes and were not suppressed using tapering.  However, the tapering does reduce the mainlobe peak which is equal to the SNRL.  Therefore, not reducing the sidelobe levels while reducing the mainlobe height results in a higher PSL.  The MMF performance of the BPSK and QPSK waveforms are shown in Table \ref{table:MMF_PSL_BQPSK}.  Both waveforms' PSLs are comparable to or better than that of the GSFM waveforms and their SNRLs are far lower than the SNRLs of the GSFM waveforms.  It is also interesting to note that the mainlobe width in time-delay did not change.  This is because the frequency tapering did not reduce the ACF sidelobes.  The PSL of the BPSK and QPSK AFs corresponded to zero-time-delay cut of the AF.  Therefore, tapering in time increased the PSL of the waveform's AF.  Overall, the MMF of the BPSK and QPSK waveforms attained lower sidelobes, SNRL, and mainlobe extent compared to the GSFM waveforms.  

\begin{table}
\begin{center}
\caption[MMF results for the GSFM waveforms employing the GSFI phase (GSFMI) and GCFI phase (GSFMII).]{MMF results for the GSFM waveforms employing the GSFI phase (GSFMI) and GCFI phase (GSFMII).  The PSL's for both waveforms are comparable but the SNRL is less for the GSFMI for TBP's of 125 and 1000.  Additionally, the GSFMII has a wider mainlobe extent for all TBP's except for 250.}
\begin{tabular}{l|lllll|lllll}
\toprule 
& \multicolumn{5}{c}{GSFM I} & \multicolumn{5}{c}{GSFM II}\\\hline

TBP & PSL & SNRL & $\Delta \tau$ & $\Delta \eta$ & $\Delta\tau\Delta\eta$
& PSL & SNRL & $\Delta \tau$ & $\Delta \eta$ & $\Delta\tau\Delta\eta$\\
\midrule
$125$  & $13.17$ &$0.71$ & $1.18$ & $1.01$ & $1.19$ & $12.86$ & $1.33$ & $1.26$ & $1.14$ & $1.44$\\
$250$  & $14.91$ &$1.49$ & $1.34$ & $1.11$ & $1.49$ & $14.52$ & $1.49$ & $1.26$ & $1.21$ & $1.52$\\
$500$  & $16.91$ &$1.78$ & $1.41$ & $1.12$ & $1.58$ & $16.53$ & $1.48$ & $1.27$ & $1.21$ & $1.54$\\
$1000$ & $17.66$ & $1.58$ & $1.27$ & $1.22$ & $1.55$& $17.81$ & $2.05$ & $1.32$ & $1.27$ & $1.67$\\
\bottomrule
\end{tabular}
\label{table:MMF_PSL_GSFM}
\end{center}
\end{table}

\begin{table}
\begin{center}
\caption[MMF results for the Costas waveform.]{MMF results for the Costas waveform.  Costas waveform attains a lower SNRL than the GSFM for all TBP's tested and its mainlobe is narrower in extent for all TBP's except for 125.  However, the Costas waveform's PSL's are higher than both GSFM's for all TBP's.}
\begin{tabular}{l|lllll}
\toprule 
& \multicolumn{5}{c}{Costas}\\\hline

TBP & PSL & SNRL & $\Delta \tau$ & $\Delta \eta$ & $\Delta\tau\Delta\eta$\\
\midrule
$125$  & $10.55$ & $0.62$ & $1.28$ & $1.02$ & $1.31$\\
$250$  & $13.17$ & $0.59$ & $1.20$ & $1.17$ & $1.40$\\
$500$  & $13.57$ & $0.56$ & $1.23$ & $1.02$ & $1.25$\\
$1000$ & $15.94$ & $1.02$ & $1.19$ & $1.25$ & $1.49$\\
\bottomrule
\end{tabular}
\label{table:MMF_PSL_Costas}
\end{center}
\end{table}

\begin{table}
\begin{center}
\caption[MMF results for the BPSK and QPSK waveforms.]{MMF results for the BPSK and QPSK waveforms.  The PSL's for both waveforms are comparable to or better than the GSFM's.  Additionally, both the BPSK and QPSK waveforms possess a lower SNRL and narrower mainlobe.}
\begin{tabular}{l|lllll|lllll}
\toprule 
& \multicolumn{5}{c}{BPSK} & \multicolumn{5}{c}{QPSK}\\\hline

TBP & PSL & SNRL & $\Delta \tau$ & $\Delta \eta$ & $\Delta\tau\Delta\eta$
& PSL & SNRL & $\Delta \tau$ & $\Delta \eta$ & $\Delta\tau\Delta\eta$\\
\midrule
$125$  & $13.95$ & $0.41$ &$1.00$ &$1.16$ &$1.16$ & $13.88$ & $0.01$ & $1.00$ & $1.02$ &$1.02$\\
$250$  & $14.85$ & $0.51$ &$1.00$ &$1.20$ &$1.20$ & $14.52$ & $0.00$ & $1.00$ & $1.00$ &$1.00$\\
$500$  & $18.16$ & $0.64$ &$1.00$ &$1.23$ &$1.23$ & $17.18$ & $0.01$ & $1.00$ & $1.01$ &$1.01$\\
$1000$ & $20.33$ & $1.01$ &$1.00$ &$1.37$ &$1.37$ & $19.09$ & $0.99$ & $1.00$ & $1.39$ &$1.39$\\
\bottomrule
\end{tabular}
\label{table:MMF_PSL_BQPSK}
\end{center}
\end{table}

\clearpage

\section{Conclusion}
\label{sec:Conclusion}
The results in the previous sections show that the GSFM not only achieves zero range-Doppler coupling in its AF mainlobe, but also achieves lower PSL's than the Costas waveform for TBPs less than 150 and the BPSK waveform for when the TBP is 50.  Additionally, a broad range of GSFM parameters $\alpha$ and $\rho$ nearly achieve the same minimal PSL's in their AF. This shows that rather than being sensitive to small changes in these parameters, a wide variety of values for $\alpha$ and $\rho$ can be used to achieve the desired AF.  This multitude of design options presents more flexibility to the waveform selection to the system's designer.  Applying a MMF to the GSFM can substantially reduce its AF sidelobes.  The MMF PSL performance of the GSFM surpasses that of the Costas waveform for all TBPs in exchange for a SNRL no more than 2.05 dB and a widened mainlobe.  The BPSK and QPSK waveforms' MMFs however attained better PSLs, SNRLs, and a narrower mainlobe than the GSFM's MMFs.  Overall, the results in this chapter show that the GSFM's AF performance is an improvement over some but not all thumbtack waveforms.  The GSFM seems particularly well suited to applications with lower TBPs and the BPSK would be a better candidate waveform for higher TBP designs.  There are a variety of currently deployed active sonar systems that utilize waveforms with a TBP below 150 \cite{Pecknold, Baggenstoss} and so the GSFM might be a strong candidate waveform for such systems.   
\vspace{1em}

Recall however that the Costas and BPSK waveforms' chips were tapered to reduce the RMS bandwidth.  This was done in order to achieve the same mainlobe width in time-delay as a GSFM with a given swept bandwidth $\Delta f$.  Without tapering, the Costas and BPSK waveforms spectral leakage was large enough to result in a RMS bandwidth that substantially exceeded that of the GSFM.  For rectangularly windowed waveforms, the PAPR is $3.0$ dB. For example, the waveforms used in Figure \ref{fig:MainlobeI}, the tapering in time increased the Costas and BPSK waveforms' PAPR to $6.2$ dB and $7.6$ dB respectively.  The QPSK, which possesses a nearly constant envelope, had a PAPR of $3.1$ dB, very close to that of a rectangularly windowed GSFM.  For a fixed peak power limit, the Costas and BPSK waveforms had $3.2$ dB and $4.6$ dB less energy respectively than a rectangular windowed GSFM of the same duration.  An additional consideration is SC, how much of the waveform's energy is concentrated in a specified band of frequencies.  The PAPR and SC provide a measure of how energy efficient each waveform is. Transmitting more energy into the medium produces a stronger target echo which in turn increases the received signal's SNR.  A thorough comparison between the PAPR and SC of the Costas, BPSK, and QPSK waveforms to the GSFM is one of the main topics of the next chapter.


\chapter{Performance Evaluation of the GSFM Waveform: Practical Considerations}
\label{ch:GSFM_Eval_Prac}
The previous chapter examined the GSFM waveform's ability to distinguish closely spaced point targets by analyzing the mainlobe and sidelobe behavior of the GSFM's AF.  The AF also plays a role in determining a waveform's ability to suppress reverberation energy, here referred throughout as Reverberation Suppression (RS).  This is accomplished by analyzing the waveform's Q-Function which is derived from the AF by \eqref{eq:QFunc}.  As was the case with the GSFM's AF shape, the GSFM's Q-Function shape is heavily influenced by the design parameters $\alpha$ and $\rho$.  This gives the waveform designer another performance characteristic to consider when choosing $\alpha$ and $\rho$.  However, the AF and Q-Function shape are not the only considerations in waveform design.  As discussed earlier in Chapters \ref{ch:Intro} and \ref{ch:signalModel}, when transmitting waveforms on piezoelectric transducers, the waveform should be spectrally contained and also possess a low PAPR.  This chapter explores these practical considerations of waveform design and again compares the GSFM's performance to that of the Costas, BPSK, and QPSK waveforms as well as some other well known PC waveforms.  Waveform performance in this chapter is evaluated by the Q-Function, SC, PAPR, and transducer replica data collected at the university's underwater test facility.  

\section{The Q-Function and Reverberation Suppression}
\label{sec:QFunc}
Reverberation refers to the unwanted echoes resulting from bubbles, fish, the sea surface and bottom, and any other acoustic scatterers present in the medium \cite{Urick}.  The Q-function evaluates the total reveberation energy from a waveform's MF tuned to particular Doppler scaling factor and therefore provides a measure of a waveform's RS performance.  Evaluating the reverberation energy for a range of Doppler scaling factors is directly computed from the waveform's AF using \eqref{eq:QFunc}.  The Q-Function is a valid model for scatterers that are uniformly distributed in range, equal in target strength, and stationary relative to the sonar system platform.  Realistic environments typically violate these assumptions, degrading the accuracy of the Q-Function model.  However, the Q-function provides a first order approximation to a waveform's RS performance.  Additionally, the Q-Function's ease of computation and direct relationship to the AF allows the waveform designer to simulate the RS performance of a collection of transmit waveforms in the absence of sea-trial data.
\vspace{1em}

There is an extensive literature dedicated to the evaluation and comparison of the RS performance of transmit waveforms.  Collins and Atkins \cite{Collins} appear to be the first in the published literature to evaluate the Q-Function of the SFM while work by Pecknold \emph{et.al} \cite{Pecknold, PecknoldI} evaluated the Q-Function for a number of waveforms including the CW, LFM, HFM, SFM, and Costas waveforms.  There also exists a number of publications dedicated to modeling reverberation and applying these models to design optimum waveforms in reverberation limited conditions.  Work by Kincaid \cite{Kincaid} explored optimum waveform design using the MF in reverberation-limited conditions.  Brill \emph{et.al} \cite{Zabal} developed a reverberation model that included the random motion of the ocean's surface and generalized the Q-Function for this model.  More recently, Newhall \cite{Newhall} developed a model for reverberation for a rough bottom and moving ocean surface that was based on perturbation theory.  Newhall then used this model to evaluate the RS of a number of waveforms.  One result of particular interest to the author was performed by Ward \cite{Ward}.  Ward analyzed the SFM's RS performance through the Q-Function and then compared the Q-Function's accuracy to a series of sea trials where the sonar platform was undergoing motion.  One of the main results of \cite{Ward} was that the Q-Function is indeed an accurate and valid model for stationary scatterers.  Additionally, Ward \cite{Ward} also showed that when the platform is undergoing motion, the reverberation had Doppler spread that the Doppler sensitive SFM would still suppress resulting in a RS level greater than that predicted by Q-Function analysis.  The results of \cite{Ward} suggest that the Q-Function model results in a more conservative evaluation of a waveform's RS performance.  This is the main reason why this dissertation uses the Q-Function to analyze RS over some of the more sophisticated models described above.
\vspace{1em}

\begin{figure}[ht]
\includegraphics[width=1.0\textwidth]{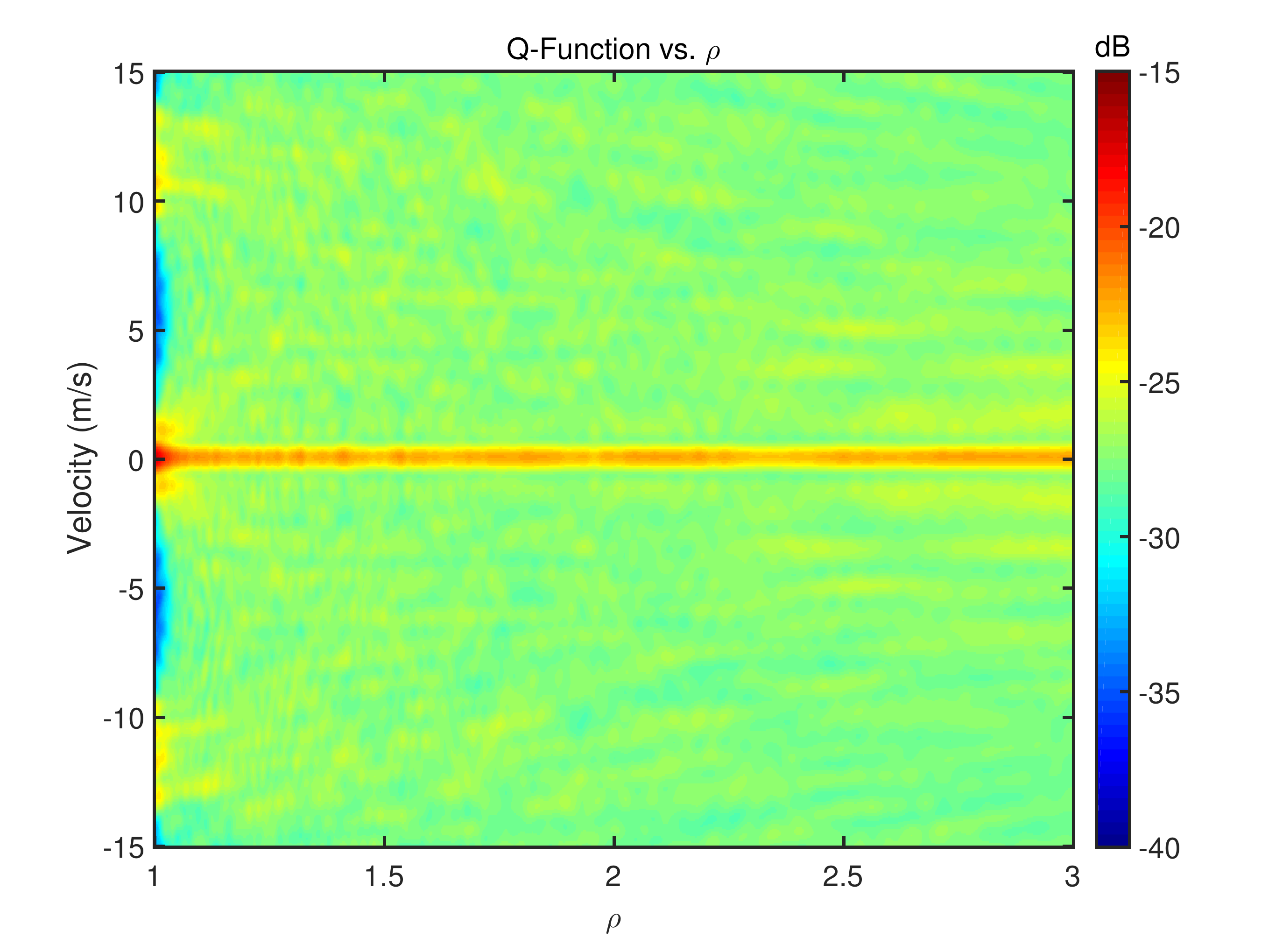}
\caption[GSFM Q-function with varying $\rho$.]{GSFM Q-function with varying $\rho$.  When $\rho = 1.0$ (i.e, an SFM waveform), there are deep notches in the Q-Function which are useful for suppressing reverberation.  As $\rho$ increases, the GSFM's AF becomes more thumbtack like and the resulting Q-Function has nearly equal energy across Doppler removing the deep notches seen in the SFM's Q-Function. }
\label{fig:QRho}
\end{figure}

\begin{figure}[ht]
\includegraphics[width=1.0\textwidth]{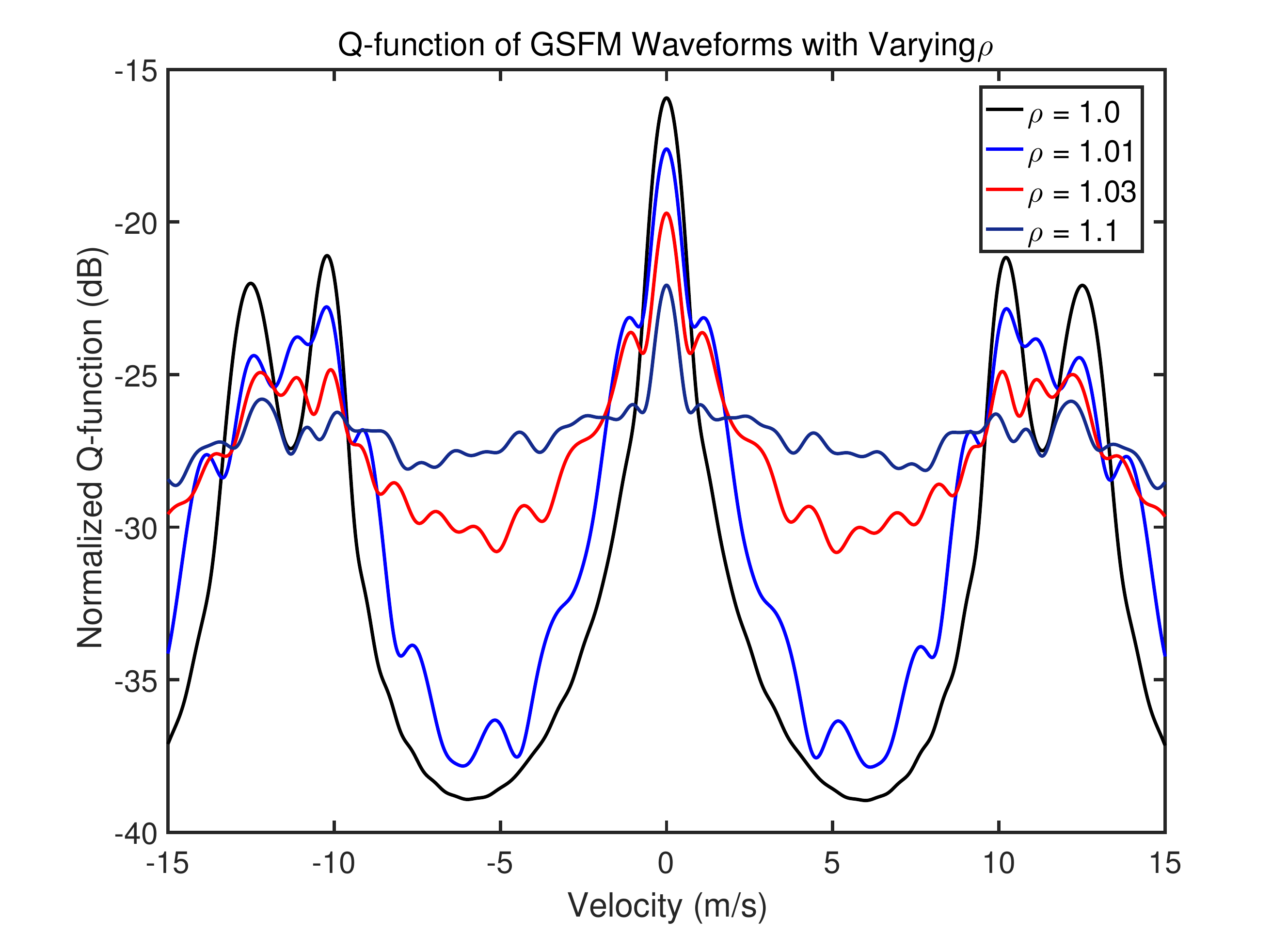}
\caption[Q-function for the GSFM for four different values of $\rho$.]{Q-function for the GSFM for four different values of $\rho$.  Each waveform has a duration $T = 0.5$ s, $f_c = 2$ KHz, and a swept bandwidth $\Delta f = 500$ Hz.  The SFM ($\rho = 1.0$) has the deepest notches.  As $\rho$ increases those notches become less deep and Q-function becomes more uniform.  From $\rho \geq 1.1$, the GSFM is largely a thumbtack waveform and thus its Q-function is nearly uniform across velocity.}
\label{fig:QFuncII}
\end{figure}

As with the AF, the GSFM's Q-function shape is largely influenced by the GSFM parameter $\rho$.  Recall that increasing $\rho$ beyond $1.0$ transitions the GSFM's BAAF from a "bed of nails" AF to a thumbtack AF.  Therefore, there must be a range of $\rho$ values where the resulting Q-function transitions from one where there are deep reverberation nulls in Doppler like the SFM's Q-Function in Figure \ref{fig:SFM_Fig} (d) to one that is nearly uniform like what is shown in Figure \ref{fig:GSFM} (d).  This is indeed the case and is illustrated in Figure \ref{fig:QRho} which displays the Q-function for a collections of GSFM waveforms where $\rho$ was varied from 1.0 to 3.0.  In fact the notches in Doppler dissipate for a value of $\rho$ that is only slightly past 1.0.  Figure \ref{fig:QFuncII} zooms in on this transition and displays the Q-functions for four GSFM waveforms with $\rho = 1.0$ (SFM), $1.01$, $1.03$, and $1.1$.  The notch in Doppler is nearly nonexistent by $\rho = 1.1$.  As $\rho$ increases, the GSFM's Q-Function notch depth decreases and its AF sidelobes decrease.  The notch depth is deepest for $\rho = 1.0$ meaning the SFM is best suited for RS.  The sidelobes are at their lowest when $\rho > 1.0$ as shown in Chapter \ref{ch:GSFM_Eval_AF} meaning the GSFM is better for resolving multiple closely spaced point targets.

\begin{figure}[ht]
\includegraphics[width=1.0\textwidth]{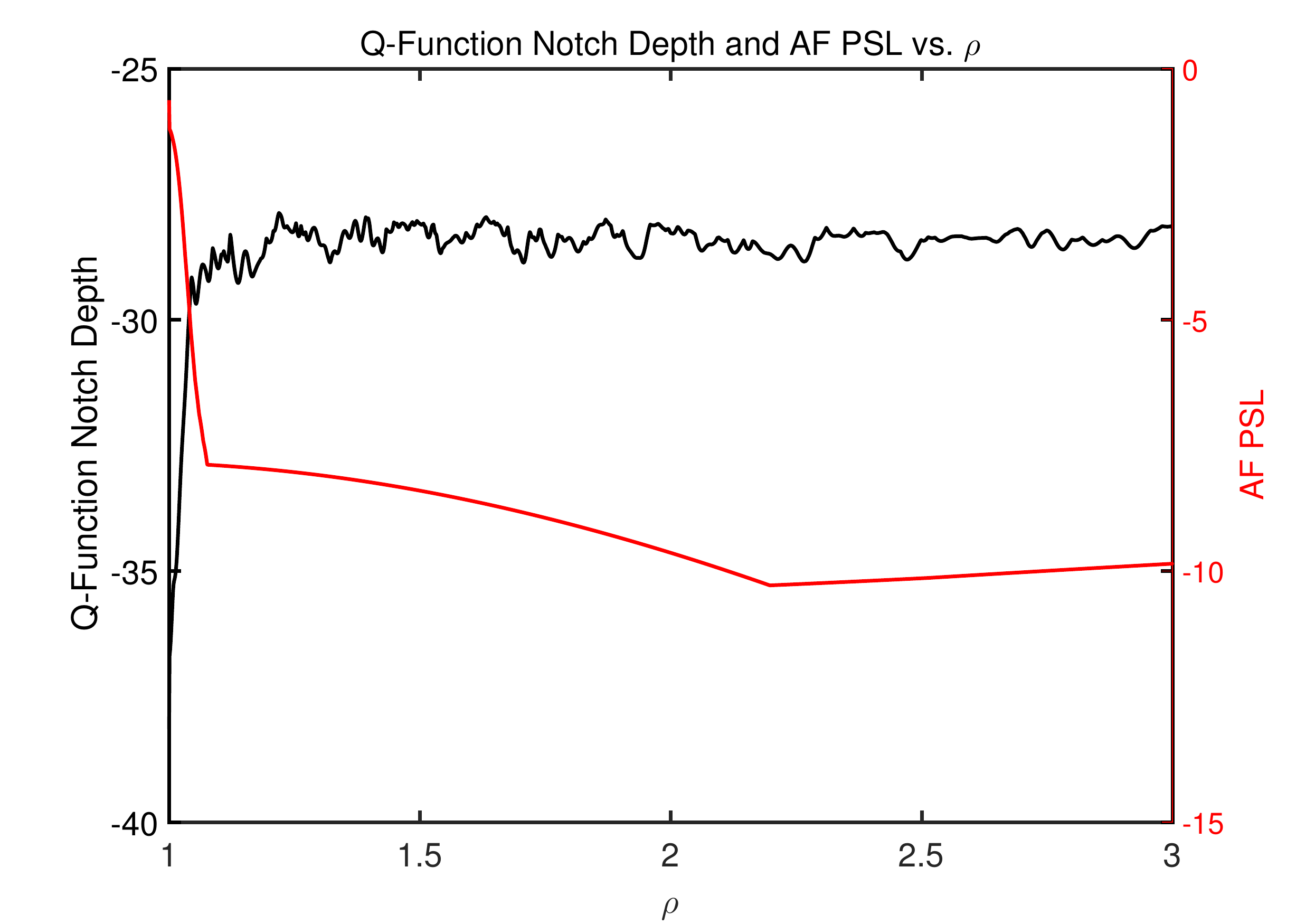}
\caption[Q-function notch depth and PSL vs. $\rho$.]{Q-function notch depth and PSL vs. $\rho$.  As $\rho$ increases, the notch depth disappears and the PSL lowers.  The GSFM waveform cannot attain low notch depth and PSL simultaneously.}
\label{fig:QNotch}
\end{figure}

\vspace{-1.5em}
The AF and Q-Function behavior in $\rho$ raises the intriguing question of whether the GSFM can smoothly tradeoff RS and low PSLs by careful selection of $\rho$.  Figure \ref{fig:QNotch} displays the GSFM's Q-Function notch depth and AF PSL as a function of $\rho$.  The notch depth is at its deepest level and highest PSL when $\rho = 1.0$.  The lowest PSL occurs when $\rho = 2.2$ and the notch is no longer clearly distinguishable from the average level of the waveform's nearly uniform Q-Function.  In the region $1.0 \leq \rho \leq 1.075$, there is a sharp transition where the notch depth increases to $\cong 10 \log_{10}\left(\Delta f\right)$, the average level of the GSFM's Q-function, and the PSL reduces from $0.6$ dB to $-7.8$ dB.  This is the same region where the GSFM waveform's AF transitions from the ``bed of nails'' shape to a thumbtack shape.  The GSFM does not have the ability to jointly suppress reverberation with a deep notch and also maintain low PSLs.  Additionally the GSFM is unable to smoothly trade-off RS notch depth and PSL by varying $\rho$.  The waveform designer is better off utilizing a SFM for RS and a GSFM with $\rho > 1.0$ for lower PSL and the ability to distinguish multiple closely spaced point targets.

\begin{figure}[ht]
\includegraphics[width=1.0\textwidth]{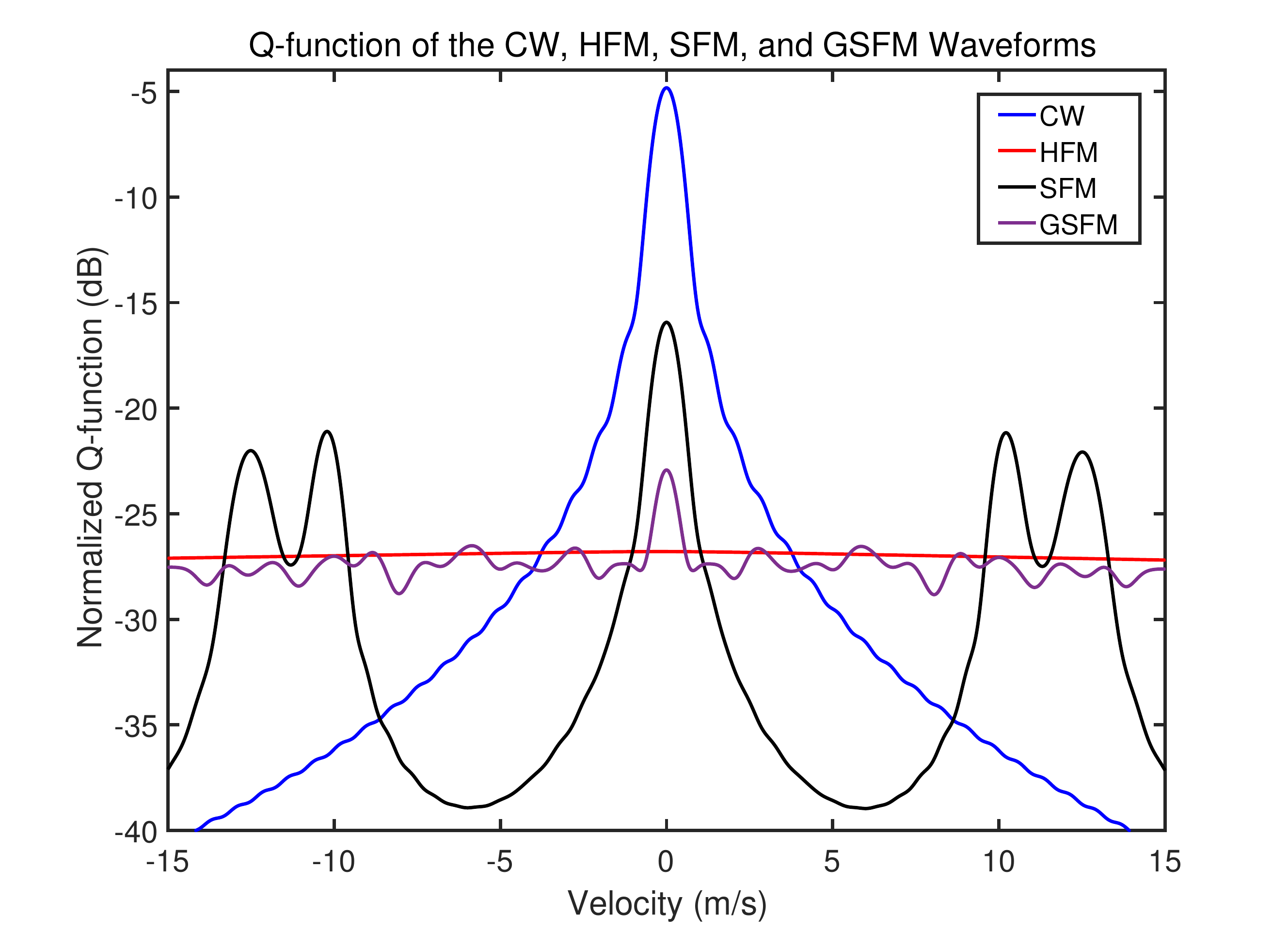}
\caption[Q-function for the CW, HFM, SFM, and GSFM waveforms.]{Q-function for the CW, HFM, SFM, and GSFM waveforms.  Each waveform has a duration $T = 0.5$ s, $f_c = 2000$ Hz, and a bandwidth $\Delta f = 500$ Hz except for the CW.  The CW is good for high velocity, the HFM is roughly the same across velocity,  the SFM has deep nulls lower than the CW at those same velocities, and the GSFM roughly resembles the HFM's Q-function.  }
\label{fig:QFuncI}
\end{figure}

The GSFM produces waveforms with substantially different Q-Functions when $\rho = 1.0$ and when $\rho > 1.0$ and it is insightful to compare these waveforms with other well known sonar waveforms.  Figure \ref{fig:QFuncI} evaluates the Q-function for the CW, HFM, SFM, and GCFI phase GSFM for $\rho = 2.0$. All four waveforms are of duration $T = 0.5$ s, center frequency $f_c = 2000$ Hz, and the three FM waveforms have a swept bandwidth $\Delta f = 500$ Hz.  The CW has the highest reverberation levels for slow targets but the lowest for very fast targets. The HFM is nearly uniform across target velocities with approximately equal to $10\log_{10}\left(\Delta f\right) = -27$ dB.  The SFM attains higher reverberation levels than the HFM for slow moving targets ($ |v| \leq 0.5$ m/s) but lower reverberation levels than the HFM for target velocities of $0.5 \leq |v| \leq 9.5$ m/s and achieves a minimum level of $-39.2$ dB at velocity $\pm 6$ m/s. The GSFM's Q-function is almost uniform across target velocities beyond $1$ m/s with an average level of $-27$ dB, similar to the HFM. The SFM is clearly best suited for a small range of target velocities as the reverberation levels are substantially lower than any of the other waveforms. However, like the HFM, the GSFM is best suited for a scenario where targets are expected to have a wide range of velocities. 

\vspace{-1.5em}
\section{PAPR and Spectral Containment}
\label{sec:PAPR}
As mentioned in Chapter \ref{ch:Intro}, the SC $\psi$ and PAPR of a waveform play important roles in maximizing the total energy in the transmitted acoustic signal projected by a real transducer.  Maximizing the energy of the acoustic signal will result in a stronger echo signal and therefore a higher received SNR.  In noise limited conditions, a higher SNR directly translates to improved probability of detection.  A sonar system utilizing PC waveforms will typically operate in a band of frequencies centered at the transducer's resonance frequency to maximize the source level and therefore total energy of the Transmitted Acoustic Signal (TAS).  Maximizing the concentration of the waveform's energy in this operational band of frequencies greatly aids in maximizing the energy in the TAS.  The PAPR measures the peak to average power in the TAS.  For waveforms with the same peak power limit and duration $T$, the PAPR provides a relative measure of each waveform's total energy.  Consider two waveforms $s_1\left(t\right)$ and $s_2\left(t\right)$ of duration $T$ with PAPR's of PAPR1 and PAPR2 and for simplicity a peak power limit of 1 Watt (W).  The total energy in each waveform is their average power multiplied by the waveform's duration or $E_1 = P_{avg1}T$ and $E_2 = P_{avg2}T$.  If PAPR1 $<$ PAPR2, then $P_{avg1} > P_{avg2}$ and therefore $E_1 > E_2$.  The lower the PAPR of a waveform, the greater total energy it will contain.  As was mentioned in Chapter \ref{ch:signalModel}, applying an amplitude tapering function to the waveform that might be employed to improve the SC will also increase the PAPR introducing a trade-off between SC and PAPR.
\vspace{1em}

One of the most well known methods of determining SC for FM waveforms is Carson's Bandwidth rule which states that $98\%$ of a FM waveform's energy resides in a band of frequencies $B = 2\left(\Delta f/2 + B_m\right)$ where $\Delta f$ is the waveform's swept bandwidth and $B_m$ is the highest frequency component of the waveform's IF function \cite{Couch}.  Similar rules exist for FSK and Phase-Coded waveforms \cite{Couch}.  Applying Carson's Bandwidth Rule for the even-symmetric phase GSFM, this band of frequencies is expressed as 
\begin{equation}
B_{GSFM} = \left(\dfrac{\Delta f}{2\alpha\rho T^{\left(\rho-1\right)}}+1\right)
2\alpha\rho T^{\left(\rho-1\right)} = \Delta f + 2\alpha\rho T^{\left(\rho-1\right)}
\label{eq:GSFM_Carson}
\end{equation}
Note that when $\rho = 1.0$, \eqref{eq:GSFM_Carson} reduces to \eqref{eq:SFM_Carson}, Carson's Bandwidth Rule for the SFM.  
\vspace{1em}

\begin{figure}[p]
\includegraphics[width=1.0\textwidth]{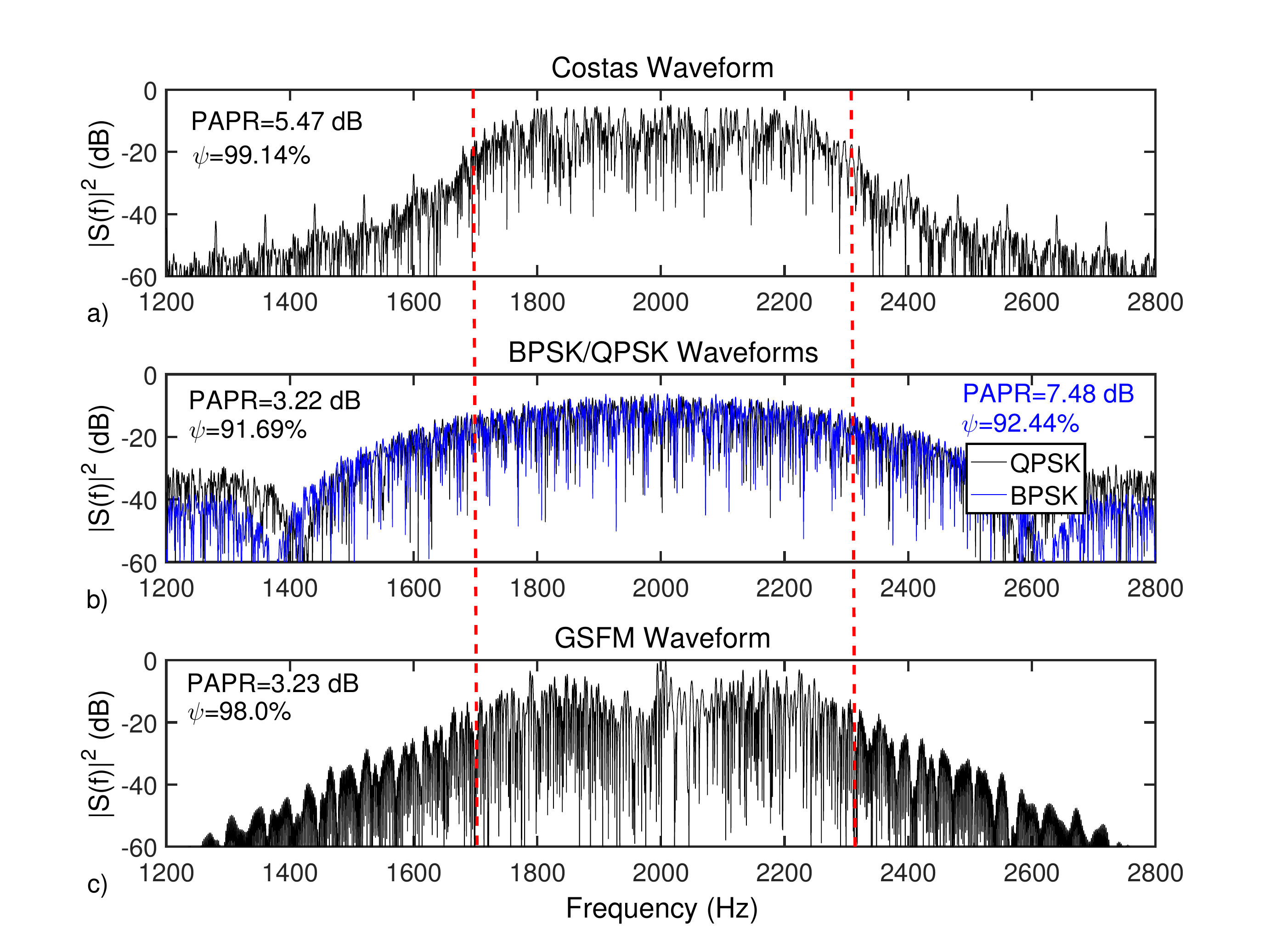}
\caption[SC of the Costas (a), BPSK and QPSK (b), and GSFM (c) waveforms.]{SC of the Costas (a), BPSK and QPSK (b), and GSFM (c) waveforms.  The GSFM attains high SC while also requiring minimal tapering resulting in a low PAPR.  None of the Costas, BPSK, or QPSK waveforms can match the GSFM in SC or PAPR.}
\label{fig:SpecConI}
\end{figure}

\begin{figure}[p]
\includegraphics[width=1.0\textwidth]{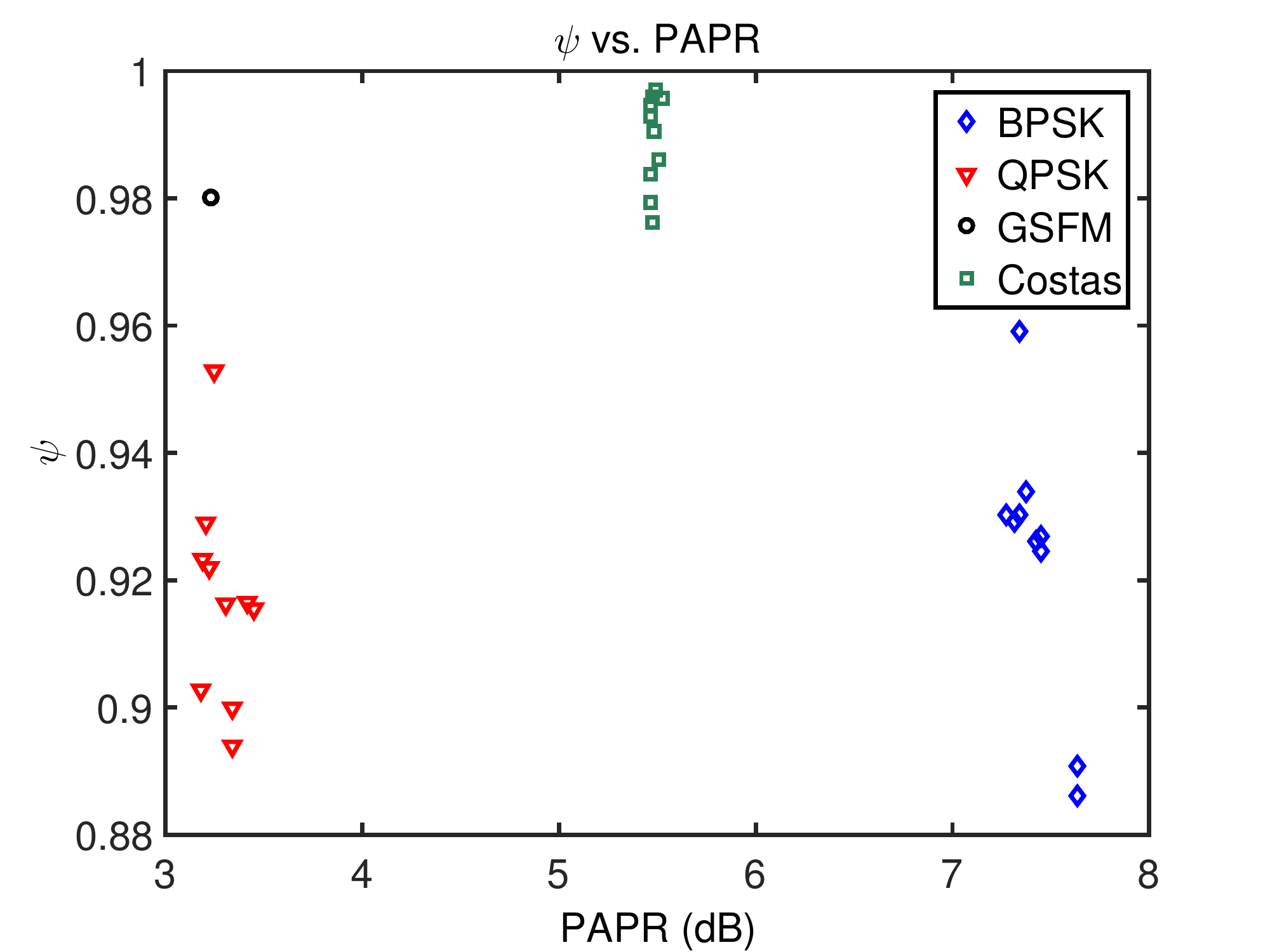}
\caption[SC of the GSFM, Costas, BPSK, and QPSK waveforms for the TBPs used in Chapter \ref{ch:GSFM_Eval_AF}.]{SC of the GSFM, Costas, BPSK, and QPSK waveforms for the TBPs used in Chapter \ref{ch:GSFM_Eval_AF}.  For each TBP, the GSFM waveform attains high SC while also requiring minimal tapering resulting in a low PAPR.  None of the Costas, BPSK, or QPSK waveforms can match the GSFM in SC or PAPR for any of the TBPs tested.}
\label{fig:SpecConII}
\end{figure}

However, a detailed analysis finds that \eqref{eq:GSFM_Carson} substantially over-estimates the $98\%$ bandwidth of the GSFM.  It is therefore more straight forward to find the $98\%$ bandwidth of the GSFM by numerically evaluating \eqref{eq:psi} for a range of $\Delta F$ values.  Figure \ref{fig:SpecConI} shows the spectrum of the GSFI GSFM, Costas, BPSK, and QPSK waveforms all with duration $T = 0.5$ s, $f_c = 2000$ Hz, and swept bandwidth $\Delta f = 500$ Hz.  The GSFM is tapered with a Tukey window with shape parameter $\alpha_T = 0.1$.  The Costas and BPSK waveforms used the same tapering as discussed in Chapter \ref{ch:GSFM_Eval_AF}.  The $98\%$ bandwidth of the GSFM was numerically determined to be $B = 632$ Hz.  That band of frequencies centered about $f_c$ is marked by the dashed red lines in Figure \ref{fig:SpecConI}.  Using that same band of frequencies, $\psi$ was computed for the other three waveforms.  In this example, the Costas waveform actually has a slightly higher SC of $99.14\%$ than the GSFM but only achieves this SC due to tapering.  As a result of the tapering, the Costas waveform has a PAPR of $5.48$ dB, substantially more than the GSFM's PAPR of $3.23$ dB.  The BPSK's SC is notably less than the GSFM's even when tapering each chip with a Hanning window.  The tapering resulted in the BPSK having a PAPR of $7.48$ dB.  Had tapering not been applied, a BPSK with the same TBP would have a SC of $\psi = 80.3\%$.  The QPSK waveform, while having a slightly better PAPR of $3.22$ dB, has a SC of $\psi = 91.69\%$, notably lower than the GSFM.  This is the general trend across TBPs as shown in Figure \ref{fig:SpecConII} which plots SC as a function of PAPR for the GSFM, Costas, BPSK, and QPSK using the TBPs tested in Chapter \ref{ch:GSFM_Eval_AF}.  These are the same waveforms from Chapter \ref{ch:GSFM_Eval_AF} that also achieved the minimal PSLs for MF processing.  The design goal is for a waveform to possess both high SC and a low PAPR which directly translates to data points that tend to the upper left corner of the figure.  The GSFM data points are closest to the upper left corner of Figure \ref{fig:SpecConII} meaning the GSFM attains high SC and a low PAPR.  None of the Costas, BPSK, or QPSK waveforms can match the same performance in both SC and PAPR of the GSFM for any of the TBPs tested. 
\clearpage

\clearpage

\clearpage

\section{Transducer Replicas}
\label{sec:Replica}
The results from the previous section show that the GSFM possesses both high SC and low PAPR, two important characteristics to consider when transmitting waveforms on piezoelectric transducers.  In addition to ensuring that a waveform has these desirable properties, a system designer will also want to record the acoustic signal that is transmitted and received by the system's transducers, referred to here as the Transducer Replica Waveform (TRW).  Recording the TRW is especially useful for calibrating active sonar systems.  During calibration, the system designer will want to verify that the TRW is an accurate representation of the transmit waveform that was designed in simulations and that the TRW maintains its AF properties.  In many cases the TRW will be used as the base MF from which other Doppler scaled MF's will be derived.  The frequency response of the transducer, whose frequency response drops off steadily beyond its resonance frequency, will attenuate the off-resonance frequency components of a waveform.  This attenuation can substantially change the AF shape of the TRW.  An equalizer filter can be designed to compensate for the transducer's frequency response and thereby remove any distortions in the TRW.  Applying an equalizer in a system with a peak power limit means attenuating the frequency components of the waveform that are at or near the resonance frequency.  This attenuation reduces the source level of the transmitted acoustic signal which reduces the echo strength which in turn reduces SNR and therefore detection performance.  If an active sonar system is operating in noise limited conditions, the designer may opt not to use an equalizer so as to maximize the source level of the transmitted acoustic signal and therefore accept any distortions in the TRW's AF shape.  These design constraints motivated the TRW experiments described in this section.  
\vspace{1em}    

The waveform transmission experiments were conducted at the university's underwater test tank facility.  The experimental setup is illustrated in Figure \ref{fig:TankSketch} below. Bridge 1 held the projector mode (transmitter) transducer and Bridge 2 held the receiver transducer. The transducers used in these tests, model numbers BT-SSS-2LF S/N 01 (transmitter) and BT-SSS-2LF S/N 02 (receiver), are prototype SideScan Sonar transducers developed by BTech Acoustics.  The resonance frequency of the two devices is $310$ KHz and the transmit/receive frequency response had a $-3$ dB bandwidth of $20$ KHz.  The long edge of the transducers were extended vertically in the test tank.  This was done because the beam patterns of these devices have a very narrow mainlobe. As a result of this, the acoustic energy is very directive vertically which prevents potential surface and bottom reflections of acoustic energy which could interfere with the signal measurements. The directive nature of the transducer coupled with the non-parallel geometry of the test tank prevented any unwanted echoes overlapping with the original transmit signal. The recorded signal data therefore had minimal interference and provided a very accurate TRW. 

\begin{figure}[ht]
\includegraphics[width=1.0\textwidth]{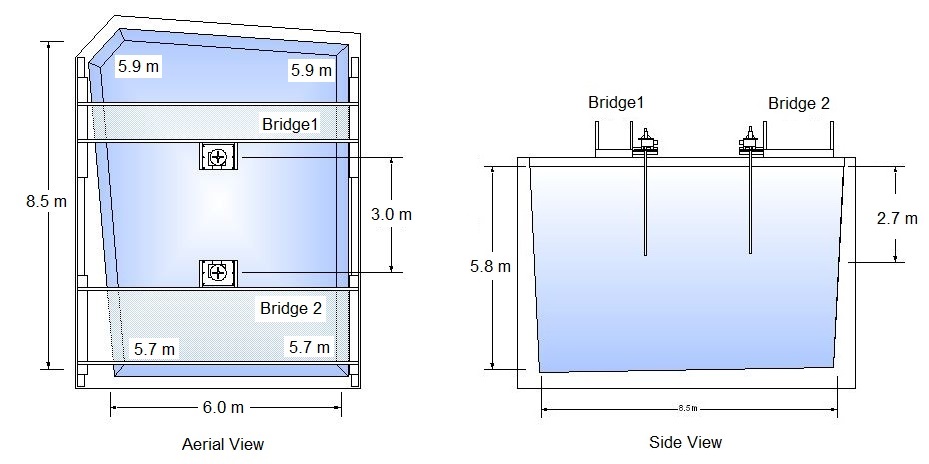}
\caption[Test Tank Experiment Setup.]{Test Tank Setup.  Bridge 1 held the transmitting transducer element and Bridge 2 held the receiving transducer element.  The transmitted signal is recorded and processed to create a replica of the original transmit waveform.  Source : http://www.umassd.edu/smast/about/smastfacilities/testtank/}
\label{fig:TankSketch}
\end{figure}

From here, the AF of the TRW is computed and compared to the original waveform.  The first objective of this experiment is to compare the AF of the original waveform and the TRW and to assess whether the AF of the TRW maintains the properties of the original design waveform generated in Matlab.  The second objective of this experiment is to compute the overall energy of each TRW.  The energy efficiency of the TRW $\tilde{E}$, compares the amount of energy in the GSFM's TRW $E_{GSFM}$ to the amount of energy in another waveform's TRW $E_w$ and is expressed in dB as 
\begin{equation}
\tilde{E} = 10\log_{10}\left(\dfrac{E_w}{E_{GSFM}} \right)
\label{eq:energyEfficiency}
\end{equation} 
The energy of the TRW \eqref{eq:energyEfficiency} is a measure that combines a waveform's SC, PAPR, and also accounts for the transducer's frequency response and therefore provides a comprehensive measure of a waveform's overall energy efficiency.
\vspace{1em}

\begin{figure}[h]
\includegraphics[width=1.0\textwidth]{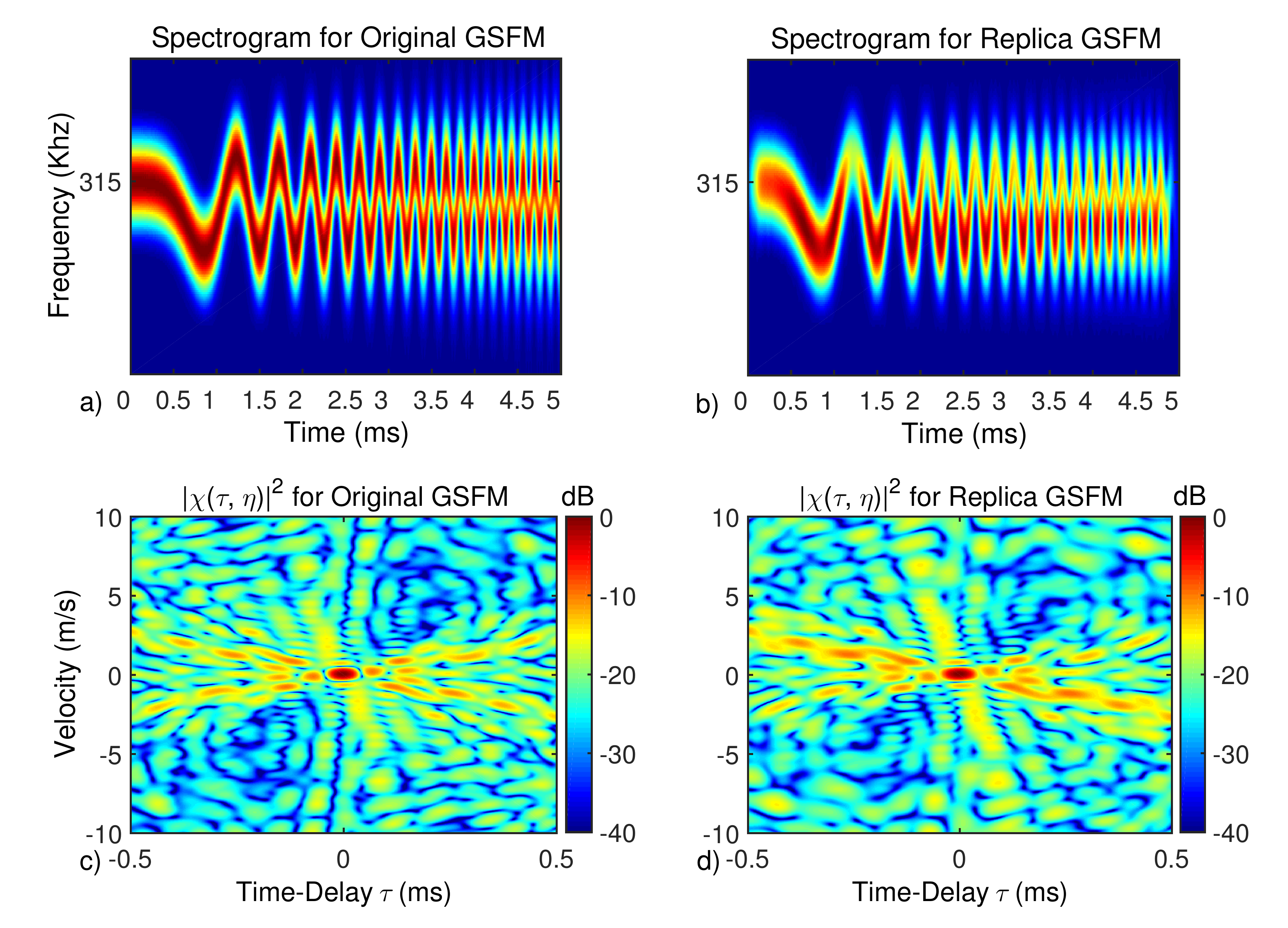}
\caption[Spectrograms of the original GSFM (a) and TRW GSFM (b) and BAAF's of the original GSFM (c) and TRW GSFM (d).]{Spectrograms of the original GSFM (a) and TRW GSFM (b) and BAAF's of the original GSFM (c) and TRW GSFM (d).  The TRW GSFM maintains the properties of the original GSFM generated in Matlab.}
\label{fig:RepComp}
\end{figure}

Figure \ref{fig:RepComp} shows the spectrogram and BAAF of a GSFM TRW with duration $T = 5$ ms, $f_c = 310$ KHz, and swept bandwidth $\Delta f = 20$ KHz and compares to the original transmit waveform generated in MATLAB.  One noticeable difference between the two spectrograms is that the TRW's high frequency components are attenuated due to the transmit/receive frequency response of the transducers.  As a consequence of this, the TRW's BAAF shows a widening of the mainlobe in time-delay.  This result is not surprising.  As was discussed in Chapter \ref{ch:GSFM_Eval_AF}, tapering the edges of a waveform's spectrum reduces the waveform's RMS bandwidth.  Therefore the relation in \eqref{eq:RangeRatio} is also reduced which translates to a slightly wider mainlobe with reduced sidelobe heights.  The widened mainlobe is clearly noticeable and visual inspection of the PSL in time-delay of the TRW's BAAF shows that the PSL was reduced $0.2$ dB.  Otherwise, the overall BAAF shape is maintained with very little noticeable differences.  

\clearpage

Table 5.1 shows waveform energy efficiency $\tilde{E}$ for a LFM, Costas, GSFM, and a PRN  \footnote{Here, the PRN waveform is a waveform composed of Gaussian random noise that is bandpass filtered to the swept bandwidth $\Delta f$ and center frequency $f_c$ of the other FM waveforms.  While not commonly employed today, the PRN was one of the first thumbtack waveforms analyzed \cite{Rihaczek}.  The author had not yet developed their analysis of BPSK and QPSK waveforms when these TRW experiments were conducted} waveforms with duration $T = 5$ ms, $f_c = 310$ KHz, and swept bandwidth $\Delta f = 20$ KHz.  All waveforms except for the LFM are thumbtack waveforms.  For the thumbtack waveforms, the GSFM is the most energy efficient.  The Costas waveform has $-3.709$ dB less energy than the GSFM and the PRN has $-8.638$ dB less energy than the GSFM.  The LFM however, is the most energy efficicent.  This is likely due to the distribution of energy in the LFM's and GSFM's spectrum.  The LFM has its energy nearly uniformly distributed across its swept bandwidth $\Delta f$ where as the GSFM has a concentration of energy at the edges of it's swept bandwidth.  If the center frequency $f_c$ is chosen to be the transducers resonance frequency, then the GSFM's has more energy concentrated at the at frequencies off resonance which results in a transmitted acoustic signal with a lower source level.  The Costas and PRN waveforms have nearly uniform spectra as well, but both have high PAPR.  The Costas waveform's chips are tapered with a Hanning window and the noiselike nature of the PRN results in widely varying amplitudes across its duration.  The GSFM's spectral energy, while not evenly distributed in frequency, is still highly concentrated in its swept bandwidth and also possesses a low PAPR which accounts for it's higher energy efficiency over Costas and PRN waveforms.
 
\begin{table}[htb]

\begin{center}
{
\caption[Energy efficiency of the LFM, Costas, GSFM, and PRN waveforms with with duration $T = 5$ ms, $f_c = 310$ KHz, and $\Delta f = 20$ KHz.]{Energy efficiency $\tilde{E}$ of the LFM, Costas, GSFM, and PRN waveforms with with duration $T = 5$ ms, $f_c = 2$ KHz, and $\Delta f = 20$ KHz.  The GSFM has the best energy efficiency of all the thumbtack waveforms.  Only the LFM has better energy efficiency than the GSFM.}
\begin{tabular}{|c||c|}\hline
Waveform  & $\tilde{E}$ (dB) \\\hline
LFM       & $0.296$		 \\\hline
Costas    & $-3.709$  		 \\\hline
GSFM      & $0.000$  		 \\\hline
PRN       & $-8.638$  		 \\\hline
\end{tabular}
}
\end{center}
\end{table} 

\section{Conclusion}
\label{sec:Conclusion}
This chapter explored the GSFM's RS, SC and PAPR performance and compared it to the Costas, BPSK, QPSK, and some other well known PC waveforms.  When the GSFM's parameter $\rho = 1.0$, it becomes the SFM and possesses deep notches in its Q-Function which greatly suppress reverberation for a range of Doppler values.  When $\rho > 1.0$, these RS notches disappear and the GSFM becomes thumbtack like.  It then supresses reverberation in a manner similar to the HFM waveform whose RS performance is nearly constant across Doppler.  The GSFM contains the vast majority of its energy in confined band of frequencies while requiring minimal amplitude tapering which results in high SC and a low PAPR, two very important considerations when transmitting waveforms on piezoelectric transducers.  The test tank  experiments showed that the GSFM's TRW largely maintains its desireable AF shape when transmitted on piezoelectric transducers and that its overall energy efficiency surpasses that of other well known thumbtack waveforms.  

\chapter{The GSFM for Continuous Active Sonar Systems}
\label{ch:GSFM_CAS}

\section{Pulsed vs. Continuous Active Sonar Systems}
\label{sec:GSFM_CAS_Intro}
The most basic challenge of radar and sonar waveform processing is to resolve multiple closely spaced targets in range and velocity and provide a constantly updated picture of the target scene.  In radar systems, these challenges are met by transmitting pulse train waveforms \cite{Rihaczek}.  In their most simple form, a pulse train is a collection of identical pulses that are transmitted periodically.  The time period between pulse transmissions is known as the Pulse Repetition Interval (PRI).  The rate at which the target scene is revisited, here called the Target Revisit Rate (TRR), is inversely proportional to the PRI. The target echoes from the transmitted pulse train are then coherently processed to detect and resolve targets in the presence of noise and clutter.  For radar systems the speed of light allows for rapid transmission of waveforms and the reception of target echoes.  A target scene can be revisited by a radar system thousands of times per second with zero ambiguities in range and velocity.  In Pulsed Active Sonar (PAS), the speed of sound ($\sim 1500$ m/s) requires wait times on the order of tens of seconds between waveform transmissions in order to avoid range ambiguities.  The long wait times of PAS mean that for given amount of time, the target is not ensonified very often resulting in low TRR's.  Low TRR's result in reduced target detection probability and large information gaps in time of the target scene which in turn degrade target tracker performance.  
\vspace{1em}

Continuous Active Sonar (CAS) systems mitigate the TRR limitations of PAS systems by continuously transmitting a long duration waveform.  Such transmit waveforms increase the duty cycle of CAS systems and therefore allows for increased TRR's.  Some of the first CAS systems used long duration $T$ (i.e. 20 seconds) LFM waveforms with large swept bandwidth $B$ (i.e. 500-1000 Hz) \cite{Krolik}.  The resulting target echoes are then processed by a bank of $P$ constant bandwidth bandpass filters that span the bandwidth of the transmit waveform.  Passing a LFM echo through each channel of the filterbank results in $P$ LFM segments of duration $T/P$ seconds in seperate frequency bands with bandwidth $B/P$.  Each filtered LFM segment is therefore shifted in time by $T/P$ seconds from the previous segment.  Each filterbank channel is then seperately dechirped in a fashion similar to FMCW radar \cite{Richards} to provide target range information.  Processing each time-shifted LFM channel seperately allows for revisiting the target scene every $T/P$ seconds.  The number of filterbank channels $P$ in a CAS system introduces a design tradeoff.  Large $P$ increases the TRR but also results in smaller LFM bands of bandwidth $B/P$.  The reduced bandwidth of each channel results in degraded range resolution and reverberation suppression.  Such systems present the system designer with a tradeoff between TRR and bandwidth.
\vspace{1em}

Recent work by Hickman and Krolik \cite{Krolik} proposed using pulse trains of frequency-hopped LFM waveforms to achieve large Instantaneous Bandwidth (IB), the bandwidth of an individual pulse, in addition to high TRR’s.  The frequency hopped LFM pulse train waveform is one example common in radar signal processing of coherently processing a train of diverse pulses to reduce range and/or Doppler ambiguities.  In addition to frequency hopping, each pulse can have a different modulation function phase coding to create pulse to pulse diversity \cite{Levanon}.  The wide variety of pulse to pulse diversity techniques raises the intriguing question of whether other types of diverse pulse trains can improve CAS systems.   This chapter expands on Hickman and Krolik's work and introduces new diverse pulse train waveforms composed of a family of GSFM waveforms that possess low cross-correlation properties even when occupying the same band of frequencies \cite{HagueIII, HagueIV}.  This work also discusses several coherent processing schemes that maintain a constant target revisit rate while adapting the Coherent Processing Interval (CPI) to be tolerant of target acceleration.  The rest of this chapter is organized as follows: Section \ref{sec:CAS_Model} describes the pulse train waveform signal model and the Ambiguity Function (AF).  Section \ref{sec:SixFlavors} describes how to create families of nearly orthogonal waveforms from the GSFM.  Section \ref{sec:CAS} discusses the coherent processing strategies, their design tradeoffs, and introduces several diverse pulse train waveform designs to consider for CAS applications.  Section \ref{sec:CAS_Results} presents proof of concept simulations to evaluate the performance of the coherent processing strategies and the diverse pulse train waveforms.  Finally, Section \ref{Conclusion} presents the conclusions.
\vspace{1em}

\section{Pulse Train Waveform Model and Ambiguity Function}
\label{sec:CAS_Model}
CAS systems are constantly transmitting waveforms into the medium and receiving echoes from targets.  CAS systems are by nature  bi-static systems (i.e., separate transmitter and receiver hardware).  This work adopts simplified system and target models assuming that the bistatic transmitter and receiver hardware are sufficiently close to be modeled as a monostatic sonar system, and the targets of interest are point targets.  This work focuses on trains of diverse pulses of the same duration $T$ transmitted with a constant (PRI) denoted as $T_{PRI}$.  The pulse train waveform is a collection of pulses transmitted sequentially and is expressed as 
\begin{equation}
s\left(t\right) = \sum_{n=1}^{N}s_n\left(t-\left(n-1\right)T_{PRI}\right)
\end{equation} 
where $N$ is the number of pulses in the pulse train, the PRI $T_{PRI}$ is equal to the individual pulse length $T_{PRI} = T$, and $s_n\left(t\right)$ is the $n^{th}$ pulse in the train expressed as 
\begin{equation}
s_n\left(t\right) = a\left(t\right) e^{j\varphi_n\left(t\right)}e^{j2\pi \left(f_c + \Delta f_n \right)t}
\end{equation} 
where, $f_c$ is the waveform carrier frequency, $\Delta f_n$ is the frequency step of the $n^{th}$ waveform in the pulse train used to apply frequency hopping to the pulse train, and $\varphi_n\left(t\right)$ is the phase modulation function of the $n^{th}$ pulse. The Instantaneous Frequency (IF) function of the $n^{th}$ pulse in the pulse train waveform $f_n\left(t\right)$ is expressed as 
\begin{equation}
f_n\left(t\right) = \dfrac{1}{2 \pi}\dfrac{\partial \varphi_n \left( t\right)}{\partial t} + \left(f_c + \Delta f_n\right)
\end{equation}

For trains of identical pulses, the maximum unambiguous range, $R_{max}$, is equal to $cT_{PRI}/2$.  The PRI must be chosen to meet some specified $R_{max}$ or risk grating lobes in the pulse train waveform's Auto-Correlation Function (ACF).  For diverse pulse trains however, $R_{max}$ is not determined by the PRI, but a term defined here as the Identical Pulse Repitition Interval (IPRI) expressed as $T_{IPRI}=NT_{PRI}$.  The maximum unambiguous range can then be expressed as $R_{max} = cT_{IPRI}/2=cNT_{PRI}/2$.  This property of diverse pulse train waveforms allows for pulse trains at a much higher PRI's than traditional identical pulse trains without suffering ACF grating lobes in range.
\vspace{1em}

Again, as discussed in Chapter \ref{ch:signalModel}, we assume MF processing and utilize the BAAF and BCAF that correlates the transmit waveforms with their Doppler scaled echoes.  The Doppler scaling factor $\eta$ in \ref{eq:eta} is accurate for target models where the target is moving with a constant velocity relative to the sonar system platform.  In realistic scenarios, a target is likely to undergo acceleration.  If the acceleration is high enough, the resulting echo is no longer strongly correlated with the waveform's MF.  This reduced correlation from an accelerating target results in a substantial SNR loss at the output of the MF.  The transmit waveform will remain acceleration tolerant so long as the following relation holds \cite{KellyWish, Kramer}
\begin{equation}
\dfrac{a}{c} < \dfrac{1}{T^2 B}
\label{eq:accelTol}
\end{equation}
where $a$ is the target's maximum acceleration.  The BAAF of the entire pulse train, here called the Broadband Composite AF (BCOAF) \cite{Rihaczek} is expressed as 
\begin{equation}
\chi\left(\tau, \eta\right) = \sum_{m=1}^{N}\sum_{n=1}^{N} \chi_{m,n}\left(\tau + \left(m-n\right)T_{PRI}, \eta\right)
\label{eq:BCOAF}
\end{equation}
where $\chi_{m,n}\left(\tau, \eta\right)$ is the BCAF between the m$^{th}$ and n$^{th}$ waveforms in the pulse train delayed in time by $\left(m-n\right)T_{PRI}$.

\section{Generating In-Band Nearly Orthogonal Waveforms}
\label{sec:SixFlavors}
\begin{figure}[p]
\centering
\includegraphics[width=1.0\textwidth]{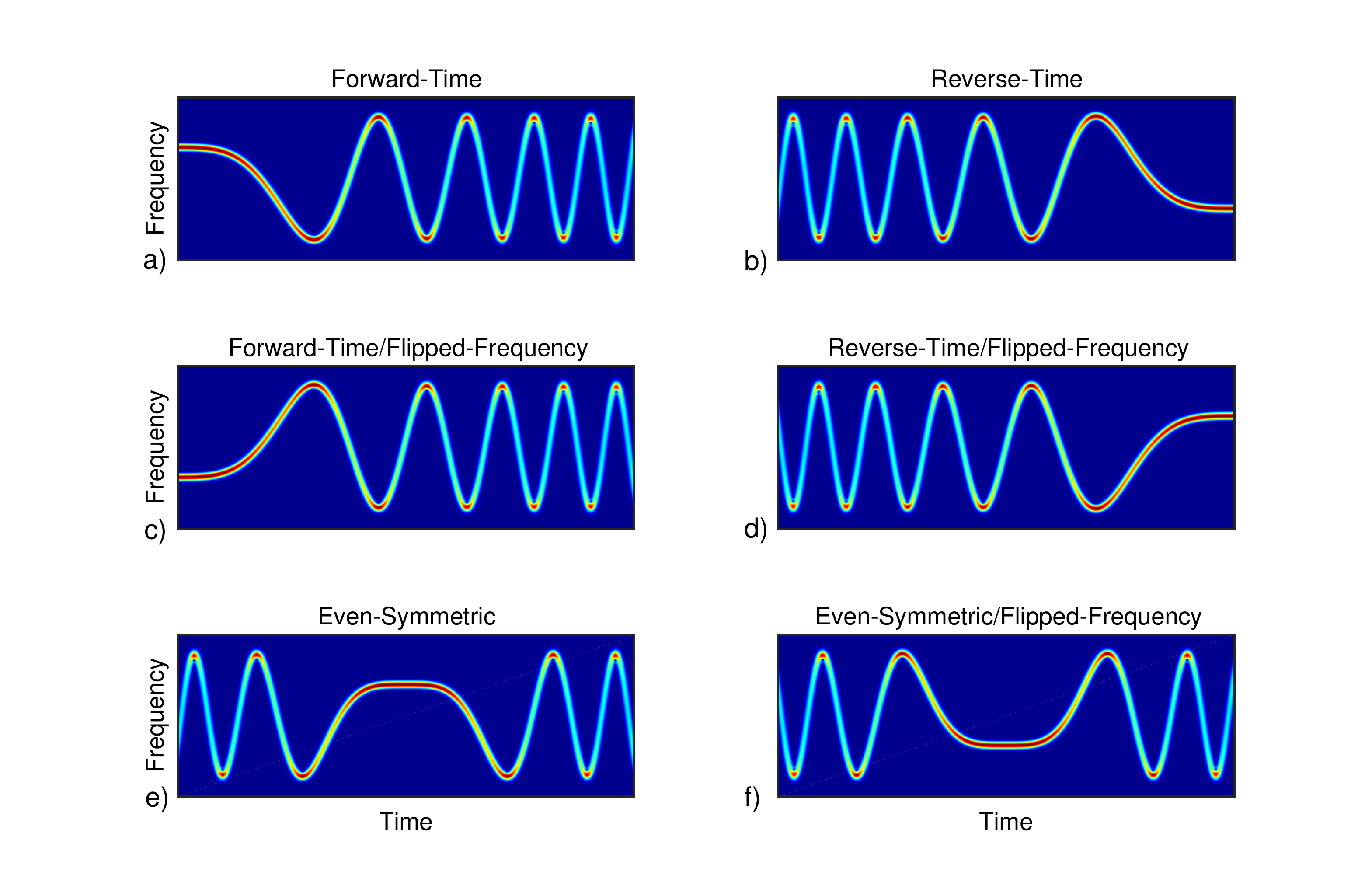}
\caption[Notional spectrograms of the six IF reflections of the GSFM.]{Notional spectrograms of the six IF reflections of the GSFM.  The first waveform (a) is the original GSFM as described in (\ref{GSFM_Phi_III}) and (\ref{GSFM_IF_III}).  The second (b) is generated by time-reversing the first waveform.  The third (c) and fourth (d) waveforms are generated by flipping the IF functions of the first and second waveforms in frequency.   The fifth waveform (e) has an even symmetric IF function.  Finally the sixth waveform (f) is the fifth waveform with its IF function flipped in frequency.}
\label{GSFM_Flavors}
\end{figure}

\begin{figure}[ht]
\centering
\includegraphics[width=1.0\textwidth]{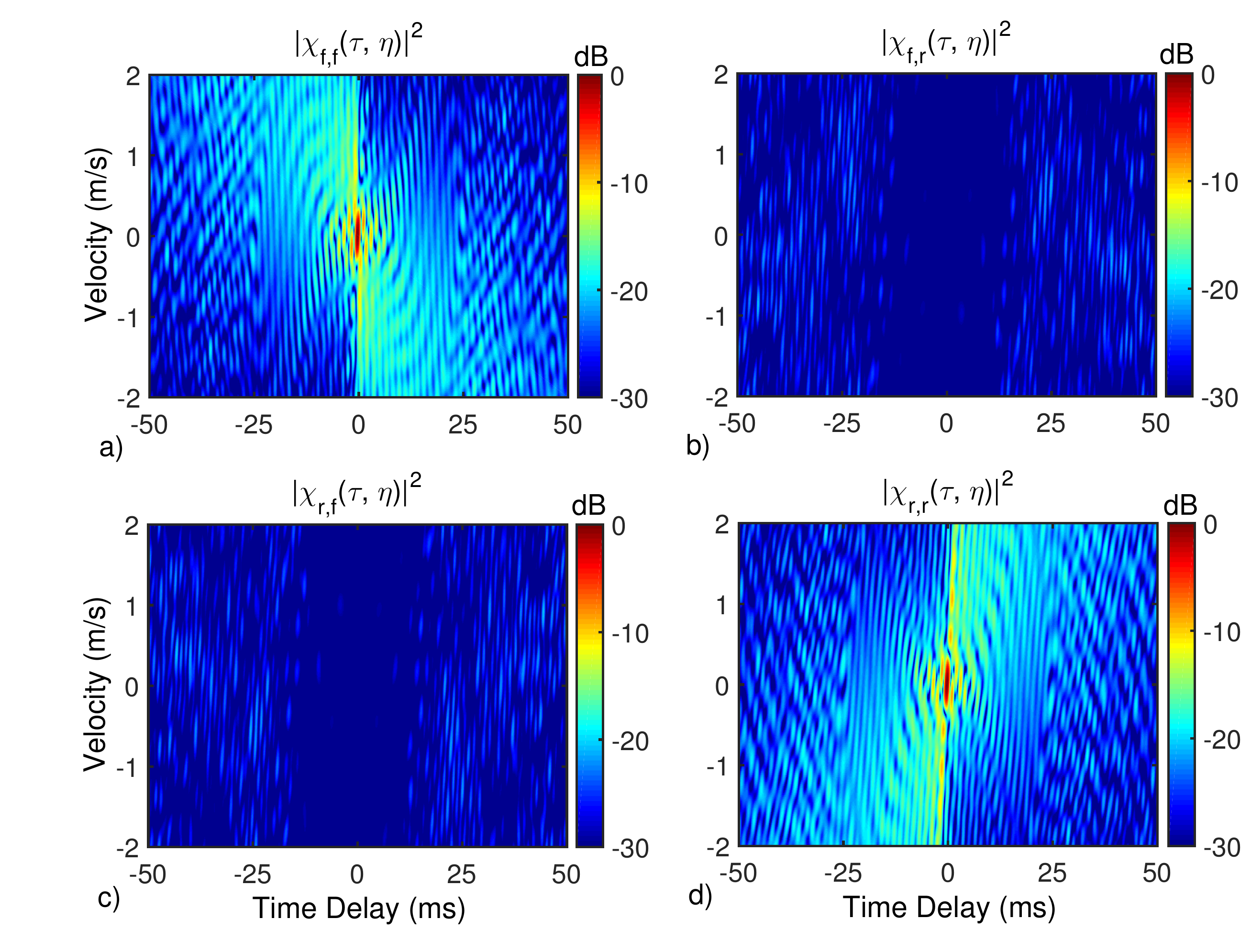}
\caption[The Broadband Auto and Cross AF's for the Forward-Time and Reversed-Time GSFM waveforms with pulse length $T = 1.0$ s, bandwidth $\Delta f = 1000$ Hz, and $f_c = 2000$ Hz.]{The Broadband Auto and Cross AF's (zoomed to $-50$ ms $\leq \tau \leq$ $50$ ms and $-2$ m/s $\leq v \leq$ $2$ m/s) for the Forward-Time and Reversed-Time GSFM waveforms with pulse length $T = 1.0$ s, bandwidth $\Delta f = 1000$ Hz, and $f_c = 2000$ Hz.  Both reflections achieve thumbtack BAAF's while their respective CAAFs spread the AF volume evenly thus attaining low cross-correlation properties between the individual waveforms.}
\label{GSFMCross}
\end{figure}

Designing GSFM waveforms with different $\alpha$ and $\rho$ values can produce a family of waveforms that occupy the same band of frequencies and are nearly orthogonal to each other.  These nearly orthogonal waveforms can then be employed in a diverse pulse train for CAS applications.  However, finding a large number of nearly orthogonal GSFM waveforms by varying $\alpha$ and $\rho$ is computationally intense and may result in a small number of such waveforms.  The number of waveforms is greatly increased by using the six different reflections of the GSFM for a single $\alpha$ and $\rho$ as seen in Figure \ref{GSFM_Flavors}.  The first waveform, known as the forward time IF GSFM and denoted as $s_f\left(t\right)$, is the original GSFM as described in (\ref{GSFM_Phi_III}) and (\ref{GSFM_IF_III}).  The second, $s_r\left(t\right)$ is generated by time-reversing the first waveform.  The third and fourth waveforms are generated by flipping the IF functions of the first and second waveforms in frequency and are denoted as $s_{\tilde{f}}$ and $s_{\tilde{r}}$.   The fifth waveform, $s_e\left(t\right)$, has an even symmetric IF function.  Finally the sixth waveform, $s_{\tilde{e}}\left(t\right)$ is the even-symmetric waveform with its IF function flipped in frequency.  The waveforms are all reflections of the complex analytic signal in (\ref{eq:ComplexExpo}) and are expressed as
\begin{equation}
  \begin{aligned}
  s_f\left(t\right) & = a\left(t\right)e^{j\varphi\left(t\right)}e^{j2\pi   f_ct} \\
  s_r\left(t\right) & = a\left(t\right)e^{j\varphi\left(T-t\right)}e^{j2\pi f_ct} \\
  s_{\tilde{f}}\left(t\right) & = a\left(t\right)e^{-j\varphi\left(t\right)}e^{j2\pi` f_ct} \\
  s_{\tilde{r}}\left(t\right) & = a\left(t\right)e^{-j\varphi\left(T-t\right)}e^{j2\pi f_ct} & \text{for } 0 \leq t \leq T, \\
  s_e\left(t\right) & = a\left(t\right)e^{j\varphi\left(t\right)}e^{j2\pi f_ct} \\
  s_{\tilde{e}}\left(t\right) & = a\left(t\right)e^{-j\varphi\left(t\right)}e^{j2\pi f_ct} & \text{for } -T/2 \leq t \leq T/2.
  \end{aligned}
\end{equation}
The six waveforms occupy the same band of frequencies, attain thumbtack BAAF’s, and are nearly orthogonal to each other.  Using $P$ different $\alpha$ and $\rho$ values produce  $6P$ in-band nearly orthogonal GSFM waveforms.  Figure \ref{GSFMCross} shows the BAAF and BCAF of two of the six reflections of the GSFM.  Both waveforms possess thumbtack BAAF's and low cross-correlation properties.  The Peak Sidelobe Level (PSL) of the BCAF determines maximum cross-correlation between two GSFM waveforms.  For two waveforms occupying the same band of frequencies, the PSL is inversely proportional to the waveform's Time Bandwidth Product (TBP) \cite{Ricker}.  Therefore, in order to minimize the cross-correlation between two GSFM waveforms, they should be designed with as large a TBP as the CAS system receiver allows.  

\clearpage

\clearpage

\vspace{-1em}
\section{CAS Processing and Pulse Train Waveform Design}
\label{sec:CAS}
This section describes several possible schemes for processing target echoes from a diverse pulse train.  This section also describes several types of diverse pulse train waveforms and compares their relative merits.

\subsection{Processing Strategies}
\label{subsec:ProceStrat}
There are several considerations that factor into processing echos from CAS transmissions.  The primary goal of CAS is to achieve a high TRR.  There are a number of other system parameters to consider.  The CPI determines the Doppler resolution. The CPI also determines the Time Bandwidth Product (TBP), also known as the Pulse Compression Gain (PCG).  The PCG determines the ratio of the output SNR of the MF over the SNR at the input to the MF.  A longer CPI translates to larger TBP and greater Doppler sensitivity.  A larger TBP controls two important performance measures.  First, a larger TBP translates to higher PCG resulting in improved detection performance.  For waveforms with a thumbtack BAAF, a larger TBP spreads the BAAF's bounded volume over a wider range of time-delay and Doppler values therefore reducing the average sidelobe levels of the thumbtack BAAF.  These reduced sidelobe levels improve the waveform's ability to detect weak targets in the presence of a much stronger target.  A larger dynamic range of target strengths can be detected with larger TBP waveforms.  However, longer CPI's also make the waveform susceptible to SNR losses due to target acceleration.  If \eqref{eq:accelTol} is violated, these losses can become severe.  Losses due to acceleration can have a negative impact on PAS waveforms \cite{Kramer}.  CAS waveforms are roughly an order of magnitude longer in duration than PAS waveforms.  Eq. \eqref{eq:accelTol} indicates that CAS waveforms are therefore roughly two orders of magnitude more sensitive to target acceleration than typical PAS waveforms.  Maintaining high TRR's, long CPI and therefore large TBP, and acceleration tolerance are the main performance considerations in this work.
\vspace{1em}

\begin{figure}
\centering
\fbox{\includegraphics[width=1.0\textwidth]{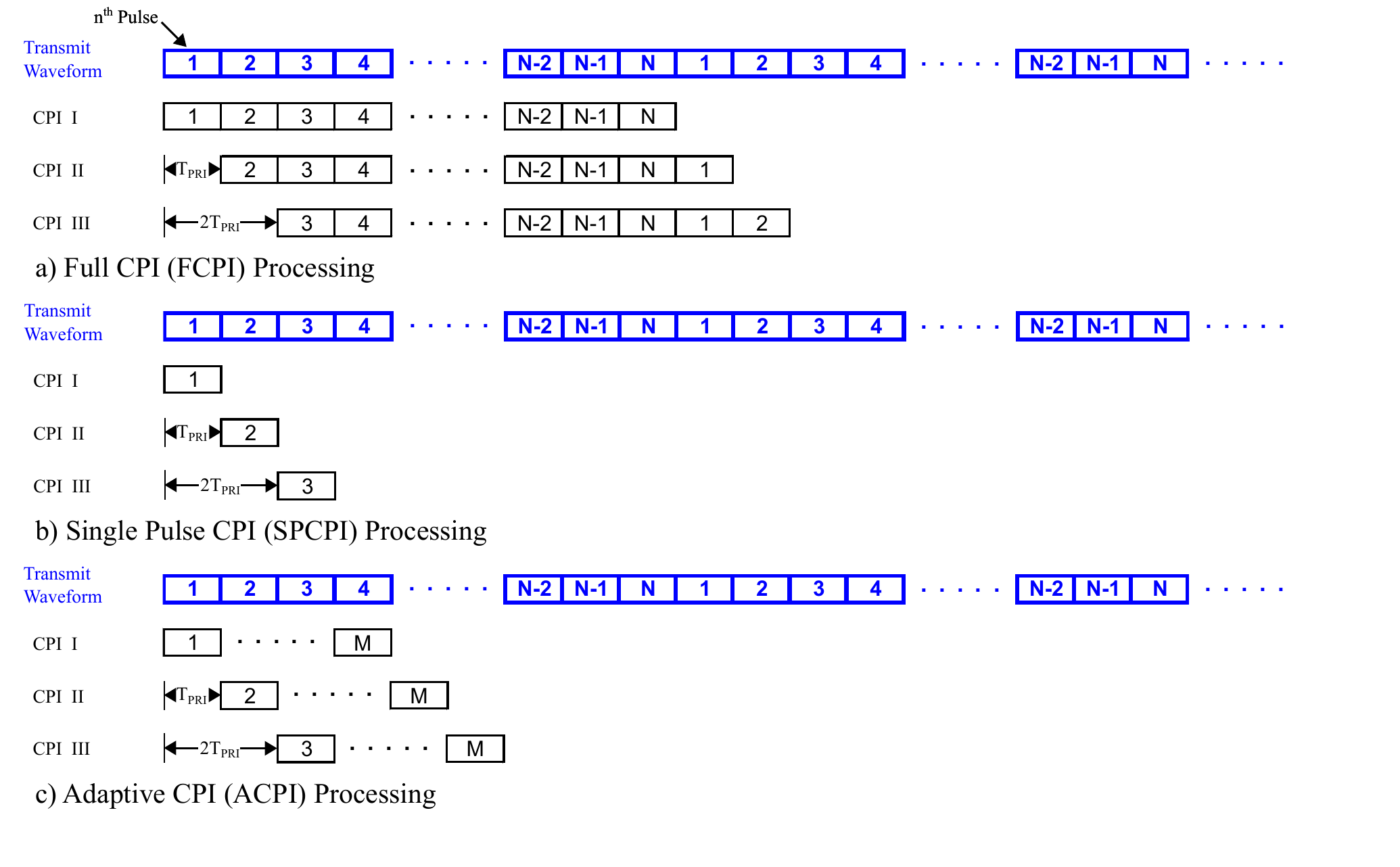}}
\caption[Illustration of the FCPI (a), SPCPI (b), and ACPI (c) processing strategies.]{Illustration of the FCPI (a), SPCPI (b), and ACPI (c) processing strategies.  All three processing strategies possess different CPI's while revisiting the target scene every $T_{PRI}$ seconds.}
\label{ProcessingSchemes}
\end{figure}

Figure \ref{ProcessingSchemes} (a) shows three seperate processing strategies that might be employed on CAS systems.  The first approach, here referred to as Full-CPI (FCPI) processing, coherently processes all $N$ pulses in the pulse train waveform and is essentially the approach proposed by Hickman and Krolik \cite{Krolik}.  A bank of MF's tuned to different Doppler scaling factors are generated for each waveform and combined contiguously in time.  For the first PRI, the MF bank is ordered from pulse 1 to pulse $N$.  For the second PRI, the MF Bank is ordered from pulse 2, 3, ... $N-1$, $N$, 1.  For each subsequent PRI, the time-contiguous MF bank is circularly shifted in time by PRI seconds.  This FCPI processing revisits the target scene every $T_{PRI}$ seconds while coherently processing a pulse train waveform with CPI $NT_{PRI}$.  This processing in turn achieves maximum PCG and TBP.  However, FCPI processing is the least tolerant of target acceleration.  The SNR loss at the output of the MF for FCPI processing could substantially hamper detection performance.  
\vspace{1em}

Target acceleration can be accounted for by designing a series of MF banks tuned to particular target velocities and accelerations.  This method, while effective, substantially increases receiver complexity \cite{Rihaczek}.  Another solution is to exploit the near-orthogonality of the GSFM waveforms for the processing strategy shown in Figure \ref{ProcessingSchemes} (b).  Each pulse can be processed separately in an approach referred to here as Single Pulse CPI (SPCPI) processing.  The PRI of SPCPI processing is still the pulse length $T$.  The individual pulses are of shorter length and therefore more tolerant of target acceleration.  However, there is a cost in employing SPCPI processing.  The CPI is substantially reduced resulting in a lower TBP.  The reduced TBP reduces the detection performance and increases the average sidelobes thus  increasing the potential to mask a weak target in the presence of a stronger one.  However, SPCPI processing will be $N^2$ times more tolerant of target acceleration.  The processing strategies in Figure \ref{ProcessingSchemes} (a) and (b) cover the extreme cases of trading off TBP with acceleration tolerance.  There may be scenarios where target may be undergoing acceleration but the CAS system may also need to detect targets with a moderate dynamic range of target strengths.  Such a scenario might require an even tradeoff between sufficient TBP and acceleration tolerance.  This is exactly the intended scenario for the third processing strategy shown in Figure \ref{ProcessingSchemes} (c).  In this case, $M < N$ pulses are coherently processed.  For the first PRI, the MF bank is ordered from pulse 1, 2, ....M.  For the next PRI, the MF bank is ordered from pulse 2, 3, .... M+1 and so on.  This adaptive CPI approach allows the system designer to choose the proper CPI for a particular scenario.  In a truly adaptive CAS system, the sonar operator will be able to change the CPI on the fly to adapt the processing to best accommodate the current target scene. 

\subsection{Pulse Train Waveforms}
\label{subsec:PulseTrainWaveforms}
\begin{figure}[ht]
\centering
\includegraphics[width=1.0\textwidth]{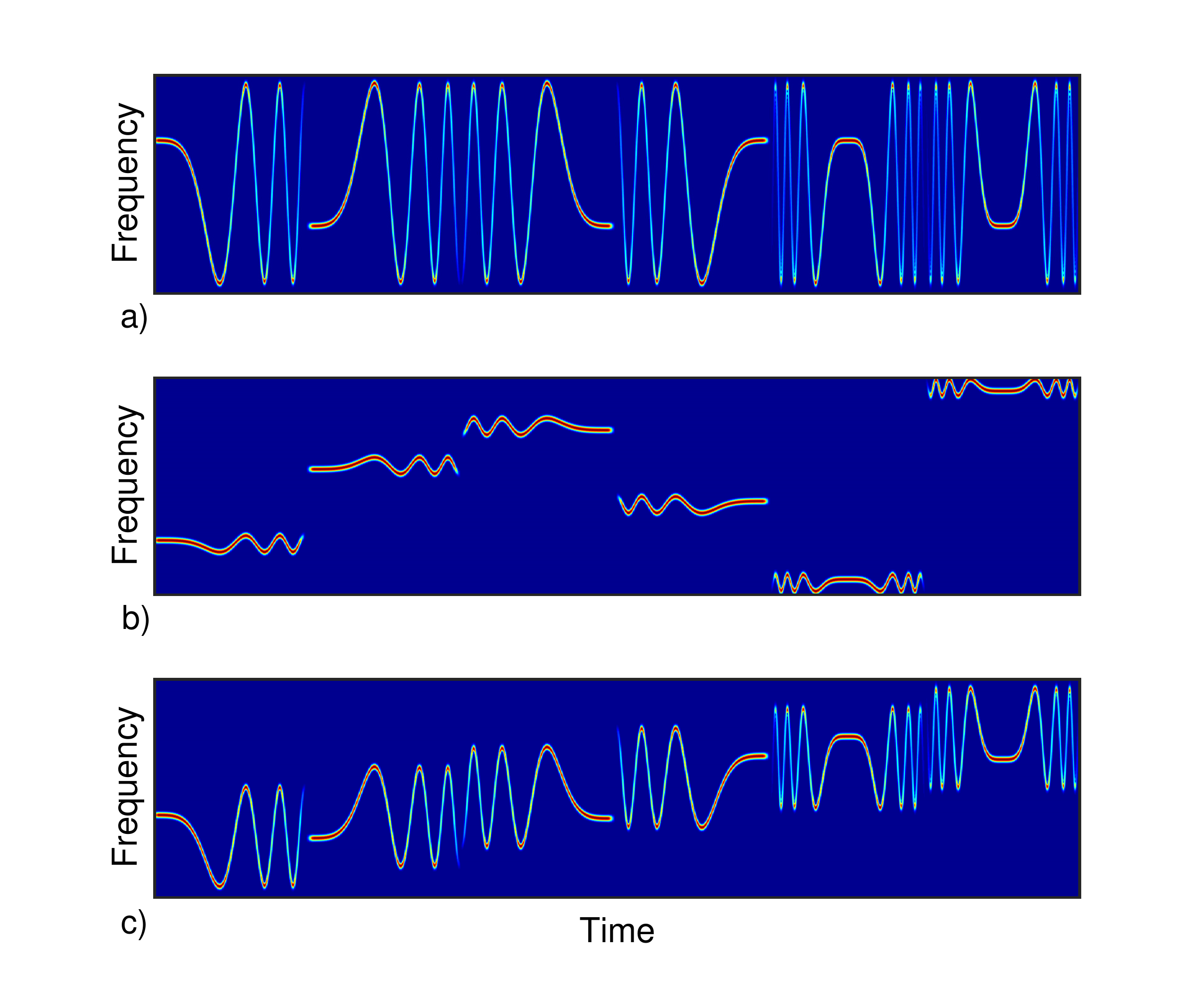}
\caption[Notional spectrograms of the FBPT (a), SBPT (b), and OBPT (c) waveforms.]{Notional spectrograms of the FBPT (a), SBPT (b), and OBPT (c) waveforms.  Both the SBPT and OBPT waveforms employ frequency hopping.  While any full frequency hopping code may be used for these pulse train waveforms, the OBPT utilizing a linear frequency hopping code is especially useful in suppressing the DBL.}
\label{fig:PulseTrains}
\end{figure}

The GSFM pulse train waveform uses two forms of pulse-to-pulse diversity by using different modulation functions derived from different $\alpha$ and $\rho$ values and changing each pulse's center frequency.  Both methods reduce the cross-correlation between each pulse thus removing ambiguities in range and Doppler and reducing cross-correlation between the pulses.  However, it is not immediately obvious which diversity method or combination of diversity methods yields the optimal pulse train waveform for CAS systems.  There are several important design considerations for choosing the proper pulse train waveform.  The first is Instantaneous Bandwidth (IB), the bandwidth of an individual pulse.  The IB determines the lower bound on the TBP for SPCPI and ACPI processing.  The second is the peak cross-correlation between two pulses in the pulse train waveform.  The lower the cross-correlation, the less mutual interference experienced by pulses in the pulse train waveform.  Cross-correlation between waveforms can play a crucial role in CAS system performance.  A CAS system will be continuously transmitting waveforms and receiving echoes from targets.  This means that the receiver will be receiving the transmitted acoustic signal from the transmitter, known as the Direct Blast (DBL), as well as echoes from targets.  Sea experiments have shown that the receiver array can be beamformed to place a null perhaps as deep as -60 dB  in the direction of the transmitter to reduce the signal from the DBL \cite{KrolikI}.  However, for cases where the receiver and transmitter are closely located, if the transmitting array transmits at high source level, then the DBL signal which reaches the receiving array will still overwhelm the steered null of the receiver beam pattern potentially masking the echoes from targets.  This potentially strong source of interference makes a strong case for introducing as much pulse-to-pulse diversity as possible to suppress DBL energy.  Lastly, it is important to discuss which processing strategies work best with a given pulse train waveform.  This work examines three different pulse train waveform designs illustrated in Figure \ref{fig:PulseTrains}.  Here, it is assumed that there is a fixed system bandwidth denoted as $B_{sys}$ that the waveforms will operate within.  The system bandwidth is determined by the frequency response of the transducers in the transmit and receive arrays. 
\vspace{1em}
 
The first pulse train design, referred to as the Full-Band Pulse Train (FBPT), uses $N$ contiguous pulses each of which utilize the full system bandwidth meaning each waveform's IB is $B_{sys}$.  The main feature of the FBPT waveform is that it possesses the largest IB and TBP for a given CPI regardless of the processing strategy employed.  Maximizing the TBP is especially important because all of the pulses occupy the same band of frequencies.  The only way to reduce the max cross-correlation between pulses for the FBPT waveform is to increase the TBP.  The TBP for MCPI processing is expressed as
\begin{equation}
TBP_{FB} = T_{PRI}B_{sys}M
\label{FB}
\end{equation}
Note that the max TBP is achieved when $M=N$ and $TBP_{FB} = T_{PRI}B_{sys}N$.
The only potential drawback to the FBPT is when the DBL is much stronger than any of the target echoes.  In that case, the DBL could mask target echoes.  Otherwise, in the absence of a strong DBL signal, the FBPT gives the best IB and TBP for any of the three processing strategies.  The second pulse train waveform, referred to as the Seperate Band Pulse Train (SBPT) waveform, divides the system bandwidth evenly between $N$ pulses resulting in a waveform bandwidth (IB) of $B_{sys}/N$.  Each pulse occupies a separate band of frequencies and their respective center frequencies can be hopped using any full hopping code \cite{Ricker}.  The SBPT waveform is well suited for any of the three processing strategies.  Additionally, the frequency diversity of the pulses results in low cross-correlation between pulses which is especially useful for suppressing the DBL.  This comes at the cost of reduced cross-correlation.  The TBP for the SBPT using MCPI processing is expressed as 
\begin{equation}
TBP_{SB} = \dfrac{T_{PRI}B_{sys}M^2}{N}
\label{SB}
\end{equation}
Note that when $M=N$ the TBP of the SBPT waveform is $T_{PRI}B_{sys}N$, the same value as the FBPT waveform.  As $M$ decreases however, the TBP is much smaller than the FBPT's TBP.  The SBPT waveform can help mitigate interference from the DBL, but can suffer from low TBP when using MCPI processing. 
\vspace{1em}

The first two pulse train designs went from using pulses that utilize the full system bandwidth to completely seperate bands.  The third pulse train waveform, the Overlapping Band Pulse Train (OBPT) waveform, exists between these two extremes.  The OBPT also utilizes frequency hopped pulses, but these pulses occupy overlapping frequency bands.  The OBPT is illustrated in Figure \ref{fig:PulseTrains} (c).  The OBPT uses pulses whose IB is a fraction of the system bandwidth expressed as $B_{sys}/K$ where $K$ is a scalar and $1 < K < N$.  Each pulse is hopped in frequency such that the entire pulse train's bandwidth equals $B_{sys}$.  For ACPI processing, the TBP of the OBPT waveform is  
\begin{equation}
TBP_{OB} = T_{PRI}\left(\dfrac{B_{sys}}{K} + \Delta f M\right) M
\label{OB}
\end{equation}
where $\Delta f$ is the minimum frequency spacing between pulses.  Again, when $M=N$, the TBP equals that of the FBPT and SBPT waveforms' TBP.  Since $K$ is always less than $N$, when $M<N$, the TBP of the OBPT waveform will always be greater than the SBPT waveform, but always smaller than the FBPT waveform.  While any frequency hopping code can in theory be applied to the pulse train, one hopping code of particular interest is the linear hop code as shown in Figure \ref{fig:PulseTrains} (c).  As will be shown in a later section, this pulse train waveform's frequency diversity suppresses the DBL and is well suited for all three processing strategies. 

\clearpage

\section{Simulations and Results}
\label{sec:CAS_Results}
This section presents a series of proof of concept simulations of the GSFM pulse train waveforms and processing strategies.  

\subsection{An Illustrative Example}
\label{subsec:IllustrativeExample}
\begin{figure}[ht]
\centering
\includegraphics[width=1.0\textwidth]{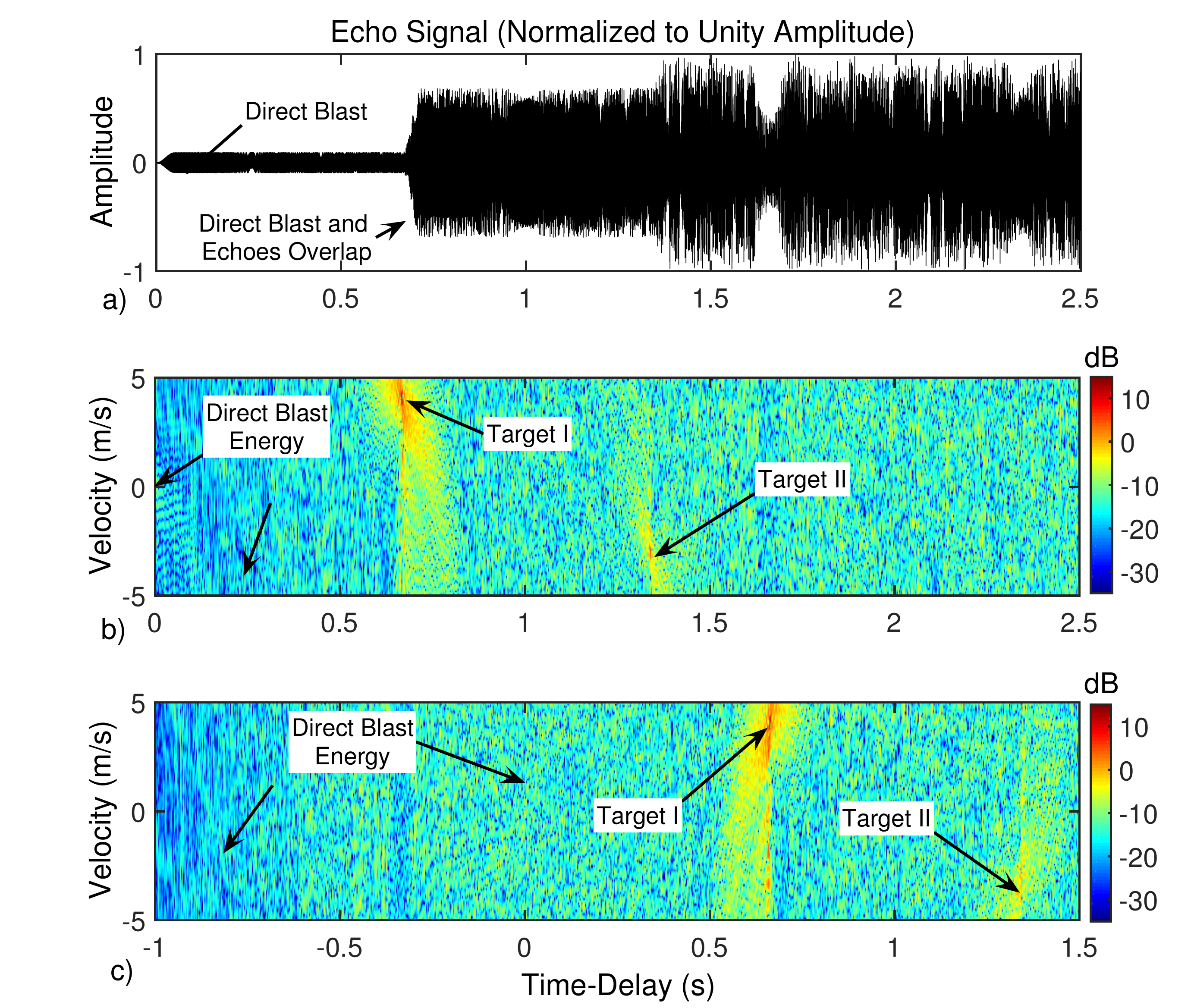}
\caption[Illustration of the CAS processing scheme.]{Illustration of the CAS processing scheme showing the echo signal (a) and the MF banks output from the first (b) and second (c) pulses normalized to the SL of the DBL at the receiver.  Note the change in time axis from (b) to (c).}
\label{POC}
\end{figure}

This simulation uses a FBPT waveform composed of two families of six GSFM waveforms each of duration $T = 1.0$ s, bandwidth $B = 900$ Hz, and center frequency $f_c = 2000$ Hz with SPCPI processing.  The resulting CAS pulse train waveform is therefore $12.0$ seconds long in duration.  The CPI and PRI are $1.0$ second and the target scene will be revisited every $1.0$ second.  This and later simulations will only include the DBL and not reverberation from the medium so as to highlight just the interference effect of the DBL on the echo signal data.  The transmitter and receiver are assumed to be spaced 10 m apart and the DBL Source Level (SL) is 185 dB (re 1$\mu$Pa @ 1m).  Additionally, the receiver array is assumed to have a null placed in the direction of the transmitter with a notch depth of 60 dB.  Assuming cylindrical spreading, the Received Level (RL) of the DBL as it reaches the receiver array is 175 dB and the beampattern notch reduces this to 115 dB.  There are two targets in this environment modeled as spherical point targets with radii of 2 m and therefore a Target Strength (TS) of 0 dB.  The first target is closing at 4 m/s with range $500$ m from the platform.  The second target is receding at $3$ m/s at range $1000$ m.  Assuming cylindrical spreading, the targets echo strengths are -54 dB and -60 dB down from the DBL respectively.  Therefore, the target echo levels are 16 dB and 10 dB above the DBL that reached the receiver array respectively.  Figure \ref{POC} shows the echo signal data and the output of the first two MF bank channels.  The DBL that reaches the array is much weaker than the target echoes.  As a result of this, both targets are clearly visible in the MF bank output. The time axis for the second MF channel is delayed by $1.0$ seconds.  This means that the target information from the second pulse in the pulse train appears $1.0$ seconds after the target information from the first pulse demonstrating revisiting the target scene every $1.0$ seconds. 

\subsection{Suppressing the Direct Blast}
\begin{figure}[ht]
\centering
\includegraphics[width=1.0\textwidth]{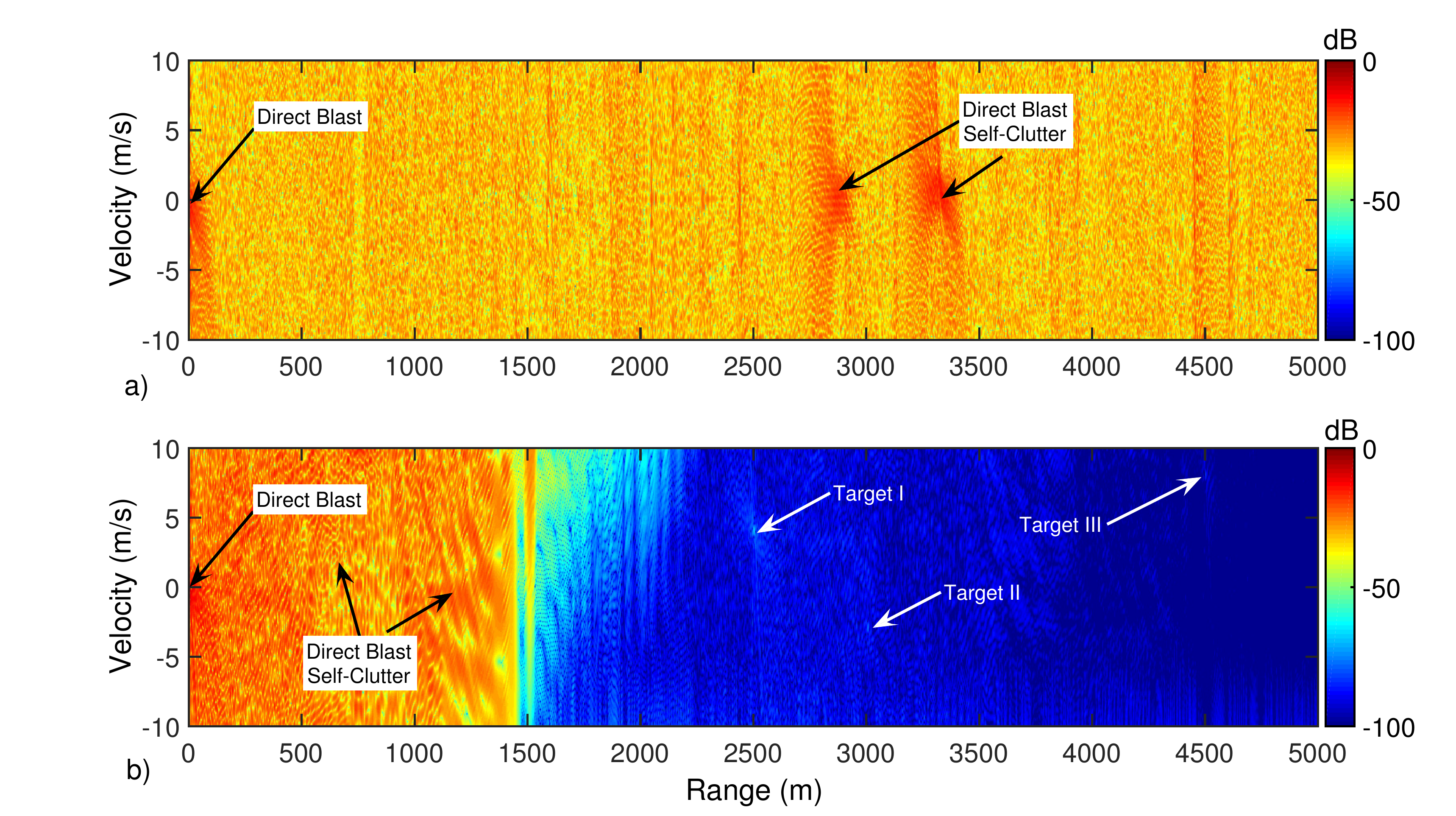}
\caption[Illustration of DBL suppression by (a) the FBPT waveform and (b) the OBPT waveform.]{Illustration of DBL suppression by (a) the FBPT waveform and (b) the OBPT waveform.  The FBPT waveform's DBL completely masks all three targets.  The OBPT waveform's DBL energy is substantially suppressed allowing to detect the three targets.}
\label{DBS}
\end{figure}

In the previous simulation, the two target echoes were much  stronger than the DBL signal at the receiver and were clearly visible.  However, a target scene may contain objects that are much smaller in size and more distant in range resulting in a substantially weaker echo signal.  Without the notch in the receiver beam pattern, a much weaker target could be completely masked by the DBL energy.  In this scenario, the transmitter and receiver are 1 m apart and the receiver array does not place a notch in the direction of the transmitter.  The transmitter's SL is 175 dB which is also the same RL of the DBL as it reaches the receiver array.  This simulation uses a OBPT waveform with SPCPI processing to compare against the FBPT waveform from the last simulation.  The OBPT waveform is composed of two families of six GSFM waveform waveforms each of duration $T = 1.0$ s and an IB of $130$ Hz.  The first pulse is centered at  $1615$ Hz and each successive pulse is stepped up in frequency by $70$ Hz.  The target scene is composed of three targets.  The first target is at a range of $2500$ m and is closing at $4.0$ m/s, with a TS of 0 dB.  The second target is at a range of $3000$ m and is receding at $-3$ m/s with a TS of $-6$ dB.  The third target is at a range of $4500$ m and is closing at $8.0$ m/s with a TS of $-10$ dB.  Again assuming cylindrical spreading, the three targets have echo levels -66 dB, -74 dB, and -83 dB below the DBL at the receiver.  Figure \ref{DBS} compares the first SPCPI MF bank outputs from both the FBPT and OBPT waveforms.  The FBPT waveform's DBL completely masks the targets.  However, the OBPT waveform's DBL energy is sufficiently suppressed so that all three targets are clearly visible.  The suppression of the DBL is a direct result of the frequency hopping employed by the OBPT waveform.

\subsection{A Comparison of all Three Processing Strategies}
\begin{figure}[ht]
\centering
\includegraphics[width=1.0\textwidth]{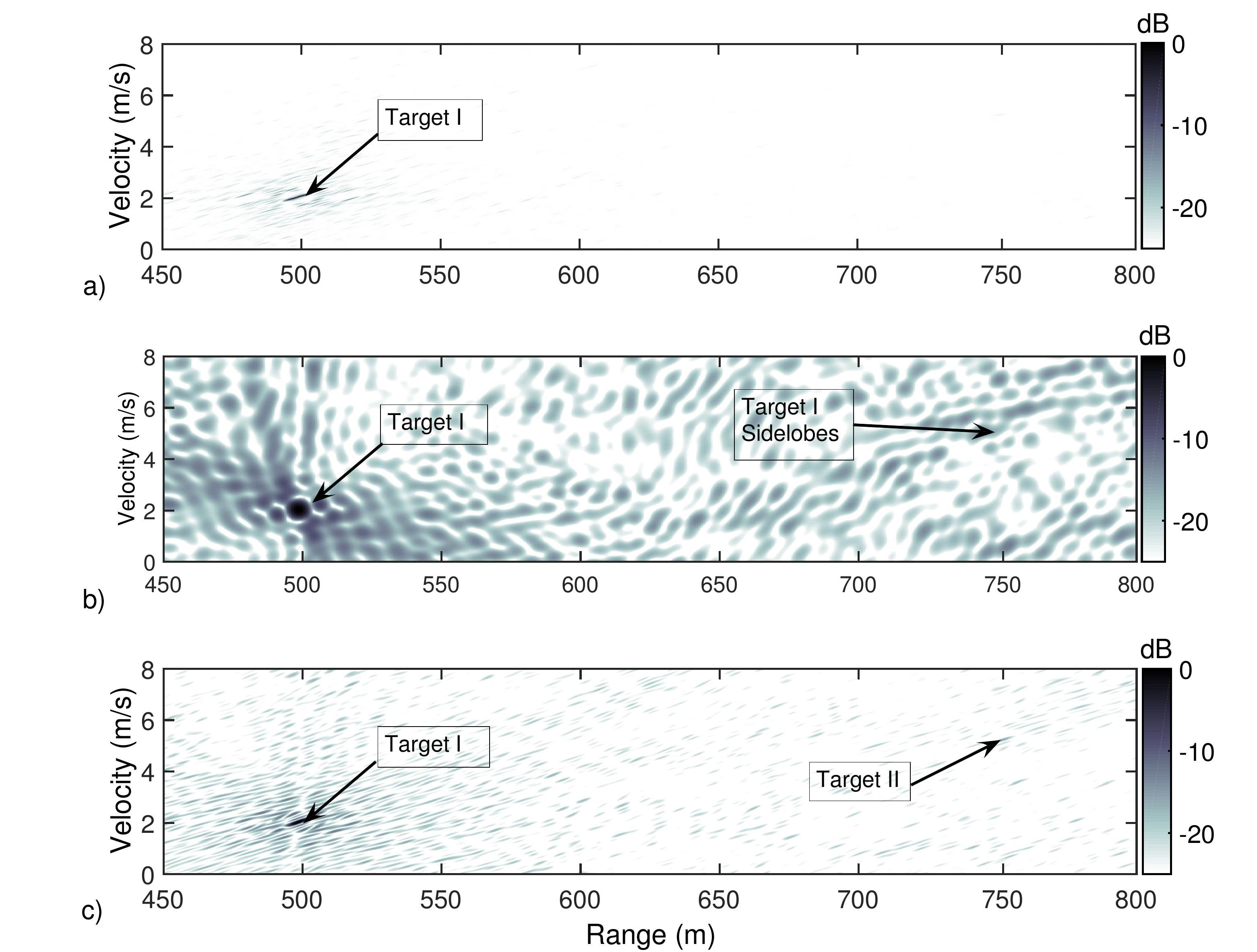}
\caption[Outputs of MF banks for (a) FCPI processing, (b) SPCPI processing, and (c) ACPI processing all normalized to the strongest target output.]{Outputs of MF banks for (a) Full CPI processing, (b) Single Pulse CPI processing, and (c) Adaptive CPI processing all normalized to the strongest target output.  Full CPI processing is too sensitive to acceleration to detect the second target.  Single Pulse CPI processing, while acceleration tolerant, had high sidelobes and masked the second target, a direct result of possessing a low Time Bandwidth Product.  The Adaptive CPI method, which coherently processes 4 pulses, possessed both sufficiently high Time Bandwidth Product and acceleration tolerance to distinguish the weak accelerating target.}
\label{fig:Acceleration}
\end{figure}

The previous two simulations have utilized SPCPI processing throughout.  This simulation uses the OBPT waveform from the previous simulation and compares each processing strategy for a new target environment.  The target scene is composed of two targets.  The first target is closing at 2 m/s with a range of 500 m and TS of 0 dB.  The second target is closing at 5 m/s and accelerating at 0.005 m/s$^2$ with a range of 750 m.  Additionally, the second target's echo level is 15 dB weaker than the first target. Figure \ref{fig:Acceleration} shows the MF bank outputs from all three processing strategies.  With FCPI processing, the first target is clearly visible.  However, the second accelerating target, while only 15 dB weaker than the first target, is not visible.  This is because FCPI processing is sensitive to target acceleration.  FCPI processing, which in this case possessed a CPI of the full 12.0 second waveform, resulted in a 10 dB SNR loss due entirely to mismatch between the MF bank and the accelerating target's echo.  The SPCPI processing method used a CPI of 1.0 second.  While the SPCPI approach is the most tolerant of acceleration, it does not detect the second target.  The TBP of the SPCPI processing is the smallest of the three processing strategies.  As a result of this, the BAAF sidelobes are higher and the first target's sidelobes completely mask the weaker second target.  The ACPI processing method coherently processes four of the twelve pulses in the pulse train waveform for a CPI of 4.0 seconds.  The ACPI method therefore possesses a moderate TBP and therefore lower sidelobe levels while also maintaining acceleration tolerance.  As a result of these properties, the ACPI method is able to distinguish the weaker accelerating target.

\clearpage

\section{Conclusion}
\label{Conclusion}
This chapter presents a proof of concept CAS system using adaptive processing strategies and pulse train waveforms composed of a family of nearly orthogonal GSFM waveforms.  The pulse train designs coupled with ACPI processing facilitate revisiting the target scene every PRI while also trading off CPI and therefore TBP for target acceleration tolerance.  The frequency hopped pulse trains substantially reduce the cross correlations between pulses which greatly aids in suppressing the DBL that would otherwise mask target echoes.  Future work will focus on evaluating these CAS pulse train waveforms and ACPI processing in experimental trials.   The data from these trials will allow for a comprehensive comparison of the GSFM pulse trains with those developed by \cite{Krolik}.


\chapter{Conclusion}
\label{ch:Conclusion}

The GSFM waveform modifies the SFM waveform to use an Instantaneous Frequency (IF) function that resembles the time-voltage characteristic of a FM chirp waveform. As a result of this the GSFM possesses a thumbtack AF.  This is a drastic improvement over the SFM which suffers from poor range resolution as the ACF contains many ambiguous peaks, a direct result of the periodicity of the SFM's IF.  The results of Chapter \ref{ch:GSFM_Eval_AF} show that the GSFM waveform achieves minimal range-Doppler coupling for single target measurements which in turn minimizes the error in jointly estimating target range and velocity. Additionally, a relatively broad range of GSFM parameters $\alpha$ and $\rho$ achieve lower PSL's than the BPSK and QPSK waveforms for a TBP of 50 and the Costas waveform for TBPs less than 125.  Utilizing an MMF for the GSFM reduces the PSL in exchange for SNRL and a wider AF mainlobe.  The PSLs were lower than the Costas waveform for all TBPs and were comparable to the BPSK and QPSK waveforms for TBPs up to 250.  
\vspace{1em}

Chapter \ref{ch:GSFM_Eval_Prac} showed that the GSFM attains both a compact spectrum and a low PAPR.  Combining these two properties results in an energy efficiency that outperforms the Costas, BPSK, and QPSK waveforms for all the TBPs tested in this dissertation.  Additionally, the transducer replicas showed that the GSFM's AF shape is largely unaltered when transmitted on a piezoelectric transducer, the most common transmit/receive device for active sonar systems.  Chapter \ref{ch:GSFM_CAS} presented a proof of concept CAS system using adaptive processing strategies and pulse train waveforms composed of a family of nearly orthogonal GSFM waveforms.  The pulse train designs coupled with  processing approach that revisits the target scene every PRI while also trading off CPI and therefore TBP for target acceleration tolerance.  The linear frequency hopped pulse trains substantially reduces the cross correlations between pulses which greatly aids in suppressing the DBL that would otherwise mask target echoes.
\vspace{1em}

There are a number of directions in which to pursue future work with the GSFM waveform for both PAS and CAS systems.  The first avenue is to evaluate the GSFM's RS performance through experimental sea trials in a manner similar to that performed by Ward in \cite{Ward} for a stationary and moving platforms.  This analysis should be performed for both monostatic and bistatic sonar sytems.  The second obvious avenue to pursue is to evaluate the pulse train waveform designs and subsequent processing for CAS systems with sea trial data.  If these experiments prove that the GSFM pulse train waveform designs and processing can successfully detect and resolve targets while revisiting the target scene often, the next stage of this work would focus on implementing these designs on practical sonar systems for the U.S. Navy.  Lastly, the high duty cycle GSFM pulse train designs attain high PRF's while also maintaining unambiguous range measurements and may also prove useful for radar and ultrasound systems.  Many Pulse-Doppler radar systems currently must trade-off high PRFs in exchange for unambiguous range measurements \cite{Richards}.  Medical ultrasound systems use pulses of short duration so as to avoid a phenomenon known as eclipsing \cite{Levanon} where the system receives echoes from objects of interest before the system has finished transmitted the waveform.  As a result of using short duration waveforms, most ultrasound waveforms are not sensitive enough to provide Doppler information on objects of interest.  Perhaps applying some of the CAS waveform designs and processing techniques will allow Ultrasound systems to extract Doppler information from objects of interest.  These avenues are all currently being pursued by the author.  
\vspace{1em}

Additionally, this dissertation generated several new fundamental research ideas that the author intends to pursue in the near future.  These ideas can be categorized into two separate topics. The first topic focuses on the idea of applying new MMF methods to reduce off-axis sidelobes in a waveform's AF.  As was mentioned in Chapter \ref{ch:GSFM_Eval_AF}, the tapering in frequency and time reduced the sidelobes in time-delay and Doppler respectively.  However, it's not guaranteed that the off axis sidelobes are suppressed using this tapering.  One idea is to analyze off-axis sidelobes by looking at "slices" of the AF in time-delay and Doppler that pass through the origin as described in \cite{Ricker}.  It may be possible to analyze these individual slices of the AF using a Fractional Fourier Transform type of analysis.  The second main topic encompasses serveral ideas that  focus on the spectrum and AF of the GSFM.  The spectrum of the GSFM \eqref{eq:GSFM_SpectrumI} is a superposition of sinc functions multiplied by GBFs whose arguments are the Fourier coefficients of the waveform's phase function.  Generally speaking, the expression in \eqref{eq:GSFM_SpectrumI} can be used to define the spectrum for any FM waveform whose phase is continuous throughout its duration and its Fourier Series exists.  A similar expression exists for the spectrum of the SFM in \eqref{eq:SFM_Spectrum} using Cylindrical Bessel Functions of the first kind and is a special case of the GSFM's spectrum.  The relation in \eqref{eq:SFM_Spectrum} was originally used to derive Carson's Bandwidth Rule for the SFM \cite{Couch}.  This suggests that perhaps a similar process exists to rigorously define Carson's Bandwidth rule for the GSFM and any general FM waveform.  Correspondingly, the NAAF and BAAF of the GSFM waveform are the NAAF and BAAF of any FM waveform whose Fourier Series exists.  An arbitrary set of coefficients can synthesize an arbitrary FM waveform whose phase and IF functions are continuous and also attain a constant envelope resulting in a low PAPR.  The Fourier coefficients of a waveform's phase play a direct role in determining overall Spectrum and AF shape as well.  This raises several intriguing questions.  First, what are the "correct" Fourier coefficients that generate a desired Spectrum/AF shape?  Secondly, how optimal are the GSFM's Fourier coefficients to attaining a thumbtack AF shape?  Lastly, can optimization methods be applied to this Fourier Series representation to synthesize novel active sonar waveforms that have specific user defined properties?  Answering these research questions may provide the ability to perform adaptive waveform optimization "in-situ" to optimize performance on the fly as the sonar system's mission requires.

\newpage
\bibliographystyle{unsrt}
\addcontentsline{toc}{chapter}{Bibliography}

\begin{appendices}

\addtocontents{toc}{\def\protect\cftchappresnum{\appendixname{} }}

\begin{appendix}
\chapter{Derivation of the Spectrum of the SFM and GSFM Waveforms}
\label{ch:appendix_A}

\section{The SFM's Spectrum}
\label{sec:SFMSPec}
Using the waveform definition in (\ref{eq:ComplexExpo}) over the interval $-T/2 \leq t \leq T/2$ with rectangular window tapering function with height $\frac{1}{\sqrt{T}}$ to ensure unit energy, the Fourier transform of the SFM waveform \eqref{eq:SFM_Phase} is expressed as
\begin{eqnarray}
S_{SFM}\left(f\right) = \dfrac{1}{\sqrt{T}}\int_{-T/2}^{T/2}e^{j\beta \sin\left(2 \pi f_m t\right)}e^{j2\pi f_c t}e^{-j2\pi f t}dt
\end{eqnarray}
Using the Jacobi-Anger expansion \cite{Abramowitz}, the expression is simplified to
\begin{equation}
S_{SFM}\left(f\right) = \dfrac{1}{\sqrt{T}}\sum_{n=-\infty}^{\infty}J_n\{\beta\} \int_{-T/2}^{T/2} e^{-j 2 \pi \left(f - f_c - f_m n \right)t}dt
\end{equation} 
Carrying out the integral and utilize the frequency shift Fourier transform property yields the result
\begin{equation}
S_{SFM}\left(f\right) = \sqrt{T}\sum_{n=-\infty}^{\infty}J_n\{\beta\} \sinc\left[\pi T \left(f - f_c - f_m n \right) \right]
\label{eq:SFM_Spec_Der}
\end{equation}

\section{The GSFM's Spectrum}
\label{sec:NonSymGSFMSpec}
This section derives the expressions for the Fourier Transforms of the GSFM waveforms with non-symmetric and even-symmetric IF functions.  Note that while these derivations use the GSFI phase GSFM in \eqref{GSFM_Phi_I}, the same analysis can be applied to the GSFM waveforms using the phase expressions of \eqref{GSFM_Phi_II} and \eqref{GSFM_Phi_III}. 

\subsection{GSFM with Non-Symmetric IF Function}
\label{subsec:GSFM_Spec_I}
For the spectrum of the non-symmetric IF GSFM waveform we use the waveform definition in (\ref{eq:ComplexExpo}) over the interval $0 \leq t \leq T$ with a rectangular window tapering function with height $\frac{1}{\sqrt{T}}$ to ensure unit energy.  The GSFM's IF and Phase functions can be represented using a Fourier Series expansion expressed as
\begin{multline}
f_{GSFM}\left(t\right) = \left(\dfrac{\Delta f}{2}\right)\sin\left(2 \pi \alpha t^{\rho}\right) = \\ \left(\dfrac{\Delta f}{2}\right) \left[ \frac{a_0}{2} + \sum_{m=1}^{\infty} a_m \cos \left(\dfrac{2\pi m t}{T} \right) + b_m \sin \left(\dfrac{2\pi m t}{T} \right)\right] + f_c
\label{eq:GSFM_IF_I}
\end{multline} 
where $a_m$ and $b_m$ are the Fourier coefficients of $f\left(t\right)$ and $T$, the pulse length, is the fundamental period of the Fourier harmonics.  Integrating \eqref{eq:GSFM_IF_I} and multiplying by $2 \pi$ yields the waveform's instantaneous phase
\begin{equation}
\varphi_{GSFM}\left(t\right) = 
\dfrac{\pi \Delta f a_0 t}{2} + A \sum_{m=1}^{\infty} \widetilde{a}_m \sin \left(\frac{2\pi m t}{T}\right) - \widetilde{b}_m \cos \left(\frac{2\pi m t}{T}\right) + 2\pi f_c t
\label{eq:GSFM_PHI_I}
\end{equation} 
where $\widetilde{a}_m = \frac{a_m}{m}$, $\widetilde{b}_m = \frac{b_m}{m}$, and $A = \dfrac{\Delta f T}{2}$. 
The Fourier transform of the GSFM waveform can now be expressed as 
\begin{multline}
S_{GSFM}\left(f\right) = \dfrac{1}{\sqrt{T}}\int_{-T/2}^{T/2} \exp\Biggl\{j\left[A \sum_{m=1}^{\infty}\widetilde{a}_m \sin\left(\dfrac{2\pi m t}{T}\right) - \widetilde{b}_m \cos \left(\frac{2\pi m t}{T}\right) \right]  \Biggr\} \\ \times e^{-j2\pi\left(f - f_c - \frac{a_0 \Delta f}{4}\right) t} dt.
\end{multline}
Utilizing a Jacobi-Anger type expansion for GBF's \cite{Lorenzutta, Dattoli, DattoliII}
\begin{multline}
\exp\Biggl\{j\left[A \sum_{m=1}^{\infty}\widetilde{a}_m \sin\left(\dfrac{2\pi m t}{T}\right) - \widetilde{b}_m \cos \left(\frac{2\pi m t}{T}\right)\right]  \Biggr\} \\ =  \sum_{n=-\infty}^{\infty}\mathcal{J}_n^{1:\infty}\{A \widetilde{a}_m; A\widetilde{b}_m \}e^{\frac{j2\pi n t}{T}}
\end{multline}
where $\mathcal{J}_n^{1:\infty}\{\}$ is the infinite dimensional GBF of the Mixed-Type \cite{DattoliII}, the integral simplifies to
\begin{equation}
S_{GSFM}\left(f\right) = \sum_{n=-\infty}^{\infty}\mathcal{J}_n^{1:\infty}\{A \widetilde{a}_m; A\widetilde{b}_m \}\int_{0}^{T}e^{-j2\pi\left(f - f_c - \frac{a_0 \Delta f}{4} - \frac{n}{T} \right) t}dt.
\end{equation}
Evaluating the integral and taking the modulus yields the result
\begin{multline}
\left|S_{GSFM}\left(f\right)\right| = \left|\sqrt{T}\sum_{n=-\infty}^{\infty} \mathcal{J}_n^{1:\infty}\Bigl\{ \frac{\widetilde{a}_m\Delta f T}{2}; \frac{\widetilde{b}_m\Delta f T}{2} \Bigr\} \right. \\ \left. \times \sinc\left[\pi T \left(f - f_c - \frac{a_0 \Delta f}{4} - \frac{ n}{T} \right)\right]\right|.
\label{eq:GSFM_Spec_Der_I}
\end{multline} 

\subsection{GSFM with Even-Symmetric IF Function}
\label{subsec:EvenSymGSFMSpec}
For the even-symmetric phase GSFM waveform we use the waveform definition in (\ref{eq:ComplexExpo}) over the interval $-T/2 \leq t \leq T/2$ with rectangular window tapering function with height $\frac{1}{\sqrt{T}}$ to ensure unit energy.  The GSFM's IF and Phase functions can be represented using a Fourier Series expansion expressed as
\begin{multline}
f_{GSFM}\left(t\right) = \left(\dfrac{\Delta f}{2}\right)\sin\left(2 \pi \alpha |t|^{\rho}\right) = \left(\dfrac{\Delta f}{2}\right) \left[ \frac{a_0}{2} + \sum_{m=1}^{\infty} a_m \cos \left(\dfrac{2\pi m t}{T} \right) \right] + f_c
\label{eq:GSFM_IF_II}
\end{multline} 
where $a_m$ are the Fourier coefficients of $f\left(t\right)$ and again $T$ defines the period of the Fourier harmonic.  Integrating \eqref{eq:GSFM_IF_II} and multiplying by $2 \pi$ yields the waveform's instantaneous phase
\begin{equation}
\varphi_{GSFM}\left(t\right) = 
\dfrac{\pi \Delta f a_0 t}{2} + A \sum_{m=1}^{\infty} \widetilde{a}_m \sin \left(\frac{2\pi m t}{T}\right) + 2\pi f_c t
\label{eq:GSFM_PHI_II}
\end{equation} 
where $\widetilde{a}_m = \frac{a_m}{m}$ and $A = \dfrac{\Delta f T}{2}$. 
The Fourier transform of the GSFM waveform can now be expressed as 
\begin{multline}
S_{GSFM}\left(f\right) = \dfrac{1}{\sqrt{T}}\int_{-T/2}^{T/2} \exp\Biggl\{j\left[A \sum_{m=1}^{\infty}\widetilde{a}_m \sin\left(\dfrac{2\pi m t}{T}\right) \right]  \Biggr\} e^{-j2\pi\left(f - f_c - \frac{a_0 \Delta f}{4}\right) t} dt
\end{multline}
Utilizing a Jacobi-Anger type expansion for GBF's \cite{Lorenzutta, Dattoli, DattoliII}
\begin{equation}
\exp\Biggl\{j\left[A \sum_{m=1}^{\infty}\widetilde{a}_m \sin\left(\dfrac{2\pi m t}{T}\right) \right]  \Biggr\} = \sum_{n=-\infty}^{\infty}J_n^{1:\infty}\{A \widetilde{a}_m \}e^{\frac{j2\pi n t}{T}}
\end{equation}
the integral simplifies to
\begin{equation}
S_{GSFM}\left(f\right) = \sum_{n=-\infty}^{\infty}J_n^{1:\infty}\{A \widetilde{a}_m\}\int_{-T/2}^{T/2}e^{-j2\pi\left(f - f_c - \frac{a_0 \Delta f }{4} - \frac{n}{T} \right) t}dt
\end{equation}

\begin{equation}
S_{GSFM}\left(f\right) = \sqrt{T}\sum_{n=-\infty}^{\infty} J_n^{1:\infty}\Bigl\{\frac{\widetilde{a}_m\Delta f T}{2}\Bigr\} \sinc\left[\pi T \left(f - f_c - \frac{a_0 \Delta f}{4} - \frac{n}{T} \right)\right]
\label{eq:GSFM_Spec_Der_II}
\end{equation} 
Reassuringly, for the case where $\rho = 1.0$ (i.e. an SFM waverform), the expressions in \eqref{eq:GSFM_Spec_Der_I} and \eqref{eq:GSFM_Spec_Der_II} collapse into the spectrum of the SFM waveform as shown in \eqref{eq:SFM_Spec_Der}.  For the case of a SFM waveform the Fourier Series for the SFM's IF function is 
\begin{equation}  a_m = \left\{
\begin{array}{ll}
      1, & m = 1 \\
      
      0, & otherwise \\
\end{array} 
\right.
\label{eq:SFM_Fourier_Series} 
\end{equation}
Noting an important identity of the GBF's \cite{Lorenzutta}
\begin{equation}
\begin{array}{ll}
J_n^{1:\infty}\{x, 0, ..., 0\} = J_n\{x\} \\
\mathcal{J}_n^{1:\infty}\{x, 0, ..., 0; 0, 0, ..., 0\} = J_n\{x\}
\label{eq:GBF_Identity}
\end{array}
\end{equation}
the GBF's become the one-dimensional Cyclindrical Bessel Function of the First Kind.  Utilizing the identity in \eqref{eq:GBF_Identity} and the fundamental harmonic to $f_m$, the expressions in \eqref{eq:GSFM_Spec_Der_I} and \eqref{eq:GSFM_Spec_Der_II} collapse back into the expression for the SFM's spectrum \eqref{eq:SFM_Spec_Der}.

\section{Fourier Series Coefficients of the GSFM's IF Function}
\label{sec:GSFM_Fourier_Coeff}
This section derives the expressions for the Fourier Series coefficients of the GSFM waveforms with non-symmetric and even-symmetric IF functions.  As with section \ref{sec:NonSymGSFMSpec}, while these derivations use the GSFI phase GSFM in \eqref{GSFM_Phi_I}, the same analysis can be applied to the GSFM waveforms using the phase expressions of \eqref{GSFM_Phi_II} and \eqref{GSFM_Phi_III}. 

\subsection{GSFM with Non-Symmetric IF Function}
\label{subsec:GSFM_Fourier_Coeff_I}
Using the Fourier Series expansion of the GSFM's IF function as shown in (\ref{eq:GSFM_IF_II}) and the even-symmetric integral formula, the Fourier coefficients $a_m$ and $b_m$ of the GSFM's IF function are expressed as 
\begin{equation}
a_0 = \dfrac{1}{T}\int_0^T\sin\left(2 \pi \alpha t^2\right)dt = \dfrac{S\{2\sqrt{\alpha}T\}}{2\sqrt{\alpha}T},
\end{equation}
\vspace{-1em}
\begin{equation}
a_m = \frac{2}{T}\int_0^{T} \sin\left(2\pi\alpha t^2\right)\cos\left(\frac{2\pi mt}{T}\right)dt,
\label{eq:am}
\end{equation}
\begin{equation}
b_m = \frac{2}{T}\int_0^{T} \sin\left(2\pi\alpha t^2\right)\sin\left(\frac{2\pi mt}{T}\right)dt
\label{eq:bm}
\end{equation}
where $S\{\}$ is the sine Fresnel integral.  Starting with the expression for $a_m$, using the trigonometric identity $\sin\theta\cos\varphi = \frac{\sin\left(\theta + \varphi\right) + \sin\left(\theta - \varphi\right)}{2}$, the integral  in \eqref{eq:am} simplifies to
\begin{equation}
a_m = \dfrac{1}{T}\int_0^{T} \sin\left(2\pi\alpha \left(t^2 + \left(\frac{m}{2T\alpha}\right)t\right)\right) + \sin\left(2\pi\alpha\left(t^2 - \left(\frac{m}{2T\alpha}\right)t \right)\right)dt.
\end{equation}
Completing the square
\begin{multline}
a_m = \dfrac{1}{T}\int_0^{T} \sin\biggl\{2\pi\alpha\left[ \left(t + \left(\frac{m}{2T\alpha}\right)\right)^2 -\left(\frac{m}{2T\alpha}\right)^2\right]\biggr\} \\ + \sin\biggl\{2\pi\alpha\left[ \left(t - \left(\frac{m}{2T\alpha}\right)\right)^2 -\left(\frac{m}{2T\alpha}\right)^2\right]\biggr\}dt.
\end{multline}
Now, using the trigonometric identity $\sin\left(\theta \pm \varphi\right) = \sin\theta\cos\varphi \pm \cos\theta\sin\varphi$, $a_m$ can be expanded as
\begin{multline}
a_m = \dfrac{1}{T}\int_0^{T} \cos\left(2\pi\alpha\left(\dfrac{m}{2T\alpha}\right)^2\right)\biggl\{\sin\left[2\pi\alpha A^2\right] + \sin\left[2\pi\alpha B^2\right] \biggr\} \\ - \sin\left(2\pi\alpha\left(\dfrac{m}{2T\alpha}\right)^2\right)\biggl\{\cos\left[2\pi\alpha A^2\right] + \cos\left[2\pi\alpha B^2\right] \biggr\}  dt
\end{multline}
where $A = \left(t+\left(\frac{m}{2T\alpha}\right)\right)$ and $B = \left(t-\left(\frac{m}{2T\alpha}\right)\right)$.  Rearranging the $2\pi\alpha A^2$ and $2\pi\alpha B^2$ terms to $\left(\pi/2\right) 4\alpha A^2$ and $\left(\pi/2\right) 4\alpha B^2$ respectively and using properties of the Fresnel Integrals, $a_m$ is then expressed as
\begin{multline}
a_m = \dfrac{1}{2\sqrt{\alpha}T}\biggl\{\cos\left(2\pi\alpha \left(\dfrac{m}{2T\alpha}\right)^2 \right)\biggl[S\bigl\{2\sqrt{\alpha} z_1 \bigr\} -  S\bigl\{2\sqrt{\alpha}z_2 \bigr\}\biggr]\biggr\}  \\  -\sin\left(2\pi\alpha \left(\dfrac{m}{2T\alpha}\right)^2 \right)\biggl[C\bigl\{2\sqrt{\alpha} z_1 \bigr\} -  C\bigl\{2\sqrt{\alpha}z_2 \bigr\} \biggr] 
\end{multline}
where $z_1 = T + \left(\frac{m}{2T \alpha}\right)$ and $z_2 = T - \left(\frac{m}{2T \alpha}\right)$.
\vspace{1em}

For the expression for $b_m$, using the trigonometric identity $\sin\theta\cos\varphi = \frac{\sin\left(\theta - \varphi\right) + \cos\left(\theta + \varphi\right)}{2}$, the integral  in \eqref{eq:bm} simplifies to
\begin{equation}
b_m = \dfrac{1}{T}\int_0^{T} \cos\left(2\pi\alpha \left(t^2 + \left(\frac{m}{2T\alpha}\right)t\right)\right) + \cos\left(2\pi\alpha\left(t^2 - \left(\frac{m}{2T\alpha}\right)t \right)\right)dt.
\end{equation}
Completing the square
\begin{multline}
b_m = \dfrac{1}{T}\int_0^{T} \cos\biggl\{2\pi\alpha\left[ \left(t + \left(\frac{m}{2T\alpha}\right)\right)^2 -\left(\frac{m}{2T\alpha}\right)^2\right]\biggr\} \\ + \cos\biggl\{2\pi\alpha\left[ \left(t - \left(\frac{m}{2T\alpha}\right)\right)^2 -\left(\frac{m}{2T\alpha}\right)^2\right]\biggr\}dt.
\end{multline}
Now, using the trigonometric identity $\cos\left(\theta \pm \varphi\right) = \cos\theta\cos\varphi \mp \sin\theta\sin\varphi$, $b_m$ can be expanded as
\begin{multline}
b_m = \dfrac{1}{T}\int_0^{T} \sin\left(2\pi\alpha\left(\dfrac{m}{2T\alpha}\right)^2\right)\biggl\{\sin\left[2\pi\alpha A^2\right] + \sin\left[2\pi\alpha B^2\right] \biggr\} \\ - \cos\left(2\pi\alpha\left(\dfrac{m}{2T\alpha}\right)^2\right)\biggl\{\cos\left[2\pi\alpha A^2\right] + \cos\left[2\pi\alpha B^2\right] \biggr\} dt
\end{multline}
where $A = \left(t+\left(\frac{m}{2T\alpha}\right)\right)$ and $B = \left(t-\left(\frac{m}{2T\alpha}\right)\right)$.  Rearranging the $2\pi\alpha A^2$ and $2\pi\alpha B^2$ terms to $\left(\pi/2\right) 4\alpha A^2$ and $\left(\pi/2\right) 4\alpha B^2$ respectively and using properties of the Fresnel Integrals, $b_m$ is finally expressed as
\begin{multline}
b_m = \dfrac{1}{2\sqrt{\alpha}T}\biggl\{\sin\left(2\pi\alpha \left(\dfrac{m}{2T\alpha}\right)^2 \right)\biggl[S\bigl\{2\sqrt{\alpha} z_1 \bigr\} - S\bigl\{2\sqrt{\alpha}z_2 \bigr\} \biggr]  \\  + \cos\left(2\pi\alpha \left(\dfrac{m}{2T\alpha}\right)^2 \right)\biggl[C\bigl\{2\sqrt{\alpha} z_1 \bigr\} -  C\bigl\{2\sqrt{\alpha}z_2 \bigr\}\biggr]\biggr\}
\end{multline}
where $z_1 = T + \left(\frac{m}{2T \alpha}\right)$ and $z_2 = T - \left(\frac{m}{2T \alpha}\right)$.

\subsection{GSFM with Even-Symmetric IF Function}
\label{subsec:GSFM_Fourier_Coeff_II}
Using the Fourier Series expansion of the GSFM's IF function as shown in (\ref{eq:GSFM_IF_II}) and the even-symmetric integral formula, the Fourier coefficients $a_m$ of the GSFM's IF function are expressed as 
\begin{equation}
a_0 = \dfrac{1}{T}\int_{-T/2}^{T/2}\sin\left(2 \pi \alpha t^2\right)dt = \dfrac{S\{\sqrt{\alpha}T\}}{\sqrt{\alpha}T},
\end{equation}
\begin{equation}
a_m = \dfrac{4}{T}\int_0^{T/2} \sin\left(2\pi\alpha t^2\right)\cos\left(\frac{2\pi mt}{T}\right)dt
\end{equation}
Using the trigonometric identity $\sin\theta\cos\varphi = \frac{\sin\left(\theta + \varphi\right) + \sin\left(\theta - \varphi\right)}{2}$, the integral simplifies to
\begin{equation}
a_m = \dfrac{2}{T}\int_0^{T/2} \sin\left(2\pi\alpha \left(t^2 + \left(\frac{m}{2T\alpha}\right)\right)t\right) + \sin\left(2\pi\alpha\left(t^2 - \left(\frac{m}{2T\alpha}\right) \right)t\right)dt.
\end{equation}
Completing the square
\begin{multline}
a_m = \dfrac{2}{T}\int_0^{T/2} \sin\biggl\{2\pi\alpha\left[ \left(t + \left(\frac{m}{2T\alpha}\right)\right)^2 -\left(\frac{m}{2T\alpha}\right)^2\right]\biggr\} \\ + \sin\biggl\{2\pi\alpha\left[ \left(t - \left(\frac{m}{2T\alpha}\right)\right)^2 -\left(\frac{m}{2T\alpha}\right)^2\right]\biggr\}dt.
\end{multline}
Now, using the trigonometric identity $\sin\left(\theta \pm \varphi\right) = \sin\theta\cos\varphi \pm \cos\theta\sin\varphi$, $a_n$ can be expanded as
\begin{multline}
a_m = \dfrac{2}{T}\int_0^{T/2} \cos\left(2\pi\alpha\left(\dfrac{m}{2T\alpha}\right)^2\right)\biggl\{\sin\left[2\pi\alpha A^2\right] + \sin\left[2\pi\alpha B^2\right] \biggr\} \\ -\sin\left(2\pi\alpha\left(\dfrac{m}{2T\alpha}\right)^2\right)\biggl\{\cos\left[2\pi\alpha A^2\right] + \cos\left[2\pi\alpha B^2\right] \biggr\}dt
\end{multline}
where $A = \left(t+\left(\frac{m}{2T\alpha}\right)\right)$ and $B = \left(t-\left(\frac{m}{2T\alpha}\right)\right)$.  Rearranging the $2\pi\alpha A^2$ and $2\pi\alpha B^2$ terms to $\left(\pi/2\right) 4\alpha A^2$ and $\left(\pi/2\right) 4\alpha B^2$ respectively and using properties of the Fresnel Integrals, $a_m$ is then expressed as
\begin{multline}
a_m = \dfrac{1}{\sqrt{\alpha}T}\biggl\{\cos\left(2\pi\alpha \left(\dfrac{m}{2T\alpha}\right)^2 \right)\biggl[S\bigl\{2\sqrt{\alpha} z_1 \bigr\} -  S\bigl\{2\sqrt{\alpha}z_2 \bigr\}\biggr]\biggr\} \\  - \sin\left(2\pi\alpha\left(\dfrac{m}{2T\alpha}\right)^2 \right)\biggl[C\bigl\{2\sqrt{\alpha} z_1 \bigr\} -  C\bigl\{2\sqrt{\alpha}z_2 \bigr\} \biggr] 
\end{multline}
where $z_1 = \left(\frac{T}{2} + \left(\frac{m}{2T\alpha}\right)\right)$ and $z_2 = \left(\frac{T}{2} - \left(\frac{m}{2T\alpha}\right)\right)$.

\section{Derivation of Carson's Bandwidth Rule for the GSFM Waveform}
\label{GSFM_Carson}
In deriving Carson's bandwidth rule for the GSFM, we use the GSFM phase and IF functions defined in \eqref{GSFM_Phi_III} and \eqref{GSFM_IF_III} respectively.  Carson's bandwidth rule \cite{Couch} states that $98\%$ of a FM waveform's energy resides in a bandwidth $B$ expressed as 
\begin{equation}
B = 2\left(\beta + 1\right)B_m = \Delta f + 2 B_m
\label{eq:Carson}
\end{equation}
where $\Delta f$ is the FM waveform's swept bandwidth, $B_m$ is the highest frequency component of the waveform's IF function, and $\beta = \Delta f / 2B_m$ is the Frequency Deviation Ratio (FDR) \cite{Couch}.  To find $B_m$ for the GSFM waveform, we need to find the highest frequency component present in \eqref{GSFM_IF_III} which for the non-symmetric IF GSFM occurs at time at $t = T$, where $T$ is the duration of the waveform.  Finding $B_m$ requires deriving the IF function of the GSFM's IF function \eqref{GSFM_IF_III}, denoted as $f_{IF}\left(t\right)$, which is expressed as  
\begin{equation}
f_{IF}\left(t\right) = \alpha \rho t^{\left(\rho-1\right)}.
\label{eq:IFIF}
\end{equation} 
Evaluating \eqref{eq:IFIF} at $t = T$ yields the result
\begin{equation}
B_m = \alpha \rho T^{\left(\rho-1\right)}.
\label{eq:B_m1}
\end{equation}    
Applying the result in \eqref{eq:B_m1} to \eqref{eq:Carson} results in a $98\%$ bandwidth of 
\begin{equation}
B = \left(\dfrac{\Delta f}{2\alpha \rho T^{\left(\rho - 1\right)}} + 1\right)2\alpha \rho T^{\left(\rho - 1\right)} = \Delta f + 2\alpha \rho T^{\left(\rho - 1\right)}.
\label{eq:GSFM_Carson_I}
\end{equation}
Note that for the case when $\rho = 1$ (i.e. an SFM), $\alpha$ becomes the SFM's modulation frequency $f_m$ and \eqref{eq:GSFM_Carson_I} becomes 
$2\left(\beta + 1\right)f_m$, Carson's bandwidth rule for the SFM waveform. 

\chapter{Derivation of the SFM NAAF and BAAF}
\label{ch:appendix_B}
This chapter derives the NAAF and BAAF for the SFM waveform.  The NAAF of the SFM derived here is a generalization of the result in Cook and Bernfield \cite{Cook} that uses an arbitrary modulation frequency $f_m$.  Cook and Berfield \cite{Cook} gives an expression for the NAAF of the SFM using a specific modulation frequency $f_m = 1/T$.  The result for the BAAF of the SFM appears to be novel.

\section{The SFM NAAF}
\label{sec:SFM_NAAF}
Using the definition of the NAAF, with the basebanded SFM waveform, the product of $s\left(t\right)$ and $s^*\left(t + \tau\right)$ is
\begin{equation}
{s\left(t\right)s^*\left(t+\tau\right)} = \frac{1}{T}e^{j\beta\left[\sin\left(2 \pi f_m t\right) - \sin\left(2\pi f_m \left(t+\tau\right)\right)  \right]}.
\label{eq:SFM_NAAF_I}
\end{equation}  
By using the sum-to-product trigonometric identity, the expression in (\ref{eq:SFM_NAAF_I}) is simplified to
\begin{equation}
{s\left(t\right)s^*\left(t+\tau\right)} = \frac{1}{T}e^{j2\beta \sin \left(- \pi f_m \tau\right) \cos \left(2\pi f_m t+ \pi f_m \tau \right)}.
\end{equation}  
The expression in  is further simplified by the Jacobi-Anger Expansion \cite{Abramowitz}
\begin{equation}
{s\left(t\right)s^*\left(t+\tau\right)} = \frac{1}{T} \sum_{n=-\infty}^{\infty} j^n J_n\{2 \beta \sin \left(- \pi f_m \tau \right) \} e^{j2 \pi f_m n t}e^{j \pi f_m n \tau}.
\end{equation}  
The NAAF can now be expressed as 
\begin{equation}  \chi \left(\tau, \phi\right) = \left\{
\begin{array}{ll}
      \frac{1}{T} \sum_{n=-\infty}^{\infty} j^n J_n\{2 \beta \sin \left(- \pi f_m \tau \right) \}e^{j \pi f_m n \tau} \times \\ \hspace{12em} \int_{\tau}^{T} e^{j 2 \pi \left(f_m n + \phi \right)t} dt & 0 \leq \tau \leq T \\
      
      \frac{1}{T} \sum_{n=-\infty}^{\infty} j^n J_n\{2 \beta \sin \left(- \pi f_m \tau \right) \}e^{j \pi f_m n \tau} \times \\ \hspace{12em}    \int_{-T}^{\tau} e^{j 2 \pi \left(f_m n + \phi \right)t} dt & -T \leq \tau < 0. \\
\end{array} 
\right. 
\end{equation}
Solving for the first integral, 
\begin{multline}
\frac{1}{T} \sum_{n=-\infty}^{\infty} j^n J_n\{2 \beta \sin \left(- \pi f_m \tau \right) \} e^{j \pi f_m n \tau}e^{j \pi \left(f_m n + \phi \right)\left(T - \tau\right)} 
\\ \times  \left[\dfrac{e^{j \pi \left(n f_m + \phi \right) \left(T-\tau\right)}- e^{-j \pi \left(n f_m + \phi \right) \left(T-\tau\right)}}{j2\pi\left(f_m n + \phi \right)}    \right].
\end{multline}
This in turn simplifies to
\begin{equation}
\dfrac{\left(T-\tau\right)}{T}\sum_{n=- \infty}^{\infty}  j^n J_n\{2 \beta \sin \left(- \pi f_m \tau \right) \} e^{j \pi f_m n \tau} e^{j \pi \left(f_m n + \phi \right) \left(T-\tau\right)} \sinc \left[\pi\left(f_m n + \phi\right)\left(T - \tau\right)\right].
\label{eq:SFM_NAAF_First}
\end{equation}
The second integral is 
\begin{multline}
-\dfrac{\left(T-\tau\right)}{T}\sum_{n=- \infty}^{\infty}  j^n J_n\{2 \beta \sin \left(- \pi f_m \tau \right) \} e^{j \pi f_m n \tau} \times ...
\\ e^{-j \pi \left(f_m n + \phi \right) \left(T-\tau\right)} \sinc \left[-\pi\left(f_m n + \phi\right)\left(T - \tau\right)\right].
\label{eq:SFM_NAAF_Second}
\end{multline}
Due to the symmetry property of the NAAF \cite{Levanon}
\begin{equation}
|\chi\left(\tau, \phi\right)| = |\chi\left(-\tau, -\phi\right)|
\end{equation}
the two integral solutions can be combined by changing $\left(T-\tau\right)$ and $-\left(T-\tau\right)$ to $\left(T-|\tau|\right)$.  Finally, taking the modulus of (\ref{eq:SFM_NAAF_First}) and (\ref{eq:SFM_NAAF_Second}) yields the absolute value of the NAAF expressed as 
\begin{equation}
|\chi\left(\tau, \phi\right)| = \dfrac{\left(T-|\tau|\right)}{T} \left| \sum_{n=- \infty}^{\infty}  J_n\{2 \beta \sin \left(\pi f_m \tau \right) \} \sinc \left[\pi\left(f_m n + \phi\right)\left(T - |\tau|\right)\right] \right|.
\label{eq:SFM_NAAF_Der}
\end{equation}

\section{The SFM BAAF}
\label{sec:SFM_BAAF}
Using the definition of the BAAF and the SFM with carrier term included, the product of $s\left(t\right)$ and $s^*\left(\eta\left(t + \tau\right)\right)$ is
\begin{equation}
{s\left(t\right)s^*\left(\eta\left(t+\tau\right)\right)} = \dfrac{\sqrt{\eta}}{T}e^{j\beta\left[\sin\left(2 \pi f_m t\right) - \sin\left(2\pi f_m \eta \left(t+\tau\right)\right)  \right]} e^{-j2 \pi \left(\eta-1\right)f_ct} e^{-j2\pi f_c \eta \tau}
\label{eq:SFM_BAAF_I}
\end{equation}  
By using the sum-to-product trigonometric identity, the expression in (\ref{eq:SFM_BAAF_I}) is simplified to
\begin{multline}
{s\left(t\right)s^*\left(\eta\left(t+\tau\right)\right)} = \dfrac{\sqrt{\eta}}{T}e^{j2\beta \left[ \sin \left(  \pi f_m \left(1-\eta\right)t - \pi f_m \eta \tau \right) + \cos \left(\pi f_m \left( 1+\eta\right)t + \pi f_m \eta \tau  \right) \right]} \times ...
\\ 
e^{-j2 \pi \left(\eta-1\right)f_ct} e^{-j2\pi f_c \eta \tau}.
\label{eq:SFM_BAAF_II}
\end{multline} 
Again, using the Jacobi-Anger expansion, (\ref{eq:SFM_BAAF_II}) now becomes
\begin{multline}
{s\left(t\right)s^*\left(\eta\left(t+\tau\right)\right)} = \dfrac{\sqrt{\eta}}{T} e^{-j2 \pi f_c \eta \tau}  \sum_{n=-\infty}^{\infty} j^n J_n\{2\beta \sin \left(  \pi f_m \left(1-\eta\right)t - \pi f_m \eta \tau \right) \} \times ...
\\ e^{j \pi f_m \eta n \tau} e^{j \pi f_m n\left(1+\eta \right) t} e^{-j2 \pi \left(\eta-1\right)f_ct}.
\label{eq:SFM_BAAF_III}
\end{multline} 
This function's integral does not have a closed form solution.  However, the expression in (\ref{eq:SFM_BAAF_III}) can be simplified to a form whose integral has a closed form solution using an approximation for the Bessel function's argument.  For the velocities encountered by realistic active sonar targets ($\pm$ 25 m/s), the Doppler scaling factor $\eta$ varies between 0.967 and 1.033 making the $\left(1-\eta\right)$ term small ($\pm$ 0.033).  Therefore, the oscillations in time $t$ in the $\sin\left(\pi f_m\left(1-\eta\right)t - \pi f_m \eta \tau\right)$ argument in (\ref{eq:SFM_BAAF_III}) are negligibly small compared to the oscillations in $\tau$.  As a result of this approximation, the dependence of time  $t$ in the $J_n\{2 \beta \sin\left(\pi f_m\left(1-\eta\right)t - \pi f_m \eta \tau\right)\}$ is removed and the only dependence of time in the function is in the exponential function argument, an easily integrable function.  Utilizing this approximation yields
\begin{multline}
{s\left(t\right)s^*\left(\eta\left(t+\tau\right)\right)} \cong \dfrac{\sqrt{\eta}}{T} e^{-j2 \pi f_c \eta \tau}  \sum_{n=-\infty}^{\infty} j^n J_n\{ 2\beta \sin\left(-\pi f_m \eta \tau \right) \} \times ...
\\ e^{j \pi f_m \eta n \tau} e^{j \pi f_m n \left(1+\eta \right) t} e^{-j2 \pi \left(\eta-1\right)f_ct}.
\end{multline} 
The BAAF can now be expressed as 
\begin{equation}  \chi \left(\tau, \eta\right) \cong \left\{
\begin{array}{ll}
      \frac{1}{T} \sum_{n=-\infty}^{\infty} j^n J_n\{ 2\beta \sin\left(-\pi f_m \eta \tau \right) \} e^{j \pi f_m n \eta \tau} \times \\ \hspace{10em} \int_{\tau}^{T} e^{-j 2 \pi \left(\left(\eta-1\right)f_c - f_m n \left(1 +\eta \right)/2  \right)t} dt & 0 \leq \tau \leq T \\
      
      \frac{1}{T} \sum_{n=-\infty}^{\infty} j^n J_n\{2 \beta \sin \left(- \pi f_m \eta \tau \right) \}e^{j \pi f_m n \eta \tau} \times \\ \hspace{10em}    \int_{-T}^{\tau}  e^{-j 2 \pi \left(\left(\eta-1\right)f_c - f_m n \left(1 +\eta \right)/2  \right)t} dt & -T \leq \tau < 0. \\
\end{array} 
\right. 
\end{equation}
Solving for the first integral,
\begin{multline}
\dfrac{\sqrt{\eta}}{T}e^{-j \pi f_c \eta \tau} \sum_{n=-\infty}^{\infty}  j^n J_n\{ 2\beta \sin\left(-\pi f_m \eta \tau \right) \} e^{-j \pi f_m \eta n \tau} e^{-j \pi \left(\left(\eta-1\right)f_c - f_m n \left(1+\eta\right) / 2 \right)\left(T+ \tau \right)} \times ...
\\ 
\dfrac{e^{j \pi \left(\left(\eta-1\right)f_c - f_m n \left(1+\eta\right) / 2 \right))\left(T- \tau \right)} - e^{-j \pi \left(\left(\eta-1\right)f_c - f_m n \left(1+\eta\right) / 2 \right)\left(T- \tau \right)}}{j 2 \pi \left(\left(\eta-1\right)f_c - f_m n \left(1+\eta\right) / 2 \right)} 
\end{multline}
which is simplified to
\begin{multline}
\dfrac{\sqrt{\eta}\left(T-\tau\right)}{T}e^{-j \pi f_c \eta \tau} \sum_{n=-\infty}^{\infty}  j^n J_n\{ 2\beta \sin\left(-\pi f_m \eta \tau \right) \} e^{j \pi f_m \eta n \tau} \times ...
\\  e^{-j \pi \left(\left(\eta-1\right)f_c - f_m n \left(1+\eta\right) / 2 \right)\left(T+ \tau \right)} \sinc \left[ \pi \left(\left(\eta-1\right)f_c - f_m n \left(1+\eta\right) / 2 \right)\left(T-\tau\right) \right].
\end{multline}
The second integral is 
\begin{multline}
\dfrac{-\sqrt{\eta}\left(\tau+T\right)}{T}e^{-j \pi f_c \eta \tau} \sum_{n=-\infty}^{\infty}  j^n J_n\{ 2\beta \sin\left(-\pi f_m \eta \tau \right) \} e^{j \pi f_m \eta n \tau} \times ...
\\  e^{-j \pi \left(f_m n \left(1+\eta\right) / 2  + \left(1-\eta\right)f_c\right)\left(T- \tau \right)} \sinc \left[ \pi \left(\left(\eta-1\right)f_c - f_m n \left(1+\eta\right) / 2 \right)\left(\tau+T\right) \right].
\end{multline}
Again, using symmetry properties of the BAAF, the two integral solutions can again be combined by changing $\left(T-\tau\right)$ and $\left(\tau+T\right)$ to $\left(T-|\tau|\right)$ which results in
\begin{multline}
|\chi\left(\tau, \eta\right)| \cong \dfrac{\sqrt{\eta}\left(T-|\tau|\right)}{T} \left|\sum_{n=- \infty}^{\infty}  J_n\{2 \beta \sin \left(\pi f_m \eta \tau \right) \} \right. \times
\\ \left. \sinc \left[ \pi\left(\left(\eta-1\right)f_c -\dfrac{ f_m n \left(1+\eta \right)}{2} \right)\left(T-|\tau|\right) \right] \right|.
\end{multline}
\end{appendix}

\chapter{Derivation of the GSFM Narrowband Ambiguity Function}
\label{ch:appendix_C}
This section derives the expressions for the NAAF and BAAF of the GSFM waveforms with non-symmetric and even-symmetric IF functions.  As with Appendix \ref{ch:appendix_A}, while these derivations use the GSFI phase GSFM in \eqref{GSFM_Phi_I}, the same analysis can be applied to the GSFM waveforms using the phase expressions of \eqref{GSFM_Phi_II} and \eqref{GSFM_Phi_III}. 

\section{NAAF of GSFM with Non-Symmetric IF Function}
\label{sec:GSFM_NAAF_Non_Sym}
Using the definition of the NAAF, with the basebanded GSFM waveform defined by using the Fourier Series expansion of the instantaneous phase given by \eqref{eq:GSFM_IF_I} and \eqref{eq:GSFM_PHI_I} respectively, the product of $s\left(t\right)$ and $s^*\left(t + \tau\right)$ is
\begin{multline}
{s\left(t\right)s^*\left(t+\tau\right)} = \frac{1}{T} e^{j \pi \Delta f a_0 \tau/2} \\ \times \exp \Biggl\{ jA\sum_{m=1}^{\infty}\widetilde{a}_m\left[\sin\left(\frac{2 \pi m t}{T}\right) - \sin\left(\frac{2\pi m \left(t+\tau\right)}{T} \right)\right] - \\
\widetilde{b}_m \left[\cos\left(\frac{2 \pi m t}{T}\right) - \cos\left(\frac{2\pi m \left(t+\tau\right)}{T} \right) \right] \Biggr\}
\label{eq:GSFM_NAAF_II_Temp}
\end{multline} 
where $A = \left(\frac{\Delta f T}{2}\right)$.  By using the sum-to-product trigonometric identities, the expression in \eqref{eq:GSFM_NAAF_II_Temp} is simplified to
\begin{multline}
{s\left(t\right)s^*\left(t+\tau\right)} = \frac{1}{T} e^{j \pi \Delta f a_0 \tau/2} \\ \times \exp \Biggl\{j2A\sum_{m=1}^{\infty}\widetilde{a}_m\sin\left(\frac{-\pi m\tau}{T}\right)\cos\left(\frac{2\pi m t}{T} +  \frac{\pi m \tau}{T} \right) \\
+\widetilde{b}_m\sin\left(\frac{-\pi m\tau}{T}\right)\sin\left(\frac{2\pi m t}{T} +  \frac{\pi m \tau}{T} \right) \Biggr\}.
\label{eq:GSFM_NAAF_III}
\end{multline} 
The expression in (\ref{eq:GSFM_NAAF_III}) is further simplified using a Jacobi-Anger type expression for Generalized Bessel Functions of the Mixed Type (GBFMT) to
\begin{multline}
{s\left(t\right)s^*\left(t+\tau\right)} = \frac{1}{T} e^{j \pi \Delta f a_0 \tau/2} \\ \times \sum_{n=-\infty}^{\infty}  \mathcal{J}_n^{1:\infty}\Biggl\{2 A \widetilde{b}_m \sin \left(\frac{-\pi m  \tau}{T}\right); 2 A \widetilde{a}_m \sin \left(\frac{-\pi m  \tau}{T}\right)\Biggr\} e^{j\frac{2\pi nt}{T}} e^{\frac{j\pi n \tau}{T}}
\end{multline} 
where $\mathcal{J}_n^{1:\infty}\Bigl\{2 A \widetilde{b}_m \sin \left(\frac{-\pi m \tau}{T}\right); 2 A \widetilde{a}_m \sin \left(\frac{-\pi m \tau}{T}\right)\Bigr\}$ is the Infinite Dimension GBFMT of the first kind \cite{DattoliII}.  The NAAF can now be expressed as 
\begin{equation}  \chi \left(\tau, \phi\right) = \left\{
\begin{array}{ll}
      \frac{e^{j \pi \Delta f a_0 \tau/2}}{T} \sum_{n=-\infty}^{\infty} j^n \mathcal{J}_n^{1:\infty}\{C\} e^{\frac{j\pi n\tau}{T}} \times  \\ \hspace{13em}   \int_{\tau}^{T} e^{j2 \pi \left[ \frac{n}{T} + \phi \right]t} dt & 0 \leq \tau \leq T \\
      
      \frac{e^{j \pi \Delta f a_0 \tau/2}}{T} \sum_{n=-\infty}^{\infty} j^n \mathcal{J}_n^{1:\infty}\{C\} e^{\frac{j\pi n\tau}{T}} \times \\  \hspace{13em} \int_{-T}^{\tau} e^{j2 \pi \left[ \frac{n}{T} + \phi \right]t} dt &  -T \leq \tau < 0. \\
\end{array} 
\right. 
\end{equation}  
where $\mathcal{J}_n^{1:\infty}\{C\} = \mathcal{J}_n^{1:\infty}\Bigl\{2 A \widetilde{b}_m \sin \left(\frac{-\pi m \tau}{T}\right); 2 A \widetilde{a}_m \sin \left(\frac{-\pi m \tau}{T}\right)\Bigr\}$.  Evaluating the integrals and using the same process described in Appendix \ref{sec:SFM_NAAF}, the NAAF of the GSFM is expressed as 
\begin{multline}
\left|\chi \left(\tau, \phi\right)\right| = \left(\dfrac{T - |\tau|}{T} \right) \\ \times \Biggl| \sum_{n=-\infty}^{\infty}\mathcal{J}_n^{1:\infty}\Biggl\{\Delta f T\widetilde{b}_m \sin \left(\frac{\pi m  \tau}{T}\right); \Delta f T \widetilde{a}_m \sin \left(\frac{\pi m  \tau}{T}\right)\Biggr\} \biggr. \\ \biggl. \times \sinc \left[\left(\frac{\pi n}{T} + \phi \right)\left(T - |\tau|\right)\right] \Biggr|.
\label{eq:GSFM_NAAF_Der_I}
\end{multline}

\section{NAAF of GSFM with Even-Symmetric IF Function}
\label{sec:GSFM_NAAF_Even_Sym}
Using the definition of the NAAF, with the basebanded GSFM waveform defined by using the Fourier Series expansion of the instantaneous phase given by \eqref{eq:GSFM_IF_II} and \eqref{eq:GSFM_PHI_II} respectively, the product of $s\left(t\right)$ and $s^*\left(t + \tau\right)$ is
\begin{multline}
{s\left(t\right)s^*\left(t+\tau\right)} = \frac{1}{T} e^{j \pi \Delta f a_0 \tau/2} \\ \times \exp \Biggl\{ jA\sum_{m=1}^{\infty}\widetilde{a}_m\sin\left(\frac{2 \pi m t}{T}\right) - \widetilde{a}_m\sin\left(\frac{2\pi m \left(t+\tau\right)}{T} \right) \Biggr\}
\label{eq:GSFM_NAAF_II}
\end{multline} 
where $A = \left(\frac{\Delta f T}{2}\right)$.  By using the sum-to-product trigonometric identity, the expression in \eqref{eq:GSFM_NAAF_II} is simplified to
\begin{multline}
{s\left(t\right)s^*\left(t+\tau\right)} = \frac{1}{T} e^{j \pi \Delta f a_0 \tau/2} \\ \times \exp \Biggl\{j2A\sum_{m=1}^{\infty}\widetilde{a}_m\sin\left(\frac{-\pi m\tau}{T}\right)\cos\left(\frac{2\pi m t}{T} +  \frac{\pi m \tau}{T} \right) \Biggr\}.
\label{eq:GSFM_NAAF_III}
\end{multline} 
The expression in (\ref{eq:GSFM_NAAF_III}) is further simplified using a Jacobi-Anger type expression for Generalized Bessel Functions (GBF) to
\begin{multline}
{s\left(t\right)s^*\left(t+\tau\right)} = \frac{1}{T} e^{j \pi \Delta f a_0 \tau/2} \\ \times \sum_{n=-\infty}^{\infty} j^n J_n^{1:\infty}\Bigl\{2 A \widetilde{a}_m \sin \left(\frac{-\pi m  \tau}{T}\right); -j, -1, j,...,j^m  \Bigr\} e^{j\frac{2\pi nt}{T}} e^{\frac{j\pi n \tau}{T}}
\end{multline} 
where $J_n^{1:\infty}\{2 A \widetilde{a}_m \sin \left(\frac{-\pi m  \tau}{T}\right); -j, -1, j,...,j^m \}$ is the Infinite Dimension/Index Cylindrical GBF of the first kind \cite{DattoliII}.  The NAAF can now be expressed as 
\begin{equation}  \chi \left(\tau, \phi\right) = \left\{
\begin{array}{ll}
      \frac{e^{j \pi \Delta f a_0 \tau/2}}{T} \sum_{n=-\infty}^{\infty} j^n J_n^{1:\infty}\{C\} e^{\frac{j\pi n\tau}{T}} \times  \\ \hspace{13em}   \int_{\tau}^{T} e^{j2 \pi \left[ \frac{n}{T} + \phi \right]t} dt & 0 \leq \tau \leq T \\
      
      \frac{e^{j \pi \Delta f a_0 \tau/2}}{T} \sum_{n=-\infty}^{\infty} j^n J_n^{1:\infty}\{C\} e^{\frac{j\pi n\tau}{T}} \times \\  \hspace{13em} \int_{-T}^{\tau} e^{j2 \pi \left[ \frac{n}{T} + \phi \right]t} dt &  -T \leq \tau < 0. \\
\end{array} 
\right. 
\end{equation}  
where $J_n^{1:\infty}\{C\} = J_n^{1:\infty}\{2 A \widetilde{a}_m \sin\left(\frac{-\pi m \tau}{T}\right) ; -j, -1, j,...,j^m  \}$.  Evaluating the integrals and using the same process described in Appendix \ref{sec:SFM_BAAF}, the NAAF of the GSFM is expressed as 
\begin{multline}
\left|\chi \left(\tau, \phi\right)\right| = \left(\dfrac{T - |\tau|}{T} \right) \Biggl| \sum_{n=-\infty}^{\infty}J_n^{1:\infty}\Bigl\{\Delta f T \widetilde{a}_m \sin\left(\frac{\pi m \tau}{T}\right); -j, -1, j,...,j^m   \Bigr\} \biggr. \\ \biggl. \times \sinc \left[\left(\frac{\pi n}{T} + \phi \right)\left(T - |\tau|\right)\right] \Biggr|.
\label{eq:GSFM_NAAF_Der_II}
\end{multline}
Again, when $\rho = 1.0$ (i.e. a SFM waveform), the resulting GSFM waveform's Fourier series is given by \eqref{eq:SFM_Fourier_Series}.  Setting the fundamental harmonic to $f_m$ and utilizing the GBF identity in \eqref{eq:GBF_Identity}, the expressions \eqref{eq:GSFM_NAAF_Der_I} and  \eqref{eq:GSFM_NAAF_Der_II} for the NAAF of the GSFM collapse back into the NAAF of the SFM given in \eqref{eq:SFM_NAAF_Der}.

\section{BAAF of GSFM with Non-Symmetric IF Function}
\label{sec:GSFM_BAAF_Even_Sym}
Using the Fourier representation of the GSFM as shown in \eqref{eq:GSFM_PHI_I} with carrier term included and the sum-to-product trigonometric identity, the product of $s\left(t\right)$ and $s^* \left(\eta\left(t+\tau\right)\right)$ is
\begin{multline}
s\left(t\right)s^* \left(\eta\left(t+\tau\right)\right) = \dfrac{\sqrt{\eta}e^{j\pi \Delta f a_0 \eta \tau/2}}{T} e^{-j 2 \pi \left[ \left(\eta - 1\right) f_c + \left(\eta-1\right) \Delta f a_0/4 \right] t}\\ \times \exp\Biggl\{2 A \sum_{m=1}^{\infty} \widetilde{a}_m \sin \left(\frac{\pi m\left(1-\eta\right)t}{T}  -  \frac{\pi m \eta \tau}{T}  \right) \cos\left(\frac{\pi m\left(1+\eta\right)t}{T}  +  \frac{\pi m \eta \tau}{T} \right) \\
-  \widetilde{b}_m \sin \left(\frac{\pi m\left(1-\eta\right)t}{T}  -  \frac{\pi m \eta \tau}{T}  \right) \cos\left(\frac{\pi m\left(1+\eta\right)t}{T}  +  \frac{\pi m \eta \tau}{T} \right)\Biggr\} 
\label{eq:GSFM_BAAF_DerII_I}
\end{multline}
where $A = \frac{\Delta f T}{2}$.  Using the Jacobi-Anger type expression for GBF's and the same approximation used in Appendix \ref{sec:SFM_BAAF}, (\ref{eq:GSFM_BAAF_DerII_I}) is simplified to 
\begin{multline}
s\left(t\right)s^* \left(\eta\left(t+\tau\right)\right) \cong \dfrac{\sqrt{\eta}e^{j\pi \Delta f a_0 \eta \tau/2}}{T} e^{-j 2 \pi \left[ \left(\eta - 1\right) f_c + \left(\eta-1\right) \Delta f a_0/4 \right] t}\\ \times
\sum_{n=-\infty}^{\infty} \mathcal{J}_n^{1:\infty}\Biggl\{2A \widetilde{a}_m \sin\left(\frac{-\pi m \eta \tau}{T}\right); -2A \widetilde{b}_m \sin\left(\frac{-\pi m \eta \tau}{T}\right) \Biggr\} \\ \times e^{j\left(\frac{\pi \left(1+\eta\right) n}{T}\right)t}  e^{\frac{j\pi \eta n \tau}{T}}.
\label{eq:GSFM_BAAF_II}
\end{multline}
The BAAF can now be expressed as 
\begin{equation}  \chi \left(\tau, \eta\right) \cong \left\{
\begin{array}{ll}
      \frac{\sqrt{\eta}e^{j \pi \Delta f a_0 \tau/2}}{T} \sum_{n=-\infty}^{\infty} j^n \mathcal{J}_n^{1:\infty}\{C\} e^{\frac{j\pi \eta n \tau}{T}} \times \\ \hspace{6em}  \int_{\tau}^{T} e^{-j 2 \pi \left[ \left(\eta - 1\right) f_c - \frac{\left(1+\eta\right) n}{2T} + \left(\eta-1\right) \Delta f a_0/4 \right] t} dt & 0 \leq \tau \leq T \\
      \frac{\sqrt{\eta}e^{j \pi \Delta f a_0 \tau/2}}{T} \sum_{n=-\infty}^{\infty} j^n \mathcal{J}_n^{1:\infty}\{C\}e^{j\pi  \left(\frac{\alpha n}{T}\right)\eta\tau} \times \\ \hspace{6em}   \int_{-T}^{\tau} e^{-j 2 \pi \left[ \left(\eta - 1\right) f_c - \frac{\left(1+\eta\right) n}{2T} + \left(\eta-1\right) \Delta f a_0/4 \right] t} dt & -T \leq \tau < 0 \\
\end{array} 
\right. 
\end{equation}
where $\mathcal{J}_n^{1:\infty}\{C\} = \mathcal{J}_n^{1:\infty}\{2A \widetilde{a}_m \sin\left(\frac{-\pi m \eta \tau}{T}\right); -2A \widetilde{b}_m \sin\left(\frac{-\pi m \eta \tau}{T}\right)\}$.  Carrying out the integrals in the same manner as the Appendix \ref{sec:SFM_BAAF}, the BAAF of the Even-Symmetric IF GSFM is expressed as
\begin{multline}
\left|\chi \left(\tau, \eta\right)\right| \cong \dfrac{\sqrt{\eta}\left(T - |\tau|\right)}{T} \times \\ \Biggl| \sum_{n=-\infty}^{\infty}\mathcal{J}_n^{1:\infty}\Biggl\{\Delta fT \widetilde{a}_m \sin\left(\frac{\pi m \eta \tau}{T}\right); \Delta fT \widetilde{b}_m \sin\left(\frac{\pi m \eta \tau}{T}\right) \Biggr\} \times \Biggr. \\ \Biggl.  \sinc \left[\pi \left( \left(\eta - 1\right) \left(f_c + \Delta f a_0/4\right) - \frac{\left(1+\eta\right) n}{2T} \right)\left(T - |\tau|\right)\right] \Biggr|.
\end{multline}

\section{BAAF of GSFM with Even-Symmetric IF Function}
\label{sec:GSFM_BAAF_Even_Sym}
Using the Fourier representation of the GSFM with carrier term included and using the sum-to-product trigonometric identity, the product of $s\left(t\right)$ and $s^* \left(\eta\left(t+\tau\right)\right)$ is
\begin{multline}
s\left(t\right)s^* \left(\eta\left(t+\tau\right)\right) = \dfrac{\sqrt{\eta}e^{j\pi \Delta f a_0 \eta \tau/2}}{T} e^{-j 2 \pi \left[ \left(\eta - 1\right) f_c + \left(\eta-1\right) \Delta f a_0/4 \right] t}\\ \times \exp\Biggl\{2 A \sum_{m=1}^{\infty} \widetilde{a}_m \sin \left(\frac{\pi m\left(1-\eta\right)t}{T}  -  \frac{\pi m \eta \tau}{T}  \right) \\ \times \cos\left(\frac{\pi m\left(1+\eta\right)t}{T}  +  \frac{\pi m \eta \tau}{T} \right)  \Biggr\}
\label{eq:GSFM_BAAF_DerII_I}
\end{multline}
where $A = \frac{\Delta f T}{2}$.  Again, using the Jacobi-Anger type expression for GBF's and the approximation used in Appendix \ref{sec:SFM_BAAF}, (\ref{eq:GSFM_BAAF_DerII_I}) is simplified to 
\begin{multline}
s\left(t\right)s^* \left(\eta\left(t+\tau\right)\right) \cong \dfrac{\sqrt{\eta}e^{j\pi \Delta f a_0 \eta \tau/2}}{T} e^{-j 2 \pi \left[ \left(\eta - 1\right) f_c + \left(\eta-1\right) \Delta f a_0/4 \right] t}\\ \times
\sum_{n=-\infty}^{\infty} j^n J_n^{1:\infty}\Biggl\{2A \widetilde{a}_m \sin\left(\frac{-\pi m \eta \tau}{T}\right); -j, -1, j,...,j^{-m}   \Biggr\} \\ \times e^{j\left(\frac{\pi \left(1+\eta\right) n}{T}\right)t}  e^{\frac{j\pi \eta n \tau}{T}}.
\label{eq:GSFM_BAAF_II}
\end{multline}
The BAAF can now be expressed as 
\begin{equation}  \chi \left(\tau, \eta\right) \cong \left\{
\begin{array}{ll}
      \frac{\sqrt{\eta}e^{j \pi \Delta f a_0 \tau/2}}{T} \sum_{n=-\infty}^{\infty} j^n J_n^{1:\infty}\{C\} e^{\frac{j\pi \eta n \tau}{T}} \times \\ \hspace{6em}  \int_{\tau}^{T} e^{-j 2 \pi \left[ \left(\eta - 1\right) f_c - \frac{\left(1+\eta\right) n}{2T} + \left(\eta-1\right) \Delta f a_0/4 \right] t} dt & 0 \leq \tau \leq T \\
      \frac{\sqrt{\eta}e^{j \pi \Delta f a_0 \tau/2}}{T} \sum_{n=-\infty}^{\infty} j^n J_n^{1:\infty}\{C\}e^{j\pi  \left(\frac{\alpha n}{T}\right)\eta\tau} \times \\ \hspace{6em}   \int_{-T}^{\tau} e^{-j 2 \pi \left[ \left(\eta - 1\right) f_c - \frac{\left(1+\eta\right) n}{2T} + \left(\eta-1\right) \Delta f a_0/4 \right] t} dt & -T \leq \tau < 0 \\
\end{array} 
\right. 
\end{equation}
where $J_n^{1:\infty}\{C\} = J_n^{1:\infty}\{2A \widetilde{a}_m \sin\left(\frac{-\pi m \eta \tau}{T}\right); -j, -1, j,...,j^{-m} \}$.  Finally, carrying out the integrals in the same manner as the Appendix \ref{sec:SFM_BAAF}, the BAAF of the Even-Symmetric IF GSFM is expressed as
\begin{multline}
\left|\chi \left(\tau, \eta\right)\right| \cong \dfrac{\sqrt{\eta}\left(T - |\tau|\right)}{T} \times \\ \Biggl| \sum_{n=-\infty}^{\infty}J_n^{1:\infty}\Biggl\{\Delta f T \widetilde{a}_m \sin\left(\frac{-\pi m \eta \tau}{T}\right); -j, -1, j,...,j^{-m} \Biggr\} \times \Biggr. \\ \Biggl. \sinc \left[\pi \left( \left(\eta - 1\right) \left(f_c + \Delta f a_0/4\right) - \frac{\left(1+\eta\right) n}{2T} \right)\left(T - |\tau|\right)\right] \Biggr|.
\end{multline}

\chapter{Mainlobe Ellipse Results}
\label{ch:appendix_D}
The EOA parameters are derived for the GSFI and GCFI phase GSFM waveforms and uses a rectangular windowed waveform $s\left(t\right)$ as seen in \eqref{eq:ComplexExpo} with the time axis defined to be $-T/2\leq t\leq T/2$.  Defining the waveform in this way produces a GSFM waveform with an even symmetric IF function which results in zero range-Doppler coupling and greatly reduces the complexity in the EOA parameter derivations.  Substituting the expression for this waveform in first the narrowband range-Doppler coupling factor produces
\begin{equation}
\gamma_N = -2\pi \Im \Biggl\{\int_{\Omega_t} ts\left(t\right)\dot{s}^*\left(t\right) dt \Biggr\}  = -2\pi \Im \Biggl\{ \int_{-T/2}^{T/2}t\left[\dot{\varphi}\left(t\right)\right]^2 dt \Biggr\}
\label{NarrowDop}
\end{equation}
Because the IF function is even-symmetric, multiplying by $t$ makes the integral in (\ref{NarrowDop}) odd-symmetric over the interval $\pm T/2$ which in turn evaluates to zero.  The broadband range-Doppler coupling factor is expressed as
\begin{multline}
\gamma_B = \int_{\Omega_t} t |\dot{s}\left(t\right)|^2 dt - \Re \Biggl\{ \int_{\Omega_t} \dot{s}\left(t\right)s^*\left(t\right)dt \times \int_{\Omega_t} -j t \dot{\varphi}\left(t\right) dt \Biggr\} =
\\
\dfrac{1}{T}\int_{-T/2}^{T/2}t\left[\dot{\varphi}\left(t\right)\right]^2 dt - \Re \Biggl\{ \dfrac{1}{T}\int_{-T/2}^{T/2}j \dot{\varphi}\left(t\right)dt \times \dfrac{1}{T}\int_{-T/2}^{T/2} jt \dot{\varphi}\left(t\right) dt \Biggr\}
\label{rangeDop1}
\end{multline}
Note that both terms of $\gamma_B$ in (\ref{rangeDop1}) have an even symmetric IF function multiplied by $t$ making the integral odd-symmetric over the interval $\pm T/2$ which again evaluates to zero.  For any even-symmetric IF function, the range-Doppler coupling factors will always be zero.

\section{EOA Parameters for the GSFI GSFM Waveform}
For the RMS bandwidth $\beta_{rms}$, we use the waveform model in \eqref{eq:ComplexExpo} minus the carrier term which reduces the complexity of the derivation.  The RMS bandwidth is then expressed as 
\begin{multline}
\beta_{rms}^2 = \int_{\Omega_t} | \dot{s}\left(t\right)|^2 dt - \left| \int_{\Omega_t} s\left(t \right) \dot{s}^*\left(t\right) dt \right| ^2\\  = \dfrac{1}{T}\int_{-T/2}^{T/2} \left[\dot{\varphi}\left( t \right)\right]^2 dt - \left| \dfrac{1}{T}\int_{-T/2}^{T/2} j \dot{\varphi}\left(t\right) dt \right| ^2.
\label{eq:RMS_BAND_App}
\end{multline}
Utilizing the definition of the GSFI GSFM waveform's phase \eqref{GSFM_Phi_I} and subsituting into \eqref{eq:RMS_BAND_App} yields 
\begin{equation}
\beta_{rms}^2 = \dfrac{\pi^2 {\Delta f}^2}{T} \int_{-T/2}^{T/2} \sin^2\left(2\pi\alpha t^{\rho}\right)  dt -
\left| \dfrac{-2j \pi\Delta f}{T} \int_0^{T/2} \sin\left(2\pi\alpha t^{\rho}\right) dt \right|^2.
\end{equation}
Because the IF function $\dot{\varphi}\left(t\right)$ is even symmetric, the integrals can be simplified by utilizing the even-symmetric integral formula and the trigonometric identity $\sin^2\theta = \frac{1 - \cos 2\theta}{2}$
\begin{multline}
\beta_{rms}^2 = \dfrac{\pi^2 {\Delta f}^2}{T} \int_0^{T/2} dt -\dfrac{\pi^2 {\Delta f}^2}{T}\int_0^{T/2} \cos\left(4\pi\alpha t^{\rho}\right)dt - \\
 \left| \dfrac{-2j \pi\Delta f}{T} \int_0^{T/2} \sin\left(2\pi\alpha t^{\rho}\right) dt \right|^2.
\end{multline}
Substituting $x_1 = 4 \pi \alpha t^{\rho}$ and $x_2 = 2 \pi \alpha t^{\rho}$ into the second and third integrals respectively, the expression simplifies to
\begin{multline}
\beta_{rms}^2 = \dfrac{\pi^2 {\Delta f}^2}{T} \int_0^{T/2} dt - \dfrac{\pi^2 {\Delta f}^2}{T}\int_0^{4 \pi \alpha {\left(T/2\right)}^{\rho}} \dfrac{1}{\rho} \left(\dfrac{1}{4 \pi \alpha} \right)^{\frac{1}{\rho}} x_1^{\left(\frac{1}{\rho}-1\right)}\cos\left(x_1\right)dx_1 - \\ \left| \dfrac{-2j \pi\Delta f}{T} \int_0^{2 \pi \alpha {\left(T/2\right)}^{\rho}} \dfrac{1}{\rho} \left(\dfrac{1}{2 \pi \alpha} \right)^{\frac{1}{\rho}} x_2^{\left(\frac{1}{\rho}-1\right)}\sin\left(x_2\right) dx_2 \right|^2.
\end{multline}
Evaluating the integrals yields the result
\begin{equation}
\beta_{rms}^2 = \dfrac{\pi^2 {\Delta f}^2}{2}\left[1 - 2\dfrac{C\{4 \pi \alpha {\left(T/2\right)}^{\rho}, 1/\rho \}}{\rho T \left(4 \pi \alpha \right)^{\frac{1}{\rho}}} -\dfrac{8 S^2\{2 \pi \alpha {\left(T/2\right)}^{\rho}, 1/\rho \}}{{\left(\rho T\right)}^2 \left(2 \pi \alpha \right)^{\frac{2}{\rho}}} \right]
\end{equation}
where $C\{4 \pi \alpha T^{\rho}, 1/\rho \}$ and $S\{2 \pi \alpha T^{\rho}, 1/\rho \}$ are the Sine/Cosine Generalized Fresnel Integrals respectively \cite{NHMF, Fresnel}.

The Doppler sensitivity factor of the GSFI GSFM waveform $\lambda_N^2$ is expressed as
\begin{equation}
\lambda_N^2 = 4\pi^2\int_{\Omega_t} \left(t-t_0\right)^2 |s\left(t\right)|^2 dt  = \dfrac{4\pi^2}{T}\int_{-T/2}^{T/2}t^2dt 
\label{lambda}
\end{equation}
where $t_0$ is the time centroid of the IF function which is zero for even-symmetric IF functions.  The expression (\ref{lambda}) evaluates to
\begin{equation}
\lambda_N^2 = \dfrac{\pi^2 T^2}{3}.
\end{equation}
The broadband Doppler sensitivity factor $\lambda_B^2$ is expressed as
\begin{multline}
\lambda_B^2 = \int_{\Omega_t} t^2 |s\left(t\right)|^2 dt - \left| \int_{\Omega_t} t s\left(t\right) \dot{s}^*\left(t\right)dt \right| ^2 = \\ 
\dfrac{1}{T}\int_{-T/2}^{T/2}t^2\left[\dot{\phi}\left(t\right)\right]^2dt - \left| \dfrac{1}{T}\int_{-T/2}^{T/2} j t \dot{\phi}\left(t\right) dt \right|^2.
\label{eq:lambaB_1}
\end{multline}
Again, because the IF function $\dot{\varphi}\left(t\right)$ is even symmetric, the first integral can be simplified by utilizing the even-symmetric integral formula.  The second integral becomes an odd-symmetric integrand evaluated over limits that are symmetric across the origin, which therefore evaluates to zero.  The expression in \eqref{eq:lambaB_1} is re-written as 
\begin{equation}
\lambda_B^2=\dfrac{2}{T}\int_{0}^{T/2}t^2\left[\pi \Delta f \sin\left(2 \pi \alpha t^{\rho} \right) + 2 \pi f_c \right]^2 dt.
\end{equation}
Expanding the square,
\begin{equation}
\lambda_B^2=\dfrac{2}{T}\int_{0}^{T/2}t^2\pi^2 {\Delta f}^2 {\sin}^2\left(2 \pi \alpha t^{\rho} \right) + 2{\pi}^2 \Delta f f_c t^2 \sin\left(2 \pi \alpha t^{\rho} \right) + 4 \pi^2 {f_c}^2 t^2  dt.
\label{BroadDopSensitivity1}
\end{equation}
Using the identity $\sin^2\theta = \frac{1 - \cos 2\theta}{2}$ and expanding the terms into separate integrals, the expression in \eqref{BroadDopSensitivity1} is re-written as
\begin{multline}
\lambda_B^2 = \dfrac{\pi^2 {\Delta f}^2}{T}\int_{0}^{T/2}t^2dt + \dfrac{8 \pi^2{f_c}^2}{T}\int_{0}^{T/2}  t^2 dt \\ - \dfrac{\pi^2 {\Delta f}^2}{T}\int_{0}^{T/2}  t^2 \cos \left(4 \pi \alpha t^{\rho} \right) dt \\ + \dfrac{4{\pi}^2 \Delta f f_c}{T}\int_{0}^{T/2}  t^2 \sin\left(2 \pi \alpha t^{\rho} \right) dt.
\end{multline}
Substituting $x_1 = 4 \pi \alpha t^{\rho}$ and $x_2 = 2 \pi \alpha t^{\rho}$ into the third and fourth integrals respectively, the expression simplifies to
\begin{multline}
\lambda_B^2 = \dfrac{\pi^2 {\Delta f}^2}{T}\int_{0}^{T/2}t^2dt + \dfrac{8 \pi^2{f_c}^2}{T}\int_{0}^{T/2} t^2 dt \\ -
\dfrac{\pi^2 {\Delta f}^2}{\rho T}\left(\dfrac{1}{4\pi\alpha}\right)^{\frac{3}{\rho}}\int_{0}^{4\pi \alpha {\left(T/2\right)}^{\rho}}  x_1^{\left(\frac{3}{\rho}-1\right)} \cos\left(x_1 \right)dx_1 \\ + \dfrac{4{\pi}^2 \Delta f f_c}{\rho T}\left(\dfrac{1}{2\pi\alpha}\right)^{\frac{3}{\rho}}\int_{0}^{2\pi \alpha {\left(T/2\right)}^{\rho}}   x_2^{\left(\frac{3}{\rho}-1\right)} \sin\left(x_2 \right)dx_2.
\end{multline}
Evaluating these integrals leads to the final result
\begin{multline}
\lambda_B^2 = \dfrac{\pi^2 f_c^2 T^2}{3} + \dfrac{\pi^2 {\Delta f}^2 T^2}{24} - \dfrac{\pi^2 {\Delta f}^2 C\{4\pi \alpha {\left(T/2\right)^{\rho}}, 3/\rho\}}{\rho T \left(4\pi \alpha\right)^{\frac{3}{p}}} \\ +
\dfrac{2 \pi \Delta f f_c S\{2\pi \alpha {\left(T/2\right)}^{\rho}, 3/\rho\}}{\rho T \left(2\pi \alpha\right)^{\frac{3}{\rho}}}.
\label{eq:lambdaB_2}
\end{multline}
The broadband Doppler sensitivity factor $\lambda_B^2$ can be shown to converge to the narrowband Doppler sensitivity factor $\lambda_N^2$ in the limit that the waveform becomes narrowband (i.e. small bandwidth compared to the center frequency $f_c$) and then invoking the narrowband assumptions used to derive the NAAF.  Defining the fractional bandwidth $\Gamma_B$ as 
\begin{equation}
\Gamma_B = \dfrac{\Delta f}{f_c}
\label{eq:Gamma_B_David}
\end{equation}
where $\Delta f$ is the waveform's bandwidth and $f_c$ is the waveform's carrier frequency and $0 < \Gamma_B \leq 1.0$.  Substiting $\Gamma_B f_c$ for $\Delta f$ into \eqref{eq:Gamma_B_David} gives
\begin{multline}
\lambda_B^2 = \dfrac{\pi^2 f_c^2 T^2}{3} + \dfrac{\pi^2 {\left(\Gamma_B f_c\right)}^2 T^2}{24} - \dfrac{\pi^2 {\left(\Gamma_B f_c\right)}^2 C\{4\pi \alpha {\left(T/2\right)^{\rho}}, 3/\rho\}}{\rho T \left(4\pi \alpha\right)^{\frac{3}{p}}} + \\
\dfrac{2 \pi \Gamma_B {f_c}^2 S\{2\pi \alpha {\left(T/2\right)}^{\rho}, 3/\rho\}}{\rho T \left(2\pi \alpha\right)^{\frac{3}{\rho}}}.
\end{multline}
Taking the limit of $\lambda_B^2$ as $\Gamma_B$ approaches zero yields
\begin{equation}
\lim_{\Gamma_B\to 0} \lambda_B^2 = \dfrac{\pi^2 {f_c}^2 T^2}{3}.
\end{equation}
Recall the Doppler sensitivity term in the Ellipse Of Ambiguity (EOA) is $\lambda_B^2 {\left(\eta - 1\right)}^2$.  The NAAF assumes that the target velocity is low compared to the speed of the medium and that $\eta \cong 1 + \frac{2v}{c}$.  The NAAF formulation also assumes that $\Gamma_B$ is small so that the Doppler effect shifts the target echo in frequency by the Doppler frequency in (6).  Invoking the NAAF assumptions on the EOA Doppler sensitivity term results in 
\begin{equation}
\dfrac{\pi^2 T^2 {f_c}^2}{3} {\left(\dfrac{2v}{c}\right)}^2 = \dfrac{\pi^2 T^2}{3} \phi^2 = \lambda_N^2 \phi^2
\end{equation}
thus resulting in the narrowband Doppler sensitivity term $\lambda_N^2$.

\section{EOA Parameters for the GCFI GSFM Waveform}
Utilizing the definition of the GCFI GSFM waveform's phase \eqref{GSFM_Phi_II} and subsituting into \eqref{eq:RMS_BAND_App} yields 
\begin{equation}
\beta_{rms}^2 = \dfrac{\pi^2 {\Delta f}^2}{T} \int_{-T/2}^{T/2} \cos^2\left(2\pi\alpha t^{\rho}\right)  dt -
\left| \dfrac{-2j \pi\Delta f}{T} \int_0^{T/2} \cos\left(2\pi\alpha t^{\rho}\right) dt \right|^2.
\end{equation}
Because the IF function $\dot{\varphi}\left(t\right)$ is even symmetric, the integrals can be simplified by utilizing the even-symmetric integral formula and the trigonometric identity $\cos^2\theta = \dfrac{1 + \cos 2\theta}{2}$ resulting in the expression
\begin{multline}
\beta_{rms}^2 = \dfrac{\pi^2 {\Delta f}^2}{T} \int_0^{T/2} dt +\dfrac{\pi^2 {\Delta f}^2}{T}\int_0^{T/2} \cos\left(4\pi\alpha t^{\rho}\right)dt \\ -
 \left| \dfrac{-2j \pi\Delta f}{T} \int_0^{T/2} \cos\left(2\pi\alpha t^{\rho}\right) dt \right|^2.
\end{multline}
Substituting $x_1 = 4 \pi \alpha t^{\rho}$ and $x_2 = 2 \pi \alpha t^{\rho}$ into the second and third integrals respectively, the expression simplifies to
\begin{multline}
\beta_{rms}^2 = \dfrac{\pi^2 {\Delta f}^2}{T} \int_0^{T/2} dt + \dfrac{\pi^2 {\Delta f}^2}{T}\int_0^{4 \pi \alpha {\left(T/2\right)}^{\rho}} \dfrac{1}{\rho} \left(\dfrac{1}{4 \pi \alpha} \right)^{\frac{1}{\rho}} x_1^{\left(\frac{1}{\rho}-1\right)}\cos\left(x_1\right)dx_1 - \\ \left| \dfrac{-2j \pi\Delta f}{T} \int_0^{2 \pi \alpha {\left(T/2\right)}^{\rho}} \dfrac{1}{\rho} \left(\dfrac{1}{2 \pi \alpha} \right)^{\frac{1}{\rho}} x_2^{\left(\frac{1}{\rho}-1\right)}\cos\left(x_2\right) dx_2 \right|^2.
\end{multline}
Evaluating the integrals yields the result
\begin{equation}
\beta_{rms}^2 = \dfrac{\pi^2 {\Delta f}^2}{2}\left[1 + 2\dfrac{C\{4 \pi \alpha {\left(T/2\right)}^{\rho}, 1/\rho \}}{\rho T \left(4 \pi \alpha \right)^{\frac{1}{\rho}}} -\dfrac{8 C^2\{2 \pi \alpha {\left(T/2\right)}^{\rho}, 1/\rho \}}{{\left(\rho T\right)}^2 \left(2 \pi \alpha \right)^{\frac{2}{\rho}}} \right]
\end{equation}
where $C\{4 \pi \alpha T^{\rho}, 1/\rho \}$ and $C\{2 \pi \alpha T^{\rho}, 1/\rho \}$ are GCFIs \cite{NHMF, Fresnel}.
\vspace{1em}

The Doppler sensitivity factor of the GCFI GSFM waveform $\lambda_N^2$ is expressed as
\begin{equation}
\lambda_N^2 = 4\pi^2\int_{\Omega_t} \left(t-t_0\right)^2 |s\left(t\right)|^2 dt  = \dfrac{4\pi^2}{T}\int_{-T/2}^{T/2}t^2dt 
\label{lambda}
\end{equation}
where $t_0$ is the time centroid of the IF function which is zero for even-symmetric IF functions.  The expression (\ref{lambda}) evaluates to
\begin{equation}
\lambda_N^2 = \dfrac{\pi^2 T^2}{3}.
\end{equation}
The broadband Doppler sensitivity factor $\lambda_B^2$ for GCFI GSFM waveform is expressed as
\begin{equation}
\lambda_B^2=\dfrac{2}{T}\int_{0}^{T/2}t^2\left[\pi \Delta f \cos\left(2 \pi \alpha t^{\rho} \right) + 2 \pi f_c \right]^2 dt.
\end{equation}
Expanding the square,
\begin{equation}
\lambda_B^2=\dfrac{2}{T}\int_{0}^{T/2}t^2\pi^2 {\Delta f}^2 {\cos}^2\left(2 \pi \alpha t^{\rho} \right) + 2{\pi}^2 \Delta f f_c t^2 \cos\left(2 \pi \alpha t^{\rho} \right) + 4 \pi^2 {f_c}^2 t^2  dt.
\label{BroadDopSensitivity}
\end{equation}
Using the identity $\sin^2\theta = \frac{1 + \cos 2\theta}{2}$ and expanding the terms into separate integrals, the expression in (\ref{BroadDopSensitivity}) is re-written as
\begin{multline}
\lambda_B^2 = \dfrac{\pi^2 {\Delta f}^2}{T}\int_{0}^{T/2}t^2dt + \dfrac{8 \pi^2{f_c}^2}{T}\int_{0}^{T/2}  t^2 dt \\ + \dfrac{\pi^2 {\Delta f}^2}{T}\int_{0}^{T/2}  t^2 \cos \left(4 \pi \alpha t^{\rho} \right) dt \\ + \dfrac{4{\pi}^2 \Delta f f_c}{T}\int_{0}^{T/2}  t^2 \cos\left(2 \pi \alpha t^{\rho} \right) dt.
\end{multline}
Substituting $x_1 = 4 \pi \alpha t^{\rho}$ and $x_2 = 2 \pi \alpha t^{\rho}$ into the third and fourth integrals respectively, the expression simplifies to
\begin{multline}
\lambda_B^2 = \dfrac{\pi^2 {\Delta f}^2}{T}\int_{0}^{T/2}t^2dt + \dfrac{8 \pi^2{f_c}^2}{T}\int_{0}^{T/2} t^2 dt \\ +
\dfrac{\pi^2 {\Delta f}^2}{\rho T}\left(\dfrac{1}{4\pi\alpha}\right)^{\frac{3}{\rho}}\int_{0}^{4\pi \alpha {\left(T/2\right)}^{\rho}}  x_1^{\left(\frac{3}{\rho}-1\right)} \cos\left(x_1 \right)dx_1 \\ + \dfrac{4{\pi}^2 \Delta f f_c}{\rho T}\left(\dfrac{1}{2\pi\alpha}\right)^{\frac{3}{\rho}}\int_{0}^{2\pi \alpha {\left(T/2\right)}^{\rho}}   x_2^{\left(\frac{3}{\rho}-1\right)} \cos\left(x_2 \right)dx_2.
\end{multline}
Evaluating these integrals leads to the final result
\begin{multline}
\lambda_B^2 = \dfrac{\pi^2 f_c^2 T^2}{3} + \dfrac{\pi^2 {\Delta f}^2 T^2}{24} - \dfrac{\pi^2 {\Delta f}^2 C\{4\pi \alpha {\left(T/2\right)^{\rho}}, 3/\rho\}}{\rho T \left(4\pi \alpha\right)^{\frac{3}{p}}} \\ +
\dfrac{2 \pi \Delta f f_c C\{2\pi \alpha {\left(T/2\right)}^{\rho}, 3/\rho\}}{\rho T \left(2\pi \alpha\right)^{\frac{3}{\rho}}}.
\label{eq:lambdaB_2}
\end{multline}
As was shown in the previous section, the broadband Doppler sensitivity factor $\lambda_B^2$ converges to the narrowband Doppler sensitivity factor $\lambda_N^2$ in the limit as the waveform becomes narrowband and then invoking the narrowband assumptions used to derive the NAAF.

\end{appendices}

\end{document}